\documentclass[10pt,journal,compsoc]{IEEEtran}
\usepackage{graphicx,subfigure,amsmath,url,multicol,makecell}
\ifCLASSOPTIONcompsoc
  \usepackage[nocompress]{cite}
\else
  \usepackage{cite}
\fi
\ifCLASSINFOpdf
\else
\fi
\begin{document}
%
\title{Recent Advances of Blockchain and Its Applications}
%
%
%
%

\author{Xiao~Li,
        Weili~Wu,~\IEEEmembership{Senior Member}
\IEEEcompsocitemizethanks{\IEEEcompsocthanksitem X. Li and W. Wu are with the Department
of Computer Science, The University of Texas at Dallas, TX, USA, 75080.\protect\\
E-mail: xiao.li@utdallas.edu and weiliwu@utdallas.edu
}
}

%
%

\markboth{}%
{Shell \MakeLowercase{\textit{et al.}}: Bare Demo of IEEEtran.cls for Computer Society Journals}
%



\IEEEtitleabstractindextext{%
\begin{abstract}
Blockchain is an emerging decentralized data collection, sharing and storage technology, which have provided abundant transparent, secure, tamper-proof, secure and robust ledger services for various real-world use cases. Recent years have witnessed notable developments of blockchain technology itself as well as blockchain-adopting applications. Most existing surveys limit the scopes on several particular issues of blockchain or applications, which are hard to depict the general picture of current giant blockchain ecosystem. In this paper, we investigate recent advances of both blockchain technology and its most active research topics in real-world applications. We first review the recent developments of consensus mechanisms and storage mechanisms in general blockchain systems. Then extensive literature is conducted on blockchain enabled IoT, edge computing, federated learning and several emerging applications including health care, COVID-19 pandemic, social network and supply chain, where detailed specific research topics are discussed in each.
Finally, we discuss the future directions, challenges and opportunities in both academia and industry.

\end{abstract}

}

\maketitle

\IEEEdisplaynontitleabstractindextext

%
\IEEEpeerreviewmaketitle

\IEEEraisesectionheading{\section{Introduction}\label{sec:introduction}}

\begin{figure}[htb!]
    \centering
    \includegraphics[width = 0.5\linewidth]{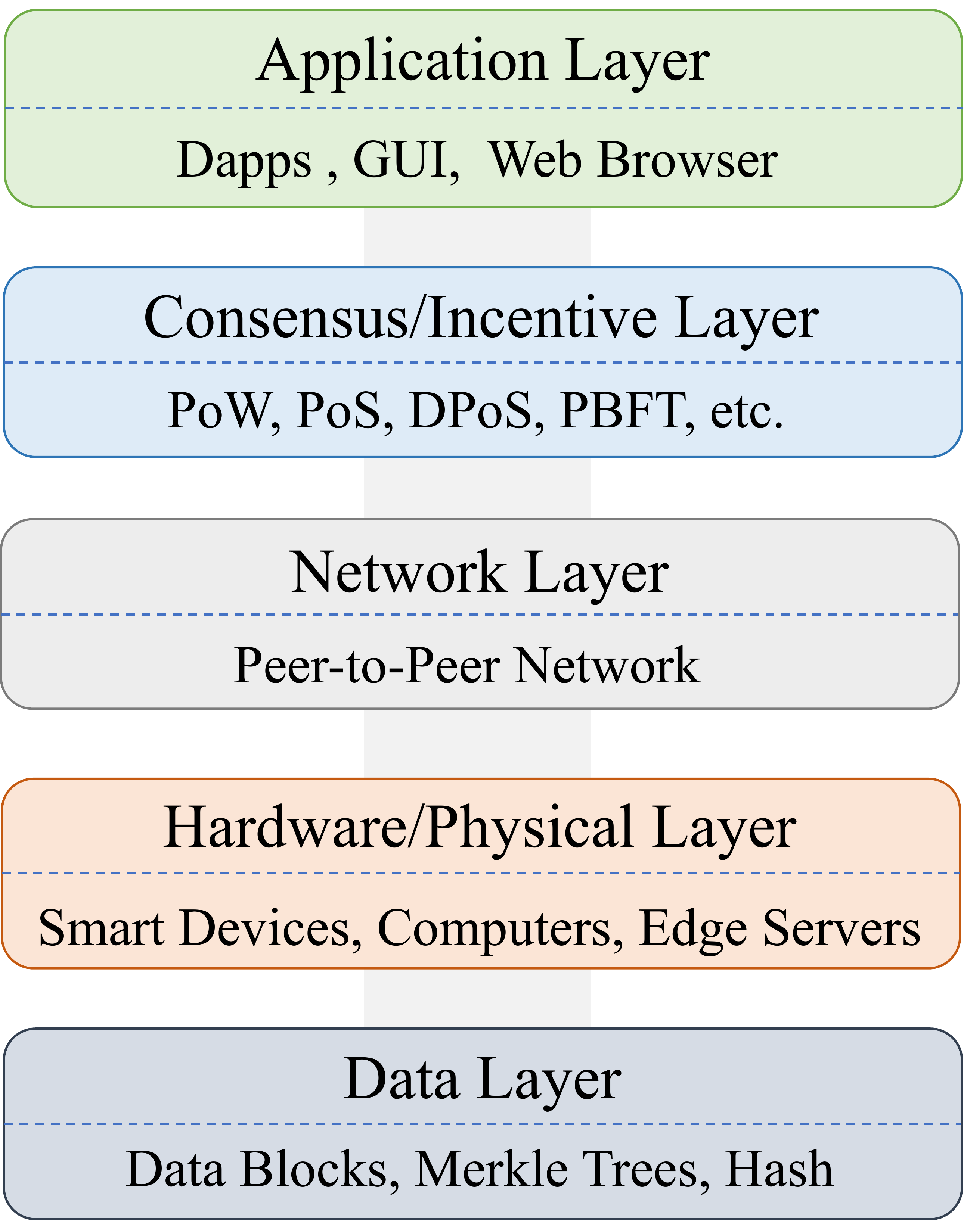}
    \caption{A typical architecture of blockchain systems.}
    \label{fig:blocksystem}
\end{figure}

\IEEEPARstart{B}{lockchain} technology is a decentralized storage technology working on peer-to-peer network. Peers, which are also called blockchain nodes, work as exactly same role and perform the same function following particular smart contracts specified in the system. In a blockchain system, every nodes keeps the whole copy of blockchain storage where transactions are packaged into blocks and every block is linked to previous block through block hash.  The actions of nodes are supervised by smart contracts defined on the blockchain, where the consensus mechanisms and incentive mechanisms are specified.
The consensus mechanisms are the core functions for maintaining system decentrality and storing blocks by carefully giving the ledgering right to one of the nodes who are called miners. Incentive mechanisms assists consensus mechanisms to distribute the working rewards and incentivize nodes to keep working honestly. A typical architecture of blockchain systems is illustrated in Figure~\ref{fig:blocksystem}.

Blockchain attracts increasing attention from both academia and industry because of its special capabilities and advantages comparing to existing conventional decentralized database storage approaches. Public blockchains, such as Bitcoin, can make the data available on every node which enables transparency to every participants. Since blockchain can work under totally anonymous setting without having to build trust among nodes, privacy of nodes can be preserved. Blockchain is tamper-proof storage, because the blocks are linked together with specific hash values that would cause a violation if any modification is made on block data. Blockchain storage is also free of single-point failure as long as the fraud users hold less than 51\%  mining power of the whole blockchain system. The comparison between conventional distributed database and blockchain is discussed in Table~\ref{tab:compare}.

\begin{table*}[htb]
\centering
\caption{The Comparison between Traditional Distributed Database Storage and Blockchain }
\label{tab:compare}
\begin{tabular}{c|c|c}
\hline
                     & Traditional Distributed Database                          & Blockchain                                                                        \\ \hline
System Management &
  \makecell[l]{Database is stored on different physical\\ places with multiple copies\\However managed by a central server} &
  \makecell[l]{Blockchain maintained by all participants\\Full copies are stored by every participant} \\ \hline
Data Accessibility   & \makecell[l]{Common participants have no access \\to whole database}        & \makecell[l]{Blockchain is public and \\accessible to all participants}                           \\\hline
Function Execution   & \makecell[l]{Central server perform data collection and calculation}     & \makecell[l]{Each participant is able to generate and \\record new data following smart contract} \\\hline
Costs                & \makecell[l]{Computation and storage cost are on central server}         & \makecell[l]{Computation and storage cost are on every participant}                             \\\hline
Communication        & \makecell[l]{Participants mainly communicate only to central server}     & \makecell[l]{Each participant broadcast updates to everyone else}                               \\\hline
Participants Privacy & \makecell[l]{Need provide information to central server\\ to build trust} & \makecell[l]{Fully functional under anonymous setting\\ with no trust been built}               \\\hline
Data Security &
  \makecell[l]{Single-point failure on central server Data can be \\tampered and destroyed if central server is breached} &
  \makecell[l]{No single-point failure Data are not able\\ to be tampered once stored on blockchain} \\ \hline
\end{tabular}
\end{table*}

Blockchain has been proven to be a remarkable success in cryptocurrency applications such as Bitcoin~\cite{nakamoto2008bitcoin}, Ethereum~\cite{wood2014ethereum}, and PeerCoin~\cite{King2012PPCoinPC}, the adoption of blockchain in many other fields is keep expanding the existing blockchain ecosystem. For instance, blockchain-enabled systems have been developed in areas of financial ledger system~\cite{Schr2021DecentralizedFO},  Internet of Things (IoT)~\cite{DBLP:journals/iotj/XuLL21, Uddin2021ASO }, edge and cloud computing~\cite{DBLP:journals/comsur/YangYSYZ19}, public administration~\cite{DBLP:conf/otm/BelchiorCV19, DBLP:conf/ecis/Belchior0V20}, healthcare~\cite{DBLP:journals/jms/HussienYUZZ19} and supply chain~\cite{DBLP:journals/mms/JabbarLHAR21}.

Current blockchain technology is still not perfect for general adoption and have many deficiencies to be improved. These deficiencies also bring troubles to blockchain-enabled applications. Researchers have devoted tremendous work on improving blockchain system with faster processing speed, more light-weight consensus mechanisms, less storage cost and lower communication bandwidth requirement. These advances of blockchain technology can benefit blockchain-enabled applications that are still at the very initial stage. However, there are nature gaps that the advances are hard to propagate among applications. It is highly demanded to connect multiple parts of the whole blockchain ecosystem together to facilitate future developments.

Existing related surveys mostly limit their scope on  specific research topics.
Wang et al~\cite{DBLP:journals/winet/WanLLW20} and Signh et al.~\cite{DBLP:journals/jsa/SinghKSCAT22} surveyed recent popular blockchain consensus mechanisms.
Zhou et al.~\cite{DBLP:journals/access/ZhouHZB20} summarize existing solutions on solving the scalability issue of blockchain. They classify the solutions into three layers: Layer0 which is about data propagation, Layer1 which is about on-chain methodologies and Layer2 which is about external off-chain solutions.
Zhang et al.~\cite{DBLP:journals/csur/ZhangXL19} and Feng et al.~\cite{DBLP:journals/jnca/FengHZKK19} investigate the security and privacy protocols of blockchain systems. Zhang et al.~\cite{DBLP:journals/csur/ZhangXL19} try to analyze how well blockchain systems supports the privacy and security requirement of transactions and conclude that only a small part of the blockchain platforms can achieve the security goals in practice. Feng et al.~\cite{DBLP:journals/jnca/FengHZKK19} summarize methodologies proposed by recent work to tackle the privacy issues in blockchain applications.

Gamage et al.~\cite{DBLP:journals/sncs/GamageWD20} introduced several blockchain applications in their survey such as supply chain, however most the mentioned applications are  special use cases of blockchain while some major applications are left behind, for instance IoT and edge/cloud computing. Huo et al.~\cite{DBLP:journals/comsur/HuoZWSCHWYL22} investigated research topics of blockchain-enabled IoT. They summarize that blockchain is mainly used in IoT for equipment safety and management, data collection and sharing,  energy trading, collaborative production and traceability.
Wang et al.~\cite{DBLP:journals/ejwcn/WangCLHX21} and Mollah et al.~\cite{DBLP:journals/iotj/MollahZNGYSLK21} conduct detailed survey about recent blockchain applications in Internet of Vehicles (IoV) which is a special instance of IoT. Blockchain-enabled IoV are usually studied with more specified use cases than general IoT~\cite{DBLP:journals/comsur/AlladiCSVGG22}, such as recent emerging electrical vehicle charging and smart parking.
Zou et al.~\cite{DBLP:journals/csur/ZouHZKWC22} extensively review blockchain developments in cloud computing and consider both Cloud as a Blockchain Service where blockchain assists could service and Blockchain as a Cloud Service where blockchain service is deployed on cloud. Liao et al.~\cite{DBLP:journals/tnsm/LiaoPZXW22} study the overlapped areas of edge computing and IoT. There are also comprehensive surveys on federated learning~\cite{DBLP:journals/soco/LiHWZLLCL22, DBLP:journals/iotj/NguyenDPPLSLNP21}, which is an emerging distributed machine learning schema to protect data providers' privacy and reduce the data transmission consumption.
Sreerakhi et al.~\cite{DBLP:journals/ijbis/SreerakhiBM22} review blockchain works in supply chain to discover the possibility of blockchain to help solve challenges including asymmetric information sharing, quality monitoring and market counterfeiting.

\begin{figure*}[htb!]
    \centering
    \includegraphics[width = 0.7\textwidth]{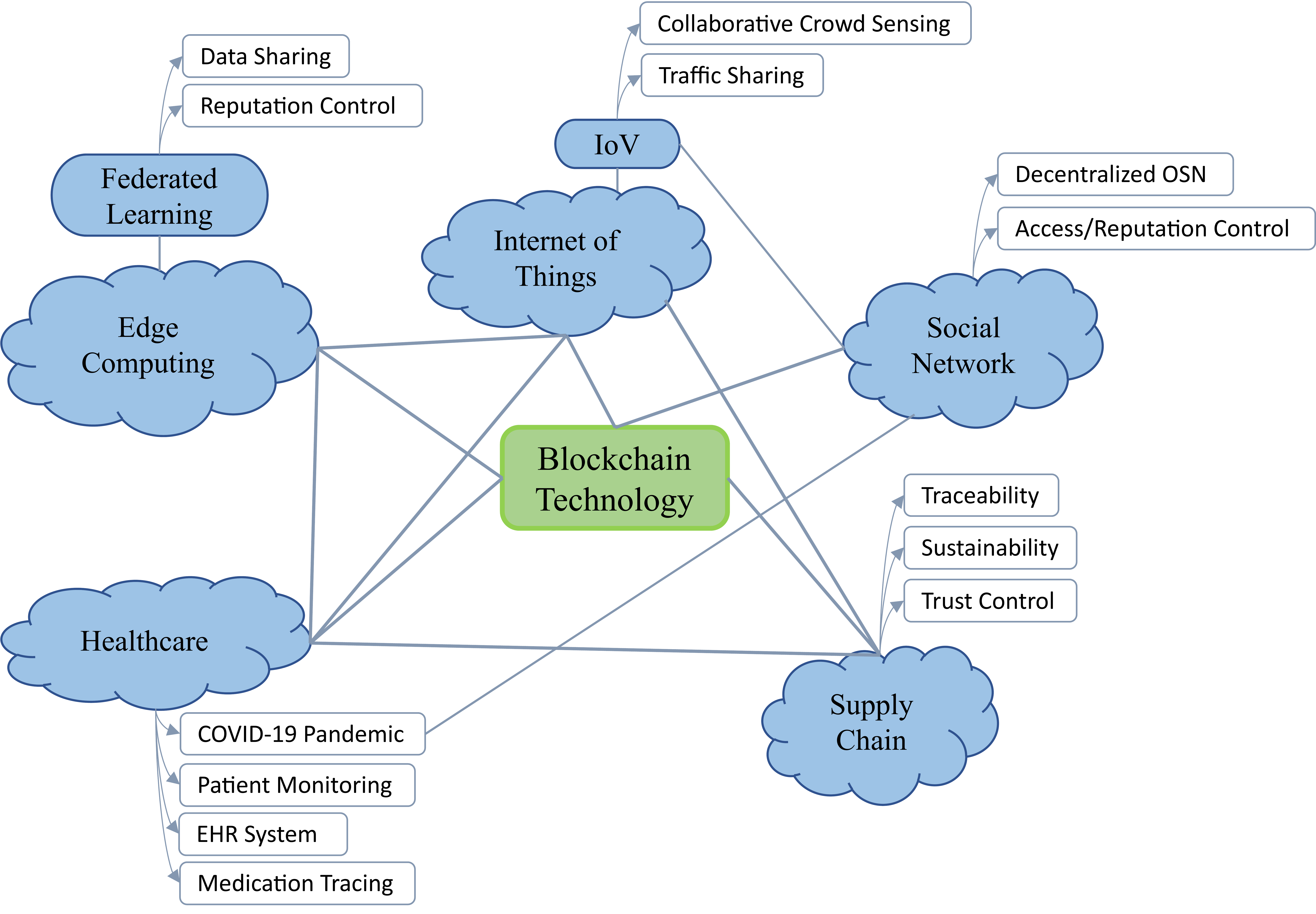}
    \caption{The blockchain ecosystem constructed in this paper.}
    \label{fig:ecosystem}
\end{figure*}

There is few survey that provides literature review for multiple research fields. In this paper, we bridge the advances of blockchain technology and it applications by expanding the reviewing scope to several most active blockchain research topics and emerging blockchain applications. An overview picture of blockchain ecosystem including blockchain and its applications is constructed as illustrated in Figure~\ref{fig:ecosystem}. We review the recent remarkable improvements of general blockchain technology. We choose IoT, edge computing, federated learning, healthcare, social network, supply chain as the most representative blockchain use cases in whole blockchain ecosystem. Extensive literature review is conducted on those selected use cases by discussing recent active research topics in each.  We enumerate several remaining issues for academia and industry to summarize the survey.

The remainder of this paper is organized as follows: In Section~\ref{sec:Gen_Blk} we first summarize the recent important developments of general blockchain technology. Then in Section~\ref{sec:IoT}, we review how blockchain can serve IoT systems and IoV which is a special use case in IoT. Next in Section~\ref{sec:ege_cmpt}, topics on blockchain-enabled edge computing and federated learning are investigated.  In Section~\ref{sec:emer_app}, we study several emerging hot topics that benefit from blockchain, including healthcare, COVID-19 pandemic, social network and supply chain. Next in Section~\ref{sec:disc}, we discuss our findings on current open issues and challenges of blockchain, then present the suggestions on future work.  Finally, this paper is concluded in Section~\ref{sec:con}.

\section{General Blockchain}
\label{sec:Gen_Blk}
Blockchain technology is explored from varied directions. Maesa et al. study blockchain from graph view~\cite{DBLP:conf/dsaa/MaesaMR16,DBLP:journals/ans/MaesaMR19}.  Chen and Liu attempt to discover communities in blockchain networks~\cite{DBLP:conf/bsci/ChenL19}. Pontiveros et al. propose a centrality measurement for Bitcoin transaction graph~\cite{DBLP:conf/icbc2/PontiverosSS19}. Li et al. discover topological and geometrical feature of Ethereum blockchain~\cite{Li2020DissectingEB}.
Banno and Shudo propose a simulation tool for simulating blockchain systems~\cite{DBLP:conf/icbc2/BannoS19}.  Liu et al. design a neural network that can automatically discover features of blockchain from the blockchain whitepapers~\cite{DBLP:journals/jpdc/LiuTBY20}.
Selfish mining problem is studied in Ethereum and Bitcoin Blockchain~\cite{DBLP:conf/icdcs/0001N19, Ritz2018TheIO, DBLP:journals/tnsm/KangCYMM21}. Chen et al. study potential phishing scam problem in financial blockchains~\cite{Chen2020PhishingSD}. Hou et al. propose SquirRL model that utilizes reinforcement learning method to analyze blockchain incentive mechanisms for vulnerabilities
such as selfish mining~\cite{DBLP:conf/ndss/HouZJDTFJ21}.

Generally, majority works are focusing on solving the scalability issue of blockchain, that is to improve the consensus efficiency, increase transaction throughput, reduce computation and communication cost as well as storage overhead.
In this section, we discuss recent advances in consensus mechanisms and storage methods for general blockchain systems.

\subsection{Consensus Mechanism}
Proof of Work (PoW) is the most popular consensus mechanism for blockchain systems. In PoW-based blockchain systems, peers invest powerful machines to solve cryptographic problems to win the right for ledgering.  Though it is proven to be secure and stable in some well-known applications, such as Bitcoin~\cite{nakamoto2008bitcoin} and Ethereum~\cite{wood2014ethereum}, it consumes extraordinary power for solving meaningless cryptographic puzzles. PoW limits the throughput of processing transactions and brings increasing computational  and storage overhead.

Proof of Stake (PoS)~\cite{King2012PPCoinPC, DBLP:conf/crypto/KiayiasRDO17}  is a popular consensus mechanism as a competitor of PoW. PoS gives the ledgering right to the peers with the probability as the contribution the peers made to the system, namely stake. PoS is less decentralized than PoW, but significantly improves the scalability and consensus efficiency. Delegated Proof of Stake (DPoS)~\cite{larimer2014delegated} is proposed to give more probability for being a miner than PoS to those who hold small amount of stakes . In DPoS consensus mechanism, whenever there is no candidate miners, every user will vote someone they trust. The weight of the vote is proportional to the stake of the voter. After voting, the peers received top $k$ votes become candidate miners. DPoS has been applied in many applications~\cite{schuh2017bitshares,DBLP:journals/pomacs/HuangWWTLZLHJ20,xu2018eos,DBLP:journals/access/GuidiMR20}.  DPoS is also a scalable and light weight consensus mechanism but not perfectly decentralized. Various modifications have been proposed for DPoS~\cite{Sun2021DTDPoSAD, DBLP:journals/isci/LiuXCMG21, DBLP:journals/access/YangZWLXZ19, DBLP:conf/mobiquitous/FanC18, Luo2018ANE}.  Xu et al.~\cite{DBLP:journals/tii/XuLK20} propose to improve the DPoS consensus by allowing nodes to vote favor, against and abstention. Then a vague value of node is calculated based on all three kind of received votes. Fuzzy value is finally derived as final score on which miners are selected.

Practical Byzantine Fault Tolerance (PBFT)~\cite{castro1999practical} is a classic byzantine fault tolerant protocol and introduced into blockchain systems as consensus mechanism~\cite{DBLP:journals/ppna/LiQL21, DBLP:journals/tpds/LiFZXCI21}. PBFT consensus mechanism commonly have five phases, namely \emph{request}, \emph{pre-prepare}, \emph{prepare}, \emph{commit} and \emph{reply}. The client sends  the message to be confirmed to a selected ``primary" at \emph{request} phase. The ``primary" then broadcast this message to all other peers (``replicas") at \emph{pre-prepare} phase. Then each ``replica" broadcasts received message to all other peers including ``primary" and other
``replicas" at  \emph{prepare} phase. Next at \emph{commit}, all peers, including the ``primary" and all ``replicas" send the message received at last phase to all other peers. Finally all peers send back the message to the ``client" at \emph{reply} phase. PBFT consensus is made through message transmission and commitment, therefore requires notable communication cost.

SCP~\cite{DBLP:journals/iacr/LuuNBZGS15}, proposed by Luu et al. constructs two-layer blockchain with committees, where one layer is for data blocks which are proposed by normal committee and another layer is for  consensus blocks which are proposed by the final designated committee in SCP to include all data blocks. The committees can make parallel PoW consensus, hence improve the efficiency and transaction throughput.
Li et al.~\cite{DBLP:journals/ijnsec/LiHGJF19} propose ISCP to promote the security level and communication efficiency of SCP . ISCP eliminates the need of final committee in SCP with a decentralized multi-partition consensus model.
Amiri et al.~\cite{DBLP:conf/icdcs/AmiriAA19} propose a novel OXII distributed diagram allowing transaction to be executed without conflict in permissioned blockchain. ParBlockchain is then proposed based on OXII diagram to achieve better transaction throughput.

In order to adapt to specific applications, various Proof of X (PoX) are developed where ``X" can be any metrics defined in those applications, such as Proof of Reputation~\cite{DBLP:conf/dasfaa/GaiWDP18}, Proof of Quality Factor~\cite{DBLP:journals/iotj/AyazSTG21}, Proof of Event~\cite{DBLP:journals/access/GuoLNS20}.
Bahri et al. study crypto-currency-free blockchain system~\cite{DBLP:conf/icdcs/BahriG19}. They propose viable permisionless non-financial Blockchain where Proof of Trust (PoT) is designed based on trust graph among peers. In PoT, peers with higher trust level can solve PoW cryptographic puzzle at lower difficulty level, thus reducing overall energy expense of PoW.

\subsection{Storage Method}
Classic blockchain systems require every peer store full copy of entire blockchain storage. This storage mechanism not only wastes enormous resources, but make system get centralized gradually. The oversized blockchain increases the bar of storage requirement for participants and also make the system hard to process data-heavy applications. With the blockchain growing in size, more and more disadvantaged nodes who can not afford the storage cost are gradually leaving the mining game.
Finally, the system becomes more and more centralized.

Blockchain sharding technology~\cite{DBLP:conf/sigmod/DangDLCLO19, DBLP:conf/ccs/LuuNZBGS16} is explored for reducing the storage overhead.
Generally, blockchain sharding is to divide peers into groups where consensus are made within each group so that transactions can be processed concurrently. Peers in each group (shard) maintain their local ledger, therefore in order to derive the full chain, an concurrency control and a commitment mechanism need to be designed~\cite{DBLP:journals/iacr/HanYLCV21}.
Zamani et al.~\cite{DBLP:conf/ccs/ZamaniM018} propose RapidChain to further reduce the communication cost while maintaining the resistance to Byzantine faults when there are less than $1/3$ fault nodes in all participated nodes. Xu and Huang develop an blockchain sharding mechanism that can tolerate $1/2$ fault nodes~\cite{DBLP:conf/sac/XuH20}. SkyChain is a dynamic sharding method enabled by deep reinforcement learning which can effectively deal with the dynamic environment in the blockchain system, i.e., joining and leaving of nodes, and malicious attacks~\cite{DBLP:conf/icpp/ZhangHQZ0C20}. Blockchain sharding technology is also developed in many domain-specific applications, such as IoT~\cite{DBLP:conf/wcnc/BandaraSRML22} and Federated Learning~\cite{DBLP:conf/bsci/MadillNLR22}.

Xu and Huang~\cite{DBLP:journals/access/XuH20a} propose segment blockchain where the whole blockchain is broken down into segments, and peers are only requires to store several segments. The whole blockchain storage can be recovered from multiple nodes' storage.
Qi et al.~\cite{DBLP:journals/tkde/QiZJZ21} propose a storage partition method namely BFT-Store for permissioned blockchain reducing the storage complexity per block from $O(n)$ to $O(1)$. Meanwhile, the data availability and data access efficiency are ensured by proposed four-phase re-encoding protocol based on PBFT and multiple replication mechanism with cache structure. The experimental results in ~\cite{DBLP:journals/tkde/QiZJZ21} show that at the same number of nodes, BFT-Store enabled blockchain can store more blocks, with remarkably lower storage overhead.


\begin{figure}[htb!]
    \centering
    \includegraphics[width = 0.8\linewidth]{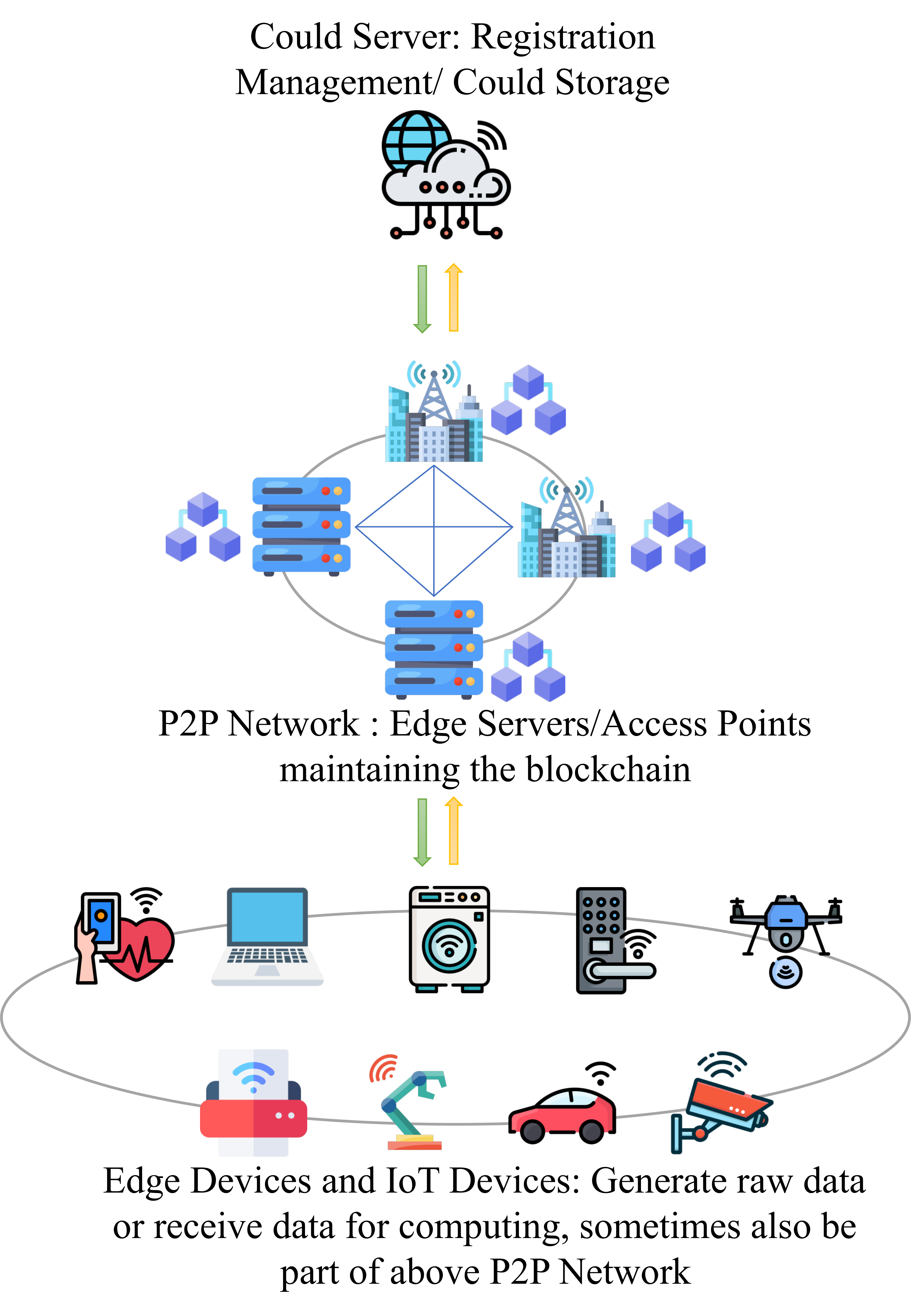}
    \caption{General Architecture of Blockchain-enabled IoT and Edge Computing systems.}
    \label{fig:IoTEdgeBlockchain}
\end{figure}

\section{Internet of Things (IoT)}
\label{sec:IoT}
Internet of things allows smart devices to connect with each other through the internet protocol for a ubiquitous data exchange~\cite{tran2022internet, alkhateeb2022hybrid}. The devices or objects working in internet of things are mostly sensors, micro-computers that can be easily compromised by malicious attacks. Blockchain technology applied in internet of things is a promising solution to improve the data integrity and security ~\cite{DBLP:journals/iotj/Novo18}. Figure~\ref{fig:IoTEdgeBlockchain} shows a typical structure of blockchain-enabled IoT and edge computing systems which will be discussed in next section.

However there are some challenges for realizing blockchain-enabled IoT systems. Due to the low computation capability,  battery life and memory storage of devices in IoT, the devices are not able to process heavy-weight consensus mechanisms like PoW~\cite{DBLP:conf/wd/MarchangIW19}. On the other side, Blockchain systems are mostly not able to produce high throughput which can not meet the demand of tremendous data generating and storage tasks in IoT systems~\cite{DBLP:journals/jnca/AggarwalCAKCZ19}.

In this section, we first review the recent blockchain works in general IoT systems, then we investigate an active special use case in IoT, namely Internet of Vehicle (IoV).

\subsection{General IoT}
Many light-weight consensus mechanism are designed to make blockchain feasible in IoT, such as credit based consensus mechanism~\cite{DBLP:conf/icdcs/HuangKC0WL19}, PoBT~\cite{DBLP:journals/iotj/BiswasSLMMW20}, PoRX~\cite{DBLP:journals/fgcs/WangLCKK20} and Proof-of-Transactions~\cite{DBLP:journals/iotj/AiC22}.
Dorri et al.~\cite{DBLP:journals/jpdc/DorriKJG19} propose a Lightweight Scalable Blockchain (LSB) for industrial IoT, where nodes are divided into clusters and managed by cluster heads. A Distributed Throughput Management (DTM) algorithm is proposed to dynamically adjust number of clusters and consensus period for maintaining high transaction throughput.
Biswas et al.~\cite{DBLP:journals/iotj/BiswasSLNW19} separate the workers in IoT network from the peers in blockchain network. The workers in IoT networks are defined as local peers who are connected to designated anchor peer representing one organization in blockchain network. Transaction commitment can be made within organization without global consensus, hence the transaction throughput is improved.

Incentive mechanisms are also studied for IoT. Ding et al.~\cite{DBLP:journals/tnse/DingGLW21} designs an incentive mechanism to motivate devices to devote more power in mining.  A two-stage Stackelberg game is formulated to find  reasonable reward pricing strategy to maximize blockchain utility.

Instead of proposing consensus mechanism, some work specifies more details and design comprehensive blockchain systems for IoT applications, such as AEChain~\cite{DBLP:journals/cem/KhanLH22}, BPAF~\cite{DBLP:journals/cem/ZhangZX22}, BET~\cite{DBLP:journals/iotj/GuoDW21} and B-MET~\cite{DBLP:journals/tgcn/GuoDW22}. Li et al.~\cite{DBLP:journals/iotj/LiFJXLW22} are the first to study the problem of extra cost caused by frequent smart contract updates in blockchain-enabled IoT system. A new smart contract architecture is proposed, namely ATOM, that can construct the bytecode of smart contract from application by directly assembling templates pre-built upon the designed Application-oriented Instruction (AoI) set rather than by compilation. Zhou et al.~\cite{DBLP:journals/iotj/ZhouFW22} aim to improve the storage efficiency of blockchain-enabled wireless communication by proposing dynamic adjusted block-assignment (DAB) contract which dynamically assigns blockchain portions to different devices.

Reinforcement learning methods are developed to optimize resource allocation in IoT networks to achieve better scalability~\cite{DBLP:journals/iotj/YangLSYSZ21, DBLP:journals/cn/WuWML21} and resource allocation~\cite{DBLP:journals/iotj/0007YSWZ20}.  Liu et al.~\cite{DBLP:journals/tii/LiuYTLS19} propose a deep reinforcement learning approach that can help maximize on-chain transaction throughput of the blockchain system by selecting the block producers and consensus algorithms as well as adjusting the block size and block interval. Yun et al.~\cite{DBLP:journals/iotj/YunGC21} propose deep Q network shard-based blockchain (DQNSB) scheme that dynamically finds the optimal throughput by selecting transaction sharding methods, also the block size and block interval. Ding et al.~\cite{ding2022pricing} introduces edge server into IoT networks, where IoT devices are able to purchase computational power from edge servers.  They derive a Stackelberg equilibrium to optimize the pricing and budget allocation.

\subsection{Internet of Vehicle}
Internet of vehicle (IoV) or Vehicular Network is a special IoT application where the IoT nodes are the mobile devices installed on vehicles. In IoV system, vehicles can share information such as road condition, traffic conjunction and accident information with other vehicles, so that vehicles are able to decide best routes or collaborating with each other on some emergency issues.

Blockchain technology brings decentralized architecture to IoV as it does to IoT. IoV usually has more strict requirements on applied blockchain system~\cite{V2022BlockchainTF}. Vehicles are moving making them can only connect to roadsides or other vehicles periodically. Vehicles are also have limited battery and computation power making them reluctant to participating in low-profit or computation-expensive tasks.

Cho et al.~\cite{DBLP:conf/icbc/ChoCH19} propose iCarChain for managing vehicle related businesses in decentralized manner. iCarChain is an initial attempt for benefiting consumers and vehicle business industry with fewer technological restrictions and more affordable expenses by decentralizing the business system. Blockchain servers as distributed storage system for IoV in~\cite{DBLP:journals/iotj/YangYLZL19} where roadside units are selected based on both PoS and PoW as miners to pack data and messages generated by vehicles.
Wang et al.~\cite{DBLP:journals/access/WangJGYCL20} propose TrafficChain which is a two-layer blockchain-enabled secure and privacy-preserving decentralized traffic information collection system. Wang et al. specially study Byzantine attack and Sybil attack on TrafficChain and propose novel LSTM based methods to defend against them. Yin et al.~\cite{DBLP:journals/iotj/YinWHDJ20} study a special case in IoV that multivehicles collaboration can be performed when a single vehicle is not able to accomplish a task. They carefully design an incentive mechanism and a task assignment algorithm to motivate vehicles to participate the tasks as well as shorten the collaborative tasks' finish time.  Hui et al.~\cite{DBLP:journals/iotj/HuiHSLCXD22} study similar collaborative crowd sensing problem that aims to motivate vehicles collaborate each other by formulating and solving a Coalition Game.

Despite some classic consensus mechanisms are adopted in IoV, such as PoS~\cite{DBLP:journals/tvt/KangXNYKZ19,DBLP:journals/iotj/YangYLZL19}, PoET~\cite{DBLP:conf/sss/ChenXSGLS17} and PBFT~\cite{DBLP:journals/cm/CebeEAAU18, DBLP:journals/vcomm/ZhangLLACCT19}, researcher are developing more scenario-specific consensus mechanisms in order to achieve better security, latency and throughput in IoV. Kang et al.~\cite{DBLP:journals/tvt/KangXNYKZ19} design a reputation-based voting scheme to improve the security of blockchain-enabled IoV. This scheme evaluates candidates' reputation using both past interactions and recommended opinions from other vehicles. Proof of Quality Factor~\cite{DBLP:journals/iotj/AyazSTG21} is proposed to bridge vehicles and edge computing servers, which allows mobile edge nodes serve as mining nodes. 
As the number of electric vehicles (EVs) increasing, Luo et al.~\cite{DBLP:journals/iotj/LuoFYS22} study the energy trading in the internet of electric vehicles. They propose to deploy blockchain server in local energy
aggregators (LEAG) to store all the trading transaction records and specify smart contacts as agents for optimal energy pricing and allocation.  Abishu et al.~\cite{ DBLP:journals/tvt/AbishuSYASL22} jointly consider PBFT and Proof of Reputation (PoR) and propose  PBFT-based PoR (PPoR). Electric vehicles are grouped in clusters according to the roadside units they connect to. PPoR will select  miners (validators) in the cluster based on their reputation value that is calculated based on evidence and opinion spaces collected from EVs in each cluster.

Reinforcement learning also plays important role in many works of blockchain-enabled IoV~\cite{DBLP:journals/cem/WangSWNR21}.  Kim and Ibrahim~\cite{Kim2022ByzantineFaultTolerantCV} design a reinforcement learning model to decide the optimal number of peers participating in consensus making to improve the latency and throughput without compromising the Byzantine fault tolerance. They connect peers in different groups through channels, and formulate the problem of choosing channels as Multi-Arm Bandit problem which is solved by the proposed reinforcement learning algorithm. Liu et al.~\cite{DBLP:conf/icc/LiuTYLS19} first propose a methodology to quantify the performance
of blockchain systems from the aspects of scalability, decentralization, latency and security, then apply deep reinforcement learning technique to select block producers, adjust
block size and block interval, in order to maximize the transaction throughput without sacrificing other properties.

\section{Edge Computing}
\label{sec:ege_cmpt}
Edge computing is a technology to allow devices at the edge of network, such as smart devices, mobile micro computers,  bases stations and network access points, to generate,  collect, transmit  and process data. Edge computing is an extension of IoT and overlaps with IoT in many applications. For example, The sensors or smart objects in IoT might  need to connect to some edge servers to complete data sharing and computing. Some devices in IoT such as electric vehicles can also be considered as edge nodes in edge computing.  In this section, we first investigate the blockchain development in general edge computing then specially discuss an emerging topic in edge computing, namely federated learning.

\subsection{General Edge Computing}

Like the deficiencies of blockchain system in many other application fields, blockchain system brings extra computation cost and communication cost to edge nodes.  Off-loading as a popular methodology to alleviate the stress of edge nodes is to move computation task to external machines, such as Edge Computing Service Provider (ESP) or Cloud Computing Service Provider (CSP)~\cite{DBLP:conf/icdcs/JiangL019}. Those service providers that are qualified to conduct the off-loaded tasks may earn some profit or reward for providing computation service. In the process of
computation offloading in edge computing, it is critical to dynamically make optimal offloading decisions to
minimize the communication delay, energy consumption spent on the devices and the throughput of data storage on blockchain~\cite{DBLP:journals/tii/QuCWLD22}.

Jiang et al.~\cite{DBLP:conf/icdcs/JiangL019} design an multi-leader multi-follower Stackelberg game to address computing resource management achieving maximized profits of service providers and the rewards of miners in the network.
Hu et al.~\cite{DBLP:journals/sensors/HuGWHL22} detail a blockchain enabled edge computing system, and propose a deep reinforcement learning algorithm to jointly optimize the computation offloading policy and block generation
strategy to maximize the scalability.
Though abundant off-loading optimization methods have been developed, it is hard to evaluate how good the outcome as well as to compare these methods. To address this issue, Qu et al.~\cite{DBLP:journals/tii/QuCWLD22} propose ChainFL, that is a lightweight simulation platform for building a test edge computing environment which also supports federated learning and blockchain technology.

Cooperative or collaborative offloading in edge computing is an extended problem over off-loading.
Feng et al.~\cite{DBLP:journals/iotj/FengYPCDZ20} utilize deep reinforcement learning algorithm to jointly optimize the cooperative offloading decision and blockchain parameters in blockchain-enabled mobile edge computing systems. Zuo et al.~\cite{DBLP:journals/iotj/ZuoJZZ21} also study cooperative mobile edge computing, they formulate the offloading optimization problem with a three-stage Stackelberg game. Cheng et al.~\cite{DBLP:journals/tcss/ChengCDGY22} use blockchain to form a authentication system for collaborative edge computing systems that achieves anonymity while avoid malicious attacks from fake IoT devices.

Xiao et al.~\cite{DBLP:journals/tcom/XiaoDJHWLP20} address selfish attack problem in edge computing that attackers use less computation resources than promised to process offloading tasks or provide faked computation results. They propose a trust mechanism to assign reputations to edge nodes, then the CPU computation resources are allocated based on the reputation.
Liu et al.~\cite{DBLP:journals/jsac/LiuGPCOH22} address the problem of the existence of low-quality data such as
missing values, inconsistent values and incorrect values due to the data heterogeneity in edge computing.  These low-quality data may not support or even slow down the computation tasks. To tackle this issue, a consortium blockchain is design in ~\cite{DBLP:journals/jsac/LiuGPCOH22} where the data quality will first be evaluated and repaired before being off-loaded.

More works incorporate blockchain system deeper with edge computing network to allow blockchain provide more reliable functions by proposing specific consensus mechanisms and comprehensive blockchain-enabled edge computing systems.
Baranwal and Kumar~\cite{DBLP:conf/percom/BaranwalK22} propose PoSP consensus mechanism that replace the hash puzzle in PoW with a service placement problem whose result can meanwhile help the resource allocation in
edge computing. Maskey et al.~\cite{DBLP:journals/cem/MaskeyBSK21} use neural networks to decide the miner's reputation instead of a heuristic computation in a blockchain-enabled vehicular edge computing environment.
 Balistri et al.~\cite{DBLP:journals/iotj/BalistriCCGRS22} embed blockchain into edge computing network in order to promote the cyber-resiliency, where edge nodes and service providers work as peers in a blockchain system. Li et al.~\cite{DBLP:journals/ipm/LiLZWL22} design a typical multi-layer blockchain enabled system, where the blockchain layer is incorporated into edge-computing layer. Similar to ~\cite{DBLP:journals/iotj/BalistriCCGRS22}, the edge computing nodes in~\cite{DBLP:journals/ipm/LiLZWL22}  are also the blockchain peers(nodes) to conduct consensus mechanism and create new blocks. Wang et al.~\cite{DBLP:journals/tpds/YuanHCZQXXY22} are the first to extend the collaborative task offloading to collaborative edge storage. They propose a blockchain system called CSEdge where a reputation based consensus mechanism called ER-BFT is designed to select edge servers based on their reputation, and a incentive mechanism is proposed to motivate edge server to help complete data offloading .

\begin{figure*}[htb!]
    \centering
    \includegraphics[width = 0.8\linewidth]{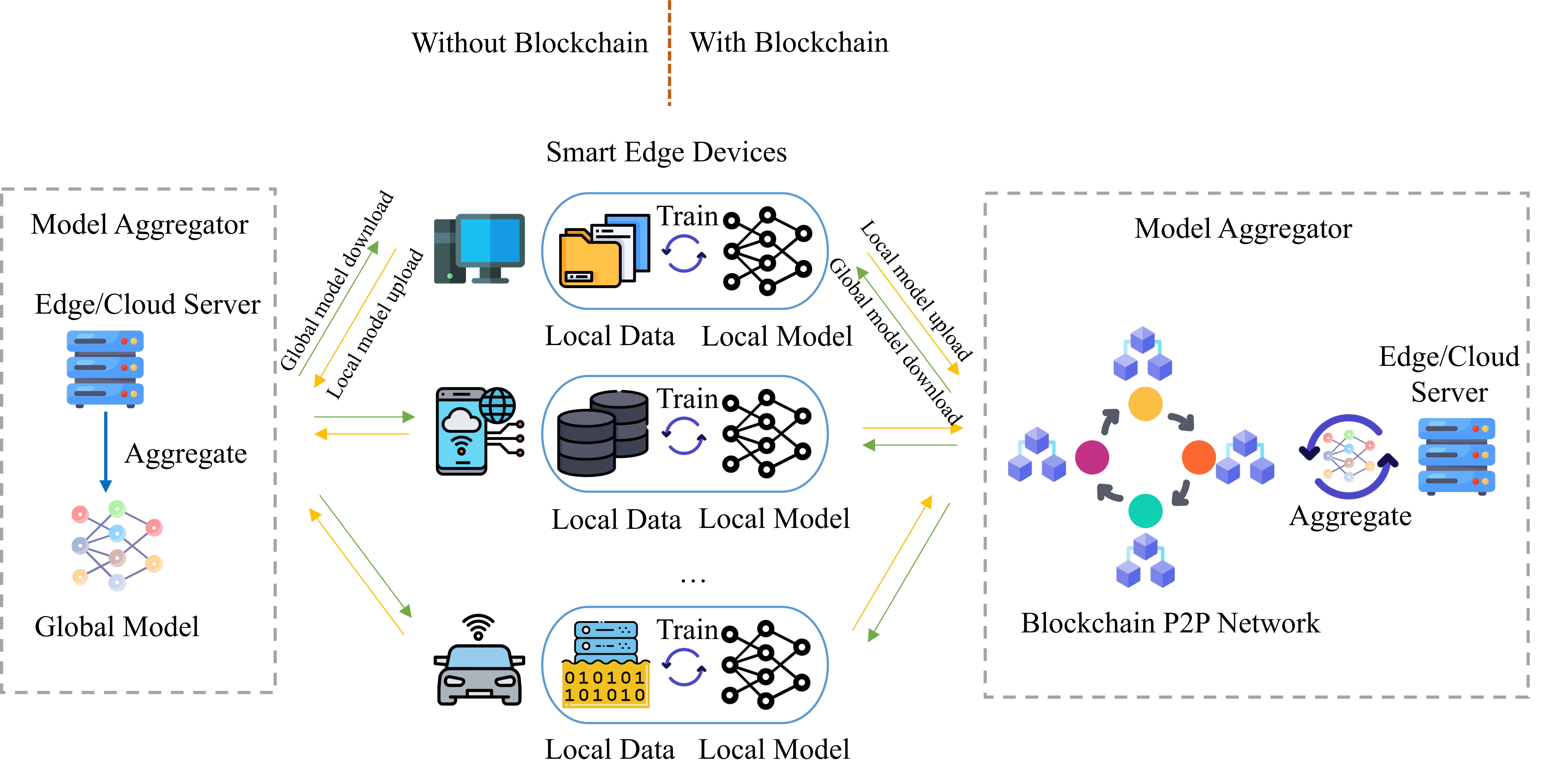}
    \caption{Comparison between Federated Learning architecture with and without blockchain.}
    \label{fig:FL}
\end{figure*}

\subsection{Federated Learning}
Federated Learning (FL), first proposed by Google~\cite{DBLP:journals/corr/KonecnyMYRSB16} is an emerging distributed machine learning schema. Instead of collect all the data first from data providers, then training a complicated machine learning model on a central computing device, federated learning allows each data provider to train a local model first and then upload the parameters to the central computing device. In federated learning schema, since data providers keep their data locally, the communication cost is saved and the privacy of data provider can be preserved.

Before federated learning is proposed, blockchain had been adopted to secure the data or model parameters in machine learning. Goel et al.~\cite{9025660} propose DeepRing which is a blockchain secured CNN model and shows significant resistance to tampering attack than ordinary models. Fu et al.~\cite{DBLP:journals/wc/FuYLLZ20} use blockchain system to secure the collective learning in IoV environment which is similar to federated learning schema.
The adoption of blockchain technology in federate learning further promotes the security of machine learning to next level~\cite{DBLP:journals/iotj/NguyenDPPLSLNP21,DBLP:journals/icl/SunWWFC22,DBLP:conf/bsci/MadillNLR22,DBLP:conf/blockchain2/RamananN20}. Based on literature review, we create a generalized blockchain-enabled federated learning architecture, and compare it with conventional federated learning in Figure~\ref{fig:FL}.

Lu et al.~\cite{DBLP:journals/tii/LuHDMZ20a} design a data sharing platform for IoT with blockchain enabled federated learning, and propose proof of training Quality (PoQ) as a light weight consensus mechanism. Lu et al.~\cite{DBLP:journals/iotj/LuHZMZ21} propose a
blockchain-enabled federated learning scheme to strengthen
communication security and data privacy protection for communication between  digital twins of IoT devices and edge network. The digital twins get the trained parameters from IoT devices instead of tedious device state information.
Peng et al.~\cite{DBLP:journals/tnse/PengXCGYGT22} propose to use blockchain to achieve verifiable and auditable federated learning framework where committee based aggregation model and a authenticated data structure are developed over blockchain system.

Li et al.~\cite{DBLP:journals/network/LiCLHZY21} propose a committee consensus to improve the consensus efficiency for a blockchain-enabled federated learning framework, called BFLC. In BFLC, the local updated model gradients and model parameters are stored on blockchain. A committee is form to evaluated the updates, and only the qualified updated will be stored in blockchain.
Qu et al.~\cite{DBLP:journals/tpds/QuWHC21} creatively combine the PoW with federated learning and proposed Proof of Federated Learning (PoFL) that instead of solving the meaningless puzzles in PoW, solving the actual tasks in the federated learning will make much less computation power waste. Nodes are gathered in pools where the PoFL is making for aggregating the desired model.

In federated learning, it is important to define suitable rewards for the worker clients who spend local computation and communication resources training local models. Otherwise, workers may be reluctant to do the training and report useless or even harmful parameters to global model aggregators.  In order to jointly satisfy the privacy, integrity, and fair incentives of blockchain-enabled federated learning, R{\"{u}}ckel et al.~\cite{DBLP:journals/cn/RuckelSH22} propose a federated learning framework that incentives each clients based on their individual contribution to the global model, uses zero-knowledge proofs to ensure data integrity and adopts local differential privacy to perturb each clients’ model update with Laplacian noise to ensure the data privacy. Gao et al.~\cite{DBLP:journals/jpdc/GaoLCXX22} propose FGFL model that assesses workers based on both contribution and reputation. They also states that it is crucial to design both an effective incentive mechanism and a reliable incentive management system to insure the fairness of incentives.

Since federated learning requires each device to upload the trained models to the aggregator, the global model may need to wait the slowest device to finally get updated. Asynchronous federated learning is then studied to deal with the delay of communication from multiple devices.  Lu et al.~\cite{DBLP:journals/tvt/LuHZMZ20} attempt to solve the asynchronous problem in IoV by optimally selecting the participating nodes through deep reinforcement learning. The models will first be aggregated within local range of vehicles asynchronously, then globally aggregated by roadside units synchronously.  Feng et al.~\cite{DBLP:journals/tc/FengZGQLY22} propose BAFL, which is a blockchain-Based Asynchronous Federated Learning Framework. In BAFL, each device is communicating with one miner, for local model uploading and global model updating. The global model can be aggregated by each device once the device decides to update it with local models. Then the consensus of global model will be reached in blockchain layer, hence avoid waiting all devices. Wang and Tsai et al.~\cite{DBLP:journals/sensors/WangT22} propose to compose a blockchain-enabled asynchronous federated learning system with multiple blockchains, where Sub-Blockchains are responsible for the model local training in multiple devices, and those Sub-Blockchains will communicate with a Main-Blockchain which is for the global model aggregation.

The distributed nature of federated learning schema make it easy to be integrated into IoT networks or edge computing networks, where sensing nodes in IoT and mobile devices can be the clients to train local models~\cite{DBLP:journals/tii/JiaZLZHL22, 9170559}. Otoum et al.~\cite{DBLP:journals/iotj/OtoumRM22} propose a blockchain-enabled FL model the decentralizes the learning process to ensure privacy and security for critical IoT infrastructure systems. Feng et al.~\cite{DBLP:journals/network/FengYGQLY22} propose a two-layer blockchain system to enable federated learning in mobile edge computing network where the first layer blockchain is for local model updates and the second layer blockchain help update global model. Ayaz et al.~\cite{DBLP:journals/tvt/AyazSTG22} propose a blockchain-enabled federated learning vehicular networks to improve the quality and efficiency of message dissemination.

\section{Emerging Applications}
\label{sec:emer_app}
In this section, we investigate four emerging research fields where blockchain is increasingly playing important roles to bring decentrality, system robustness and security to related applications.

\subsection{Healthcare}
The potential of blockchain technology in healthcare has shown and been discussed as a revolution for over 5 years~\cite{DBLP:conf/healthcom/Mettler16, DBLP:journals/jms/HussienYUZZ19, DBLP:journals/jnca/McGhinCLH19}. Traditional healthcare systems are suffering from single point of failures and information leakage by cybersecurity attacks~\cite{Yaqoob2021BlockchainFH}, as well as lacking transparency, trustful traceability, immutability, audit, privacy, and security~\cite{DBLP:journals/jms/YueWJLJ16}. Blockchain technology provides promising solutions to tackle above issues that can decentralize the storage and permission management, keep data traceable, verifiable and immutable~\cite{agbo2019blockchain, haleem2021blockchain}.

Person health information (PHI) nowadays is usually digitized into Electrical Health/Medical Record (EHR/EMR) and stores in healthcare authorities' databases, such as hospitals, health insurance companies or medical laboratories. People may have their PHI in multiple healthcare authorities. Though the information are private, people are not able to manage the abuse of their own information~\cite{rahmadika2018blockchain}. Blockchain enabled decentralized healthcare information management systems are proposed to tackle this issue~\cite{DBLP:journals/access/JaimanU20,rahmadika2018blockchain, DBLP:conf/dasfaa/LiuLAFM22}. Soni and Singh~\cite{Soni2021BlockchainbasedS} provide a general mapping from blockchain technology to medical processes and  discuss the ability of blockchain to enable access control, secure devices, identity protection and cost reduction. Zaabar et al.~\cite{DBLP:journals/cn/ZaabarCJAA21} propose HealthBlock which is a six-layer blockchain-enabled health information management system. Blockchain works in a layer in HealthBlock to manage the access from multiple parties in other layers. Bhattacharya et al.~\cite{DBLP:journals/tnse/BhattacharyaTBT21} propose Blockchain-Based Deep Learning as-a-Service (BinDaaS) system that first adopts blockchain to securely store the collected information using lattice based signature generation and verifying operations, then applies deep learning technology to produce valuable prediction service, such as patient future disease prediction. Zhang et al.~\cite{DBLP:journals/cn/ZhangYL22} uses pairing-based cryptography to generate temper-proof EHR which is further packaged into transactions in blocks. They also design secure payment protocols  between patients and healthcare providers through smart contracts in blockchain. Chelladurai and Pandian focus on improving the data access speed among multiple parties with proposed Modified Merkle Tree data structure in blockchain~\cite{DBLP:journals/jaihc/ChelladuraiP22}. Wu et al.~\cite{DBLP:journals/titb/WuWNZ22} propose multi-level smart contracts to achieve dynamic access control that allows different access rights for different scenarios. They propose a Privacy Attribute Classification algorithm to classify medical records into different privacy levels, then the access right can be matched.

Another helpful blockchain use case in healthcare is medication tracing. Counterfeit medications have bring unneglectable public health concern and sever impact on treatment outcomes due to insufficient, incorrect, erroneous ingredients, falsified information or wrong labeling~\cite{DBLP:journals/hij/UddinSJPE21}. Blockchain as a powerful distributed data storage method that can manage accessibility, ensure data transparency and immutability is hence an proactive approach to track, detect, and manage counterfeits in healthcare supply chain~\cite{DBLP:journals/access/MusamihSJADAE21 ,DBLP:conf/sac/KambiloZGSKV22}. Musamih et al.~\cite{DBLP:journals/access/MusamihSJADAE21} implement an blockchain-enabled healthcare supply chain system with Ethereum. They designed on-chain and and off-chain structure where the actual healthcare data are store in off-chain low-cost decentralized storage system, blockchain is response for storing the logs and interact with off-chain resources. Abbas et al.~\cite{10.1145/3477314.3507118} design Couch-DB where a machine-learning model is built upon the blockchain system to provide drug recommendation to customers.

Some miscellaneous topics in healthcare are studied with blockchain. Liu et al.~\cite{Liu2019ABS} and  Mendoza{-}Tello et al.~\cite{DBLP:conf/riiforum/Mendoza-TelloMM20} propose to use blockchain to avoid healthcare insurance fraud.  Blockchain can also be used to secure the channel of remote patient monitoring~\cite{9581059}. Pighini et al.~\cite{DBLP:conf/primelife/PighiniVMMMGCC21} implement SynCare ecosystem with blockchain and cloud service that allows patients to directly send data to healthcare professionals without concern of data leakage so that the patient can be securely remotely monitored. Many other patient monitoring systems require participation of IoT devices, such as smart sensors, meters or network access points, which will be discussed later. Blockchain is also introduced into clinical trails which are usually with a larger flow of information and more confidential data from more parties~\cite{Wong2019PrototypeOR}. Wong et al.~\cite{Wong2019PrototypeOR} and Albanese et al.~\cite{ DBLP:journals/jaihc/AlbaneseCSC20} have implement prototype of blockchain system for clinical trail data management.

Blockchain bridges healthcare with various other research fields. As mentioned above, IoT devices are widely used in healthcare, such as monitoring the status of patients, sensing  important parameters for treatment or surgery~\cite{DBLP:journals/fgcs/SinghRATY22}. Ali et al.~\cite{DBLP:journals/sensors/AliAHPFKTZ22} propose an efficient blockchain system for IoT-incorporated healthcare applications where a secure search algorithm is designed to encrypt and anonymously search the data stored in blockchain. Hossein et al.~\cite{DBLP:journals/comcom/HosseinEDKC21} propose two-chain structured blockchain system, namely BCHealth for IoT healthcare applications that allows data owners to personalize the access policies over their healthcare data. In BCHealth, one chain stores access policies and the other chain stores data transactions.

In healthcare, artificial intelligence models provides valuable predictions and analysis for diagnosis. However due to privacy, healthcare providers are reluctant to share their data for a common AI task.  Federated learning is hence introduced into Healthcare with blockchain to protect the data and AI models in healthcare.
Aich et al.~\cite{DBLP:conf/icact/AichSKACJ022} propose a general framework to incorporate federated learning with multiple healthcare providers, where blockchain works as the intermediate platform for transmitting data from healthcare providers to federated learning AI task.

\subsection{Special Healthcare Case Study: COVID-19 Pandemic}
COVID-19 pandemic has lasted for over 3 years. Researchers have developed abundant approaches contributing to the prevention of virus spread. In this section, we review the literature of this special use case in healthcare, and discuss how Blockchain can benefit the recovery from pandemics.

As special use case of healthcare, there are blockchain-enabled EHR management system specially designed for COVID-19 pandemic~\cite{DBLP:books/sp/21/TorkyDH21,DBLP:journals/connection/YaoJLC22}. Tan et al.~\cite{DBLP:journals/tnse/TanYSYWL22} propose a traceable COVID-19 record sharing system powered by blockchain where a security game (IND-CPA) is built in the system to achieve attack resistance. Aslan et al. ~\cite{DBLP:journals/itc/AslanA21} evaluate the possibility of world-wide COVID-19 information sharing among countries with DApps (Decentralized Applications) on blockchain system, but the idea is initial and of high level, no implementation is provided. Abid et al. ~\cite{DBLP:journals/spe/AbidCKJ22} propose NovidChain which is a blockchain system to replace the central server that stores test/vaccine certificates of users. NovidChain works as a bridge for certificate issuer, holder and verifier, and is evaluated to be secure, scalable and low-cost extensively in the paper. However NovidChain is set to be private and manged by governments or healthcare institutions, which brings centralization and privacy concerns to NovidChain.

Contact tracing as one of the most effective ways to defeat pandemic has been developed in many countries~\cite{DBLP:journals/jcn/TahirTSRK21}.  Contact tracing requires people to share their private contact history, sometime even including sensitive information such as GPS coordinates or medical history~\cite{DBLP:journals/access/RicciMFF21, DBLP:conf/wcre/KassabD21}. Most initial attempts of blockchain-powered contact tracing approaches are of high level, treat blockchain naively as external storage or with no simulation provided to illustrate the effectiveness, such as BeepTrace~\cite{DBLP:journals/iotj/XuZOFBI21}, Arifeen et al.~\cite{arifeen2020blockchain} and Choudhury et al.~\cite{DBLP:journals/isjgp/ChoudhuryGG21}.
Hasan et al.~\cite{DBLP:journals/access/HasanSJYOE21} propose to use blockchain to record participants' GPS coordinates and trigger proof of location to conduct contact tracing and risk alert. In the proposed system, external oracles are adopted to conduct contact tracing algorithm, and blockchain works as a bridge from external oracles to involved parties, including testing center and patients. Torkye et al.~\cite{DBLP:journals/informatics/TorkyGSH21} also use blockchain to securely bridge contact tracing-concerned parties and propose to use specific code patterns to encode peoples locations, so that only people who have been to the same place can be identified as contact cases while protecting privacy. However, this method is not able to reflect accurate contact history. Peng et al. focus on contact data verification to ensure data integrity, and propose $P^2$B to realize a blockchain system~\cite{DBLP:conf/sigmod/PengXWHXC21}.

Most of above contact tracing approaches assume people are willing to join the contact tracing system and share their contact history.  However in real world people may be reluctant to use such system or act reluctantly after joining the contact tracing system, which lowers the effectiveness of contact tracing. Incentive mechanisms play important roles in blockchain systems that motivate people or participants to perform contact tracing function honestly and actively. Naren et al.~\cite{DBLP:journals/iotm/NarenTHCKG21} analyzed importance of incentive mechanisms, but no specific method is proposed to solve the mechanism.  Lv et al.~\cite{DBLP:journals/tnse/LvWJCQZ22} considers large-scale contact tracing with the help of IoT and proposed ByChain where an artificial potential field-based incentive
allocation mechanism is proposed to motivate IoT witnesses to maximize monitoring coverage.

Alansari et al.~\cite{DBLP:journals/iotj/AlansariBMAAA22} extend contact tracing with two other subsystems to perform public places access control and safe-places recommendation, respectively. All three subsystems are incorporated with consortium blockchain to manage the data access and storage.

Blockchain technology also helps COVID-19 vaccine control and management~\cite{DBLP:journals/access/RicciMFF21}. Considering fragile biological substances, which should be take special care during transmission and distribution, Robit et al.~\cite{DBLP:journals/jicts/RotbiMG22} discuss a concept of blockchain-enabled automatic vaccine lots management to promote the transparency and immutability of management data. Musamih et al.~\cite{DBLP:journals/access/MusamihJSHYA21} implement a prototype of blockchain system on Ethereum to help track the vaccine during delivery from raw material supplier to the beneficiary. The blockchain is responsible for storing logs and events generated by smart contracts, and record delivery events of the COVID-19 vaccine. The management of vaccine is a special case of supply chain management, we will investigate more blockchain works on supply chain in Section~\ref{sec:supp}.

\subsection{Social Network}
Social network has become an indispensable part of our daily lives. Users of social media, such as Facebook, Twitter and Weibo set up their profiles and make posts. The huge amount of data generated by users are managed by the social media providers which are sometimes not reliable. For instance, Facebook has several data leakage incidents recent years. Users have no control of their data even some data are of high privacy concern.
To solve the single-point failure problem, ensure data security and preserve necessary privacy in public social network, blockchain technology is discovered to be one possible solution~\cite{DBLP:journals/isci/LaxRF21,DBLP:journals/percom/Guidi20} .

Jiang et al.~\cite{DBLP:journals/tcss/JiangZ19} design a blockchain-based decentralized social network, where blockchain serves as a replacement centralized server to allow user registration, user posting, adding friends or commenting with the help of smart contracts. In the evaluation, the authors prove each post users made will cost around 1.137 USD, which makes the system not a budget solution. The data stored on blockchain is not modifiable, therefore how users update their registration information and posts is a remaining problem. Zhang et al.~\cite{DBLP:journals/connection/ZhangYSWLS21} propose another blockchain-based social network, namely BPP. They also propose a privacy preserving searching algorithm in BPP. However, BPP is not fully decentralized, blockchain works as external storage system to assist social network providers.
Most current designs or implementations of blockchain in social network system are still at very fist stage that blockchain only takes a minor part of the system  and have many deficiencies such as scalabity and throughput problems. With the development of blockchain system, handful fully decentralized social networks are able to come to surface, such as Steemit~\footnote{\url{ https://steemit.com}} and FORESTING~\footnote{\url{https://foresting.io/}} which are both implemented with Steem Blockchain~\cite{kiayias2018puff}. The diagram of social network without a central server requires all peers to maintain the whole network, which gives much more rights to peers than centralized diagram. With no proper user behavior control mechanism, malicious users might make the blockchain-enabled social network more vulnerable than traditional ones.

Gu et al.~\cite{DBLP:journals/access/GuWJ19} study privacy concern during resource sharing, such as moves, songs or pictures within social network communities. Access control of the resource is achieved by blockchain, and peers within the communities are motivated by a smart contract to help disseminate the resources. Rahman et al.~\cite{DBLP:journals/jpdc/RahmanGB20} also study access control problem in social network with blockchain. They design four smart contracts, namely Access Control Contract for controlling the access to the resources in the network, Reputation Contract for calculating reputation score of users, Inspector Contract for monitoring user behavior and  Registrar Contract for verifying user identities. Zhang el al.~\cite{DBLP:conf/infocom/ZhangSLNLXZ22} extend the resource sharing within one community to the sharing across multiple social networks. Their proposed framework achieves consistent consensus on photo dissemination control across independent and disparate social network platforms.
Yan et al.~\cite{DBLP:journals/toit/YanPFY21} propose Social-Chain to solve the trust issue in Pervasive Social Networking (PSN) where there is usually lacking a a centralized party to perform information collection, social data aggregation, and trust evaluation. Social-Chain is able to efficiently store the trust evaluation of users into blockchain with the proposed Proof-of-Trust. Guo et al.~\cite{DBLP:conf/bsci/GuoXWZ22} study user reputation evaluation task in social network where existing methods are facing the issue caused by fake comments posted by adversaries. They propose a consortium blockchain-based method that users are the peers in the blockchain and a behavior game model mechanism is developed to motivate peers to work honestly.

Chen et al.~\cite{DBLP:journals/isci/ChenXLWH19} propose DEPLEST, a blockchain-based distributed database system. They focus on solving the storage cost if all users of social network stores whole copies of database in blockchain-based social network. They propose each user only need to store a part whole data, and the size of storage on each device is fixed. To save the storage cost of synchronized blockchain, only sensitive data will be encrypted and secured through blockchain, non-sensitive data will be stored in traditional external database. Proof-of-Communication is also proposed to save the time for appending a new block. Nguyen et al.~\cite{DBLP:conf/sigir/NguyenBAND22} propose SoChainDB, which is a general database framework to facilitate blockchain-based social networks for collecting data generated in the network. SochainDB provides a efficient pipeline to crawl and formalize distributed data storage in blockchain-based social network, and fills in the gap between conventional social network engineers and blockchain developers.

Ochoa et al.~\cite{DBLP:conf/quatic/OchoaMSGFL19} propose FakeChain to detect fake news in social network. They assume each node in social network is also a node in blockchain, and when each node publish some news, the news will be stored in blockchain. Then, with the tractability of blockchain, the source of fake news can be easily detected.

\begin{figure*}[htb!]
    \centering
    \includegraphics[width = 0.75\linewidth]{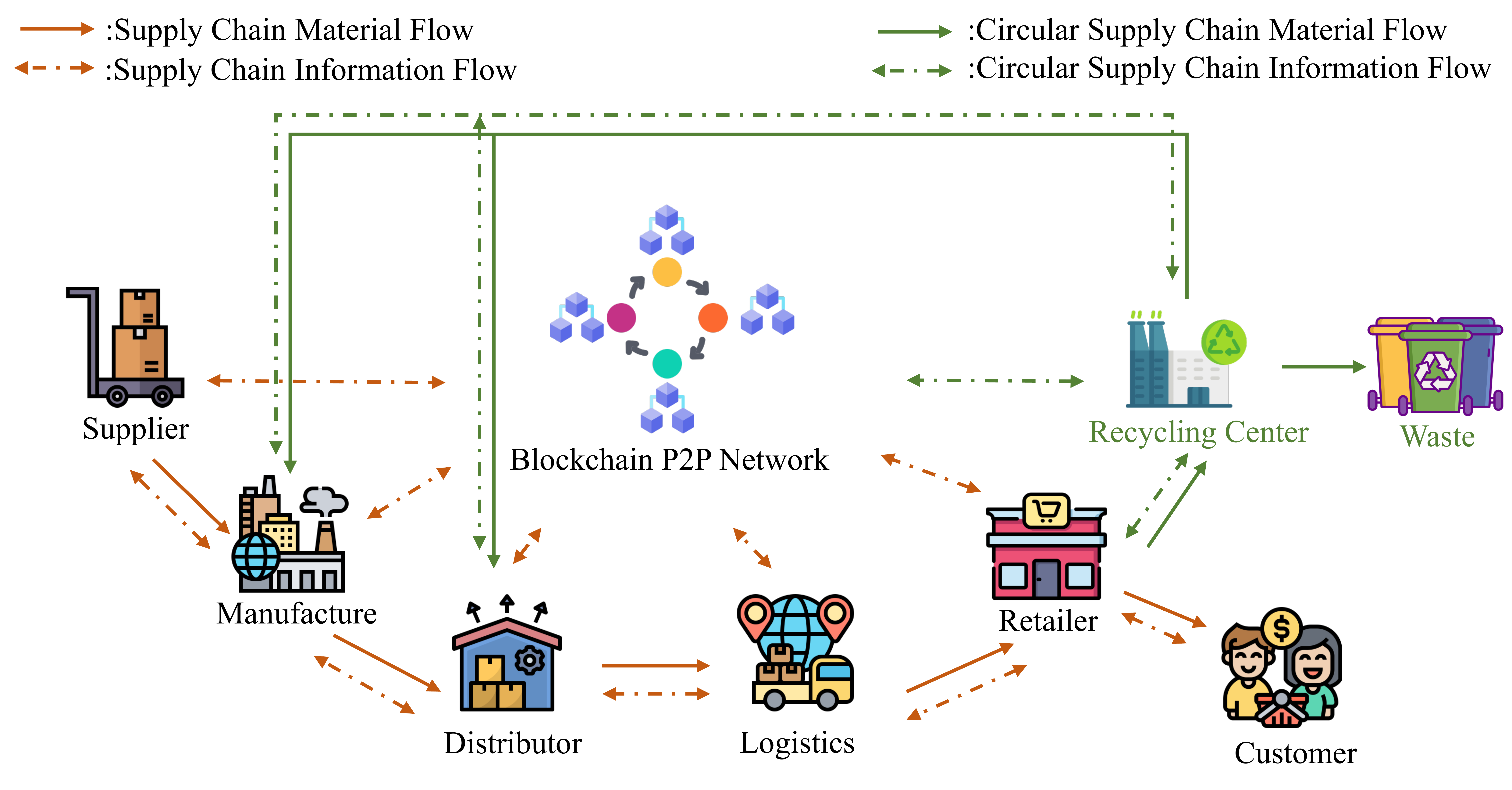}
    \caption{Blockchain-enabled (circular) supply chain.}
    \label{fig:sup_chain}
\end{figure*}

\subsection{Supply Chain}
\label{sec:supp}
Supply chain management is a vital component of industry. Good traceability of products allows manufactures and retailers to identify the source of parts and raw materials, as well as avoid business fraud with transparent product information. With the increasing complexity of global supply chain networks, traditional supply chain management approaches are facing challenges to match the requirements of efficiency, accessibility, transparency and security~\cite{lambert2017issues,ivanov2019impact}.

Blockchain can be a revolutionary technique in supply chain to provide effective product tracing, transparent information sharing and reliable attack resistance. Many companies have announced blockchain projects in their supply chain for better tractability, such as Walmart~\footnote{\url{https://one.walmart.com/content/globaltechindia/en_in/Tech-insights/blog/Blockchain-in-the-food-supply-chain.html}}, Toyota and Alibaba~\cite{DBLP:journals/itpro/KshetriL19}. Queiroz et al.~\cite{Queiroz2021BlockchainAI} systematically analyzed the barrier for the adoption of blockchain in supply chain with a real-world empirical study in the Brazilian OSCM (Operations and Supply Chain Management) context. Through questionnaire, they found out that the performance expectancy may also constitute an impediment to adoption of blockchain.  In Wu's theoretical analysis~\cite{Wu2021AnAO} blockchain technology is able to elevate the profit of supply chain, though the profit may differ when different parties lead the construction of the blockchain system.

One of the most fundamental property that supply chains ought to have is traceability. Abundant blockchain-enabled supply chains have been investigated and designed for many specific use cases to improve traceability, including fresh produces and foods ~\cite{Wu2021AnAO, Casino2021BlockchainbasedFS, DBLP:journals/computers/SekuloskaE22, DBLP:journals/computers/SekuloskaE22}, animal products~\cite{marinello2017development}, agriculture~\cite{kamble2020modeling} and healthcare~\cite{DBLP:journals/access/MusamihJSHYA21, DBLP:journals/candie/OmarDJSOA22}. Yakubu et al.~\cite{DBLP:journals/peerj-cs/YakubuLYKM22} propose RiceChain to provide traceable rice supply chain, achieving at most 25\% lower tracing latency than existing work. Caro et al. ~\cite{Caro2018BlockchainbasedTI} propose AgriBlockIoT and implement with Ethereum and Hyperledger Sawtooth, which is a fully-decentralized traceable supply chain for agriculture and food with blockchain and IoT. In AgriBlockIoT, IoT devices deployed in the supply chain processes are working nodes of blockchain, and the whole blockchain is maintained on cloud.
To evaluate the traceability level of blockchain-enabled supply chain, Dasaklis et al.~\cite{Dasaklis2019DefiningGL} define the granularity levels of traceability. They design a smart contract to collect necessary information for conduct traceability classification based on the existing traceability granularity standard\footnote{\url{https://www.gs1.org/docs/tl/T_L_Keys_Implementation_Guideline.pdf}}.

Kouhizadenh et al.~\cite{Kouhizadeh2021BlockchainTA} and Saberi et al.~\cite{Saberi2019BlockchainTA} analyze that blockchain could help verify audit and certificate sustainability in supply chain which is emphasized of great importance recent years, yet hardly can be achieved in most supply chain systems. The key idea of sustainability of supply chain is to save energy, build environmental-friendly products in sustainable manner. However, in addition to the benefits brought by blockchain for supporting sustainability, blockchain itself may introduce extra overhead and energy consumption for maintaining functionality of smart contracts and miners. Some other problems such as scalability, communication deficiency and unprecedented security issues may also present~\cite{Rana2021BlockchainTF}.

The interactions between multiple partners in the supply chain rely on the trust among them. Therefore trust management plays the crucial rule to build and maintain a supply chain. Beside the above blockchain-based tamper-proof and audible supply chain is proposed, Malik et al.~\cite{DBLP:conf/blockchain2/MalikDKJ19} consider the trust evaluation problem for parties in the supply chain. In the propose TrustChain, a reputation evaluation algorithm is developed for calculating the ratings of product sellers in the blockchain-based supply chain. AlRakhami et al.~\cite{AlRakhami2021ABT} bring both blockchain and IoT into trust management in supply chain. In their work, IoT devices work as data collector and relayor to transmit cryptographically
edited essential data to blockchain. However, incorporating IoT devices may introducing potential security threats into supply chain that unauthorized or uncontrolled IoT devices from malicious parties may access and tamper the sensitive data. Song et al.~\cite{Song2021ASS} propose a robust blockchain based IoT enabled supply chain management framework, where a registration module is designed for enforcing registration policies on all participants and inspection module is designed for monitoring, analyzing and judging misbehaviour of participants.

Circular supply chain extends the one-way supply that from manufacture to products with three extra process, namely recycle, remanufacture and redistribute. Different from the simple forward manner in common supply chains, circular supply chains usually have more complicated product information flow among multiple parties that may back and forth when redistributing and recycling. Figure~\ref{fig:sup_chain} illustrates the information flow and material flow in regular supply chain and circular supply chain. Centobelli et al.~\cite{Centobelli2021BlockchainTF} design a Triple Retry blockchain framework with multiple smart contracts for executing different processes.

Raj et al.~\cite{DBLP:journals/candie/RajJRP22} propose to use blockchain to solve the payment delay issue in supply chain especially when the participants locating in different place of the globe. With smart contracts enforced on participants as well as the authenticity and tamper-proof nature, blockchain is able to achieve trustable information and payment confirming which alleviates the transaction delays in supply chain.

Despite abundant blockchain-based supply chains are designed and implemented, the profiting outcome with the adoption of blockchain technology is still question to be answered. Researchers may result in opposite conclusions when modeling and scenario settings differ. Zhou et al.~\cite{DBLP:journals/ecra/ZhouLZS22} and Sun et al.~\cite{DBLP:journals/apjor/SunXS22} conduct detailed analysis and proof on the impact of blockchain in two-echelon supply chain where only one supplier and one retailer exist. Zhou et al.~\cite{DBLP:journals/ecra/ZhouLZS22} point out that, blockchain enabled supply chain not necessarily superior than non-blockchain ones, which is closely  related to the  reliability of information and the transparency cost of products. Based on full equilibrium  results  through the Stackelberg game  between the supplier and the retailer designed in the paper, the authors conclude that the retailer can tolerant higher blockchain adopting cost than supplier in most cases especially when the product cost is high, while supplier can tolerant a higher adopting cost only when both the product cost and consumers’ willing to pay are low. Zhou et al.~\cite{DBLP:journals/ecra/ZhouLZS22} thinks both supplier and retailer will benefit from blockchain technology when the adopting cost is low enough, while Sun et al.~\cite{DBLP:journals/apjor/SunXS22} concludes blockchain technology can always improves the supply chain profit no matter the status of market demand.

\begin{figure*}[htb!]
    \centering
    \includegraphics[width = 0.7\linewidth]{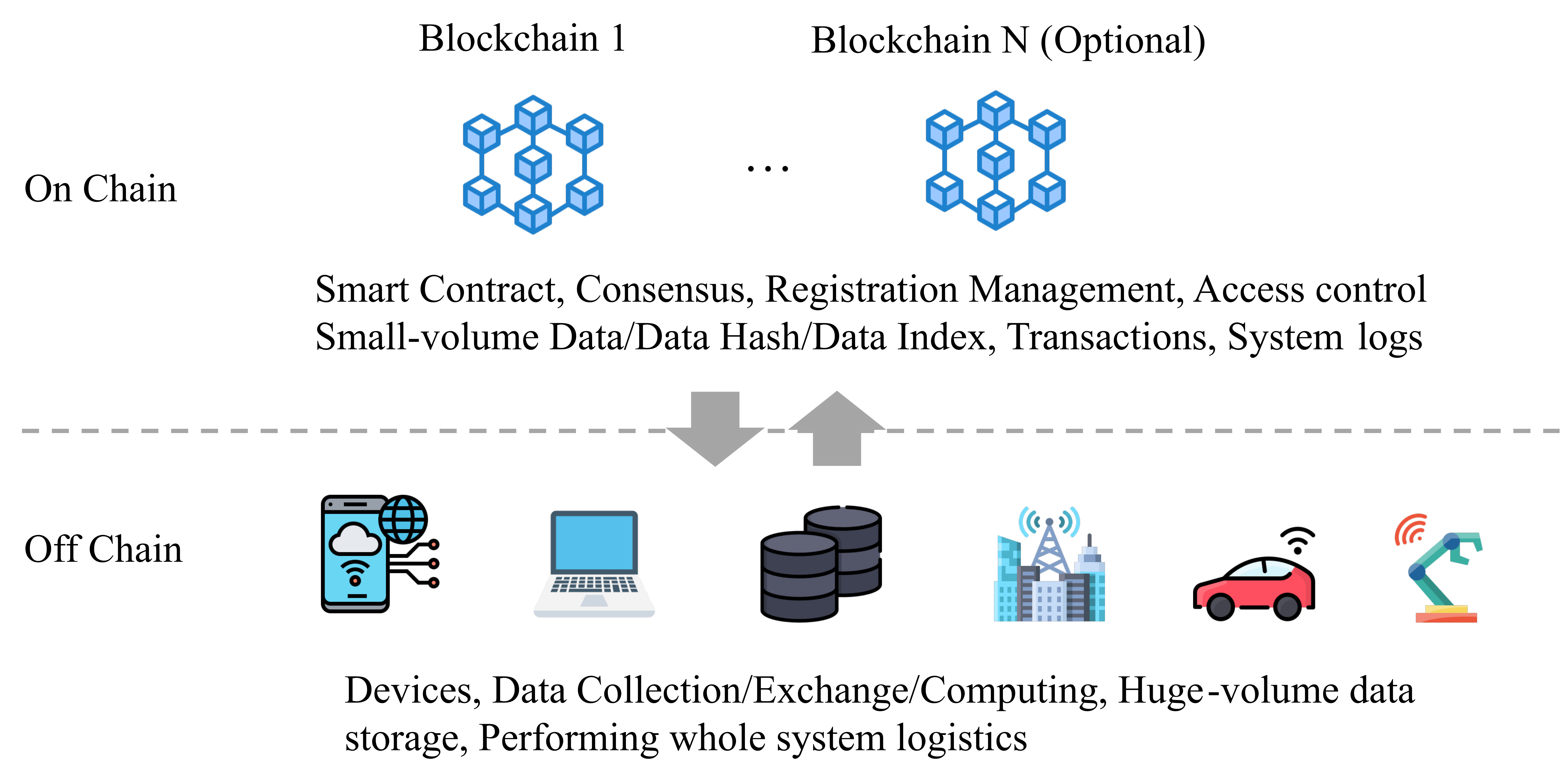}
    \caption{The On-Chain/Off-chain architecture of blockchain applications.}
    \label{fig:onoffchain}
\end{figure*}

\section{Discussion}
\label{sec:disc}
In above extensive literature review, we have investigated and discussed recent advances of blockchain and its applications as well as providing some suggestions for some specific research topics.  In this section, we make further conclusions and more general foresight of future academic or industry work on Blockchain.

Blockchain systems are constructed upon peer-to-peer networks, where smart contracts and consensus mechanisms are enforced on every participants to achieve transparency. Some consensus mechanisms, especially Proof-of-Work and its variants, will consume notable computation resources. Though many other consensus mechanisms are proposed, such as Proof of Stake or PBFT which significantly reduce computation cost, they also bring extra communication resource consumption. The communication or communication consumption are not affordable to many lightweight applications. On the other hand, applications with resource-limited devices are often not able to consistently perform stable communications or computation tasks. For example, in a IoT network, the out-door smart meters or smart sensors not only have limited computation resources, and but may face challenges to collect data or loss network connection due to bad weather, which will fail effective communications. To mitigate this problem and make blockchain effective in these use cases, future work may consider to simultaneously optimize blockchain consensus mechanism and communication schema.

Another obstacle to the adoption of blockchain technology is the scalability issue. In other words, the transaction throughput, the transaction processing latency and the storage cost of blockchain can not all satisfy the demands of many use cases. We have mentioned many remarkable works above that improves the scalability. Efficient consensus mechanisms, such as PoS,  DPoS and community-wise consensus mechanisms are developed to package transactions into blocks in much shorter time than PoW. In order to mitigate the total storage consumption meanwhile to keep the robustness and tamper-proof ability, literature proposes blockchain sharding methods and storing actual data at external databases while only keeping the data index on blockchain. However, it is hard to achieve the perfect balance among three key properties: decentralization, security, and scalability~\cite{DBLP:journals/access/ZhouHZB20}. Most above-mentioned works are able to optimize one or two of them under given particular application scenarios, and only handful works are trying to optimize all three properties at the same time~\cite{DBLP:journals/tii/LiuYTLS19}, which are still initial attempts under many constraints, such as only limited number of consensus mechanisms are considered and the block size is assumed to be only discrete numbers. New consensus mechanisms are highly demanded by optimizing all three properties under much more general settings and use cases.

The industry is no doubt a critical rule in the adoption of blockchain technology from theory to practice. The main concern of companies to adopt blockchain in IoT services, healthcare systems or supply chain systems is if blockchain will bring more profit than the cost to build it. The cost of blockchain technology are seldom analyzed in existing literature.  We have mentioned several works that theoretically analyze the potential benefits and cost of blockchain in supply chain~\cite{DBLP:journals/ecra/ZhouLZS22, DBLP:journals/apjor/SunXS22, Wu2021AnAO}. Further extensive research can be done in many other applications such as IoT, edge computing and  healthcare. In addition, blockchain simulation tools are highly desired for helping evaluate the performance of blockchains system, which provides intuitive results for industry to understand the performance and cost of blockchain systems.

Apparently, lower the cost of blockchain will promote the adoption of blockchain in industry. The cost to build a blockchain generally comes from the development of blockchain client for each working node, the computation resource to perform consensus mechanism, the storage resource to store the blocks and the cost of incentive mechanism to reward working nodes if necessary. The cost for developing blockchain client is mostly decided by software engineering market price. Therefore better consensus mechanisms and corresponding incentive mechanism take great weights in lowering the cost for blockchain industry.

Blockchain is a third-party free, non-trust built, distributed data management approach. The adoption of blockchain technology not only brings benefits, but also potential risks and security weakness due to anonymity. Though most popular consensus mechanisms are proved able to resist dishonest users when the ratio is under 51\% or 1/3, the resistance to cyber-security attacks such as registration attack, data leakage and encryption break-through is still a question. Different from the most famous successful blockchain system such as Bitcoin and Ethereum which run on high-end computers, the blockchain applications in IoT, edge computing,  healthcare and supply chain usually involves tremendous edge devices, such as mobile smart phones, IoT smart devices or network access points, which can be easily compromised. Future academy and industry may work together to study those external cyber-security attacks in blockchain systems.

In many existing blockchain-enabled applications, we found that blockchain serves as an external distributed storage approach to simply replace the traditional storage instead of being specially designed to be integrated into application logistics.  In this architecture, blockchain assists computation server by providing authentication control, data indexing, system logging without defining specific consensus mechanisms and incentive mechanisms. In other words, the participants or peers in the blockchain can not actively compute and generate data, while only passively take the data given by the computation server.
Some works take On-chain/Off-chain architecture for their applications as illustrated in Figure~\ref{fig:onoffchain}, that uses blockchain for making important consensus and storing important system logs or transaction, while still keeps the logistics and massive data on off-chain devices. Though this architecture is good for alleviating consumption of blockchain system in term of computation and storage resources, the blockchain does not directly participate in system logistics.
We believe the blockchain-enabled applications can put more functions on blockchain through smart contracts to decentralize the computation power and take full advantages of blockchain technology.

It is also worth-mentioning that the advantages of blockchain are not necessarily the benefits for applications sometimes. For example, blockchain is tamper-proof, that everything stored on blockchain can never be modified or deleted in anyway, otherwise the chain rule will be broken due to the uniqueness of block hash. For the application of social network, it is common that people can leave the social network and need to erase all the social records. However blockchain-enabled social networks are hard to achieve this as long as there is any data of users stored on blockchain.
Another example is that blockchain is decentralized that ideally requires every participants to store the full copy of
blockchain. However, in real world, many use cases mentioned above can not satisfy the ideal situation that the storage cost will soon become unaffordable if the blockchain stores all data. In addition, in some use cases, such as healthcare, it is not always secure to allow everyone holds full copy of data, since some sensitive information of patients may not be supposed to be accessible to some particular parties. We suggest future work may develop variants of current popular blockchain systems to meet the demands of particular use cases.

\section{Conclusion}
\label{sec:con}
In this paper, we created an overview picture of blockchain ecosystem by reviewing the recent advances of blockchain technology as well as the most active blockchain applications overlapped with each other. With the steep expanding of the whole blockchain ecosystem, it is of great meaning to review the development in the most noticeable parts in the ecosystem. We first reviewed the recent studies on general blockchain technology, then the blockchain-enabled applications, including IoT, IoV, edge computing, federated learning, healthcare, COVID-19, social network and supply chain. With the extensive review, we suggested several foreseeing problems for future developments of blockchain ecosystem, including the dilemma to achieve balance among scalability, security, dencentrality and cost, the external security risks outside blockchain from cyber-attack in industry, the ignorance of blockchain smart contracts and the unexpected disadvantages caused by blockchain inevitable properties. This paper toward weaving the core part of current blockchain ecosystem from both academic research to frontier industry applications. Therefore we expect this survey could be helpful for future researchers developing more and better blockchain-enabled applications.

\bibliographystyle{IEEEtran}
\bibliography{ref}

\begin{thebibliography}{100}
\providecommand{\url}[1]{#1}
\csname url@samestyle\endcsname
\providecommand{\newblock}{\relax}
\providecommand{\bibinfo}[2]{#2}
\providecommand{\BIBentrySTDinterwordspacing}{\spaceskip=0pt\relax}
\providecommand{\BIBentryALTinterwordstretchfactor}{4}
\providecommand{\BIBentryALTinterwordspacing}{\spaceskip=\fontdimen2\font plus
\BIBentryALTinterwordstretchfactor\fontdimen3\font minus
  \fontdimen4\font\relax}
\providecommand{\BIBforeignlanguage}[2]{{%
\expandafter\ifx\csname l@#1\endcsname\relax
\typeout{** WARNING: IEEEtran.bst: No hyphenation pattern has been}%
\typeout{** loaded for the language `#1'. Using the pattern for}%
\typeout{** the default language instead.}%
\else
\language=\csname l@#1\endcsname
\fi
#2}}
\providecommand{\BIBdecl}{\relax}
\BIBdecl

\bibitem{nakamoto2008bitcoin}
S.~Nakamoto, ``Bitcoin: A peer-to-peer electronic cash system,''
  \emph{Decentralized Business Review}, p. 21260, 2008.

\bibitem{wood2014ethereum}
G.~Wood \emph{et~al.}, ``Ethereum: A secure decentralised generalised
  transaction ledger,'' \emph{Ethereum project yellow paper}, vol. 151, no.
  2014, pp. 1--32, 2014.

\bibitem{King2012PPCoinPC}
S.~King and S.~Nadal, ``Ppcoin: Peer-to-peer crypto-currency with
  proof-of-stake,'' 2012.

\bibitem{Schr2021DecentralizedFO}
F.~Sch{\"a}r, ``Decentralized finance: On blockchain- and smart contract-based
  financial markets,'' \emph{Review}, 2021.

\bibitem{DBLP:journals/iotj/XuLL21}
\BIBentryALTinterwordspacing
L.~D. Xu, Y.~Lu, and L.~Li, ``Embedding blockchain technology into iot for
  security: {A} survey,'' \emph{{IEEE} Internet Things J.}, vol.~8, no.~13, pp.
  10\,452--10\,473, 2021. [Online]. Available:
  \url{https://doi.org/10.1109/JIOT.2021.3060508}
\BIBentrySTDinterwordspacing

\bibitem{Uddin2021ASO}
M.~A. Uddin, A.~Stranieri, I.~Gondal, and V.~Balasubramanian, ``A survey on the
  adoption of blockchain in iot: Challenges and solutions,'' 2021.

\bibitem{DBLP:journals/comsur/YangYSYZ19}
\BIBentryALTinterwordspacing
R.~Yang, F.~R. Yu, P.~Si, Z.~Yang, and Y.~Zhang, ``Integrated blockchain and
  edge computing systems: {A} survey, some research issues and challenges,''
  \emph{{IEEE} Commun. Surv. Tutorials}, vol.~21, no.~2, pp. 1508--1532, 2019.
  [Online]. Available: \url{https://doi.org/10.1109/COMST.2019.2894727}
\BIBentrySTDinterwordspacing

\bibitem{DBLP:conf/otm/BelchiorCV19}
\BIBentryALTinterwordspacing
R.~Belchior, M.~Correia, and A.~Vasconcelos, ``Justicechain: Using blockchain
  to protect justice logs,'' in \emph{On the Move to Meaningful Internet
  Systems: {OTM} 2019 Conferences - Confederated International Conferences:
  CoopIS, ODBASE, C{\&}TC 2019, Rhodes, Greece, October 21-25, 2019,
  Proceedings}, ser. Lecture Notes in Computer Science, H.~Panetto,
  C.~Debruyne, M.~Hepp, D.~Lewis, C.~A. Ardagna, and R.~Meersman, Eds., vol.
  11877.\hskip 1em plus 0.5em minus 0.4em\relax Springer, 2019, pp. 318--325.
  [Online]. Available: \url{https://doi.org/10.1007/978-3-030-33246-4\_21}
\BIBentrySTDinterwordspacing

\bibitem{DBLP:conf/ecis/Belchior0V20}
\BIBentryALTinterwordspacing
------, ``Towards secure, decentralized, and automatic audits with
  blockchain,'' in \emph{28th European Conference on Information Systems -
  Liberty, Equality, and Fraternity in a Digitizing World, {ECIS} 2020,
  Marrakech, Morocco, June 15-17, 2020}, F.~Rowe, R.~E. Amrani, M.~Limayem,
  S.~Newell, N.~Pouloudi, E.~van Heck, and A.~E. Quammah, Eds., 2020. [Online].
  Available: \url{https://aisel.aisnet.org/ecis2020\_rp/68}
\BIBentrySTDinterwordspacing

\bibitem{DBLP:journals/jms/HussienYUZZ19}
\BIBentryALTinterwordspacing
H.~M. Hussien, S.~M. Yasin, N.~I. Udzir, A.~A. Zaidan, and B.~B. Zaidan, ``A
  systematic review for enabling of develop a blockchain technology in
  healthcare application: Taxonomy, substantially analysis, motivations,
  challenges, recommendations and future direction,'' \emph{J. Medical Syst.},
  vol.~43, no.~10, pp. 320:1--320:35, 2019. [Online]. Available:
  \url{https://doi.org/10.1007/s10916-019-1445-8}
\BIBentrySTDinterwordspacing

\bibitem{DBLP:journals/mms/JabbarLHAR21}
\BIBentryALTinterwordspacing
S.~Jabbar, H.~Lloyd, M.~Hammoudeh, B.~Adebisi, and U.~Raza,
  ``Blockchain-enabled supply chain: analysis, challenges, and future
  directions,'' \emph{Multim. Syst.}, vol.~27, no.~4, pp. 787--806, 2021.
  [Online]. Available: \url{https://doi.org/10.1007/s00530-020-00687-0}
\BIBentrySTDinterwordspacing

\bibitem{DBLP:journals/winet/WanLLW20}
\BIBentryALTinterwordspacing
S.~Wan, M.~Li, G.~Liu, and C.~Wang, ``Recent advances in consensus protocols
  for blockchain: a survey,'' \emph{Wirel. Networks}, vol.~26, no.~8, pp.
  5579--5593, 2020. [Online]. Available:
  \url{https://doi.org/10.1007/s11276-019-02195-0}
\BIBentrySTDinterwordspacing

\bibitem{DBLP:journals/jsa/SinghKSCAT22}
\BIBentryALTinterwordspacing
A.~Singh, G.~Kumar, R.~Saha, M.~Conti, M.~Alazab, and R.~Thomas, ``A survey and
  taxonomy of consensus protocols for blockchains,'' \emph{J. Syst. Archit.},
  vol. 127, p. 102503, 2022. [Online]. Available:
  \url{https://doi.org/10.1016/j.sysarc.2022.102503}
\BIBentrySTDinterwordspacing

\bibitem{DBLP:journals/access/ZhouHZB20}
\BIBentryALTinterwordspacing
Q.~Zhou, H.~Huang, Z.~Zheng, and J.~Bian, ``Solutions to scalability of
  blockchain: {A} survey,'' \emph{{IEEE} Access}, vol.~8, pp. 16\,440--16\,455,
  2020. [Online]. Available: \url{https://doi.org/10.1109/ACCESS.2020.2967218}
\BIBentrySTDinterwordspacing

\bibitem{DBLP:journals/csur/ZhangXL19}
\BIBentryALTinterwordspacing
R.~Zhang, R.~Xue, and L.~Liu, ``Security and privacy on blockchain,''
  \emph{{ACM} Comput. Surv.}, vol.~52, no.~3, pp. 51:1--51:34, 2019. [Online].
  Available: \url{https://doi.org/10.1145/3316481}
\BIBentrySTDinterwordspacing

\bibitem{DBLP:journals/jnca/FengHZKK19}
\BIBentryALTinterwordspacing
Q.~Feng, D.~He, S.~Zeadally, M.~K. Khan, and N.~Kumar, ``A survey on privacy
  protection in blockchain system,'' \emph{J. Netw. Comput. Appl.}, vol. 126,
  pp. 45--58, 2019. [Online]. Available:
  \url{https://doi.org/10.1016/j.jnca.2018.10.020}
\BIBentrySTDinterwordspacing

\bibitem{DBLP:journals/sncs/GamageWD20}
\BIBentryALTinterwordspacing
H.~T.~M. Gamage, H.~D. Weerasinghe, and N.~G.~J. Dias, ``A survey on blockchain
  technology concepts, applications, and issues,'' \emph{{SN} Comput. Sci.},
  vol.~1, no.~2, p. 114, 2020. [Online]. Available:
  \url{https://doi.org/10.1007/s42979-020-00123-0}
\BIBentrySTDinterwordspacing

\bibitem{DBLP:journals/comsur/HuoZWSCHWYL22}
\BIBentryALTinterwordspacing
R.~Huo, S.~Zeng, Z.~Wang, J.~Shang, W.~Chen, T.~Huang, S.~Wang, F.~R. Yu, and
  Y.~Liu, ``A comprehensive survey on blockchain in industrial internet of
  things: Motivations, research progresses, and future challenges,''
  \emph{{IEEE} Commun. Surv. Tutorials}, vol.~24, no.~1, pp. 88--122, 2022.
  [Online]. Available: \url{https://doi.org/10.1109/COMST.2022.3141490}
\BIBentrySTDinterwordspacing

\bibitem{DBLP:journals/ejwcn/WangCLHX21}
\BIBentryALTinterwordspacing
C.~Wang, X.~Cheng, J.~Li, Y.~He, and K.~Xiao, ``A survey: applications of
  blockchain in the internet of vehicles,'' \emph{{EURASIP} J. Wirel. Commun.
  Netw.}, vol. 2021, no.~1, p.~77, 2021. [Online]. Available:
  \url{https://doi.org/10.1186/s13638-021-01958-8}
\BIBentrySTDinterwordspacing

\bibitem{DBLP:journals/iotj/MollahZNGYSLK21}
\BIBentryALTinterwordspacing
M.~B. Mollah, J.~Zhao, D.~Niyato, Y.~L. Guan, C.~Yuen, S.~Sun, K.~Lam, and
  L.~H. Koh, ``Blockchain for the internet of vehicles towards intelligent
  transportation systems: {A} survey,'' \emph{{IEEE} Internet Things J.},
  vol.~8, no.~6, pp. 4157--4185, 2021. [Online]. Available:
  \url{https://doi.org/10.1109/JIOT.2020.3028368}
\BIBentrySTDinterwordspacing

\bibitem{DBLP:journals/comsur/AlladiCSVGG22}
\BIBentryALTinterwordspacing
T.~Alladi, V.~Chamola, N.~Sahu, V.~Venkatesh, A.~Goyal, and M.~Guizani, ``A
  comprehensive survey on the applications of blockchain for securing vehicular
  networks,'' \emph{{IEEE} Commun. Surv. Tutorials}, vol.~24, no.~2, pp.
  1212--1239, 2022. [Online]. Available:
  \url{https://doi.org/10.1109/COMST.2022.3160925}
\BIBentrySTDinterwordspacing

\bibitem{DBLP:journals/csur/ZouHZKWC22}
\BIBentryALTinterwordspacing
J.~Zou, D.~He, S.~Zeadally, N.~Kumar, H.~Wang, and K.~R. Choo, ``Integrated
  blockchain and cloud computing systems: {A} systematic survey, solutions, and
  challenges,'' \emph{{ACM} Comput. Surv.}, vol.~54, no.~8, pp. 160:1--160:36,
  2022. [Online]. Available: \url{https://doi.org/10.1145/3456628}
\BIBentrySTDinterwordspacing

\bibitem{DBLP:journals/tnsm/LiaoPZXW22}
\BIBentryALTinterwordspacing
Z.~Liao, X.~Pang, J.~Zhang, B.~Xiong, and J.~Wang, ``Blockchain on security and
  forensics management in edge computing for iot: {A} comprehensive survey,''
  \emph{{IEEE} Trans. Netw. Serv. Manag.}, vol.~19, no.~2, pp. 1159--1175,
  2022. [Online]. Available: \url{https://doi.org/10.1109/TNSM.2021.3122147}
\BIBentrySTDinterwordspacing

\bibitem{DBLP:journals/soco/LiHWZLLCL22}
\BIBentryALTinterwordspacing
D.~Li, D.~Han, T.~Weng, Z.~Zheng, H.~Li, H.~Liu, A.~Castiglione, and K.~Li,
  ``Blockchain for federated learning toward secure distributed machine
  learning systems: a systemic survey,'' \emph{Soft Comput.}, vol.~26, no.~9,
  pp. 4423--4440, 2022. [Online]. Available:
  \url{https://doi.org/10.1007/s00500-021-06496-5}
\BIBentrySTDinterwordspacing

\bibitem{DBLP:journals/iotj/NguyenDPPLSLNP21}
\BIBentryALTinterwordspacing
D.~C. Nguyen, M.~Ding, Q.~Pham, P.~N. Pathirana, L.~B. Le, A.~Seneviratne,
  J.~Li, D.~Niyato, and H.~V. Poor, ``Federated learning meets blockchain in
  edge computing: Opportunities and challenges,'' \emph{{IEEE} Internet Things
  J.}, vol.~8, no.~16, pp. 12\,806--12\,825, 2021. [Online]. Available:
  \url{https://doi.org/10.1109/JIOT.2021.3072611}
\BIBentrySTDinterwordspacing

\bibitem{DBLP:journals/ijbis/SreerakhiBM22}
\BIBentryALTinterwordspacing
V.~Sreerakhi, N.~Balagopal, and A.~Mohan, ``Transforming supply chain network
  and logistics using blockchain - a survey,'' \emph{Int. J. Bus. Inf. Syst.},
  vol.~39, no.~2, pp. 193--218, 2022. [Online]. Available:
  \url{https://doi.org/10.1504/IJBIS.2022.121430}
\BIBentrySTDinterwordspacing

\bibitem{DBLP:conf/dsaa/MaesaMR16}
\BIBentryALTinterwordspacing
D.~D.~F. Maesa, A.~Marino, and L.~Ricci, ``Uncovering the bitcoin blockchain:
  An analysis of the full users graph,'' in \emph{2016 {IEEE} International
  Conference on Data Science and Advanced Analytics, {DSAA} 2016, Montreal, QC,
  Canada, October 17-19, 2016}.\hskip 1em plus 0.5em minus 0.4em\relax {IEEE},
  2016, pp. 537--546. [Online]. Available:
  \url{https://doi.org/10.1109/DSAA.2016.52}
\BIBentrySTDinterwordspacing

\bibitem{DBLP:journals/ans/MaesaMR19}
\BIBentryALTinterwordspacing
------, ``The bow tie structure of the bitcoin users graph,'' \emph{Appl. Netw.
  Sci.}, vol.~4, no.~1, pp. 56:1--56:22, 2019. [Online]. Available:
  \url{https://doi.org/10.1007/s41109-019-0163-y}
\BIBentrySTDinterwordspacing

\bibitem{DBLP:conf/bsci/ChenL19}
\BIBentryALTinterwordspacing
Y.~Chen and J.~Liu, ``Distributed community detection over blockchain networks
  based on structural entropy,'' in \emph{Proceedings of the 2019 {ACM}
  International Symposium on Blockchain and Secure Critical Infrastructure,
  {BSCI} 2019, Auckland, New Zealand, July 8, 2019}, K.~Gai, K.~R. Choo, D.~He,
  and S.~Wen, Eds.\hskip 1em plus 0.5em minus 0.4em\relax {ACM}, 2019, pp.
  3--12. [Online]. Available: \url{https://doi.org/10.1145/3327960.3332381}
\BIBentrySTDinterwordspacing

\bibitem{DBLP:conf/icbc2/PontiverosSS19}
\BIBentryALTinterwordspacing
B.~B.~F. Pontiveros, M.~Steichen, and R.~State, ``Mint centrality: {A}
  centrality measure for the bitcoin transaction graph,'' in \emph{{IEEE}
  International Conference on Blockchain and Cryptocurrency, {ICBC} 2019,
  Seoul, Korea (South), May 14-17, 2019}.\hskip 1em plus 0.5em minus
  0.4em\relax {IEEE}, 2019, pp. 159--162. [Online]. Available:
  \url{https://doi.org/10.1109/BLOC.2019.8751401}
\BIBentrySTDinterwordspacing

\bibitem{Li2020DissectingEB}
Y.~Li, U.~Islambekov, C.~G. Akcora, E.~Smirnova, Y.~R. Gel, and
  M.~Kantarcioglu, ``Dissecting ethereum blockchain analytics: What we learn
  from topology and geometry of ethereum graph,'' in \emph{SDM}, 2020.

\bibitem{DBLP:conf/icbc2/BannoS19}
\BIBentryALTinterwordspacing
R.~Banno and K.~Shudo, ``Simulating a blockchain network with simblock,'' in
  \emph{{IEEE} International Conference on Blockchain and Cryptocurrency,
  {ICBC} 2019, Seoul, Korea (South), May 14-17, 2019}.\hskip 1em plus 0.5em
  minus 0.4em\relax {IEEE}, 2019, pp. 3--4. [Online]. Available:
  \url{https://doi.org/10.1109/BLOC.2019.8751431}
\BIBentrySTDinterwordspacing

\bibitem{DBLP:journals/jpdc/LiuTBY20}
\BIBentryALTinterwordspacing
L.~Liu, W.~Tsai, M.~Z.~A. Bhuiyan, and D.~Yang, ``Automatic blockchain
  whitepapers analysis via heterogeneous graph neural network,'' \emph{J.
  Parallel Distributed Comput.}, vol. 145, pp. 1--12, 2020. [Online].
  Available: \url{https://doi.org/10.1016/j.jpdc.2020.05.014}
\BIBentrySTDinterwordspacing

\bibitem{DBLP:conf/icdcs/0001N19}
\BIBentryALTinterwordspacing
C.~Feng and J.~Niu, ``Selfish mining in ethereum,'' in \emph{39th {IEEE}
  International Conference on Distributed Computing Systems, {ICDCS} 2019,
  Dallas, TX, USA, July 7-10, 2019}.\hskip 1em plus 0.5em minus 0.4em\relax
  {IEEE}, 2019, pp. 1306--1316. [Online]. Available:
  \url{https://doi.org/10.1109/ICDCS.2019.00131}
\BIBentrySTDinterwordspacing

\bibitem{Ritz2018TheIO}
F.~Ritz and A.~Zugenmaier, ``The impact of uncle rewards on selfish mining in
  ethereum,'' \emph{2018 IEEE European Symposium on Security and Privacy
  Workshops (EuroS\&PW)}, pp. 50--57, 2018.

\bibitem{DBLP:journals/tnsm/KangCYMM21}
\BIBentryALTinterwordspacing
H.~Kang, X.~Chang, R.~Yang, J.~V. Misic, and V.~B. Misic, ``Understanding
  selfish mining in imperfect bitcoin and ethereum networks with extended
  forks,'' \emph{{IEEE} Trans. Netw. Serv. Manag.}, vol.~18, no.~3, pp.
  3079--3091, 2021. [Online]. Available:
  \url{https://doi.org/10.1109/TNSM.2021.3073414}
\BIBentrySTDinterwordspacing

\bibitem{Chen2020PhishingSD}
W.~Chen, X.~Guo, Z.~Chen, Z.~Zheng, and Y.~Lu, ``Phishing scam detection on
  ethereum: Towards financial security for blockchain ecosystem,'' in
  \emph{IJCAI}, 2020.

\bibitem{DBLP:conf/ndss/HouZJDTFJ21}
C.~Hou, M.~Zhou, Y.~Ji, P.~Daian, F.~Tram{\`{e}}r, G.~Fanti, and A.~Juels,
  ``Squirrl: Automating attack analysis on blockchain incentive mechanisms with
  deep reinforcement learning,'' in \emph{28th Annual Network and Distributed
  System Security Symposium, {NDSS} 2021, virtually, February 21-25,
  2021}.\hskip 1em plus 0.5em minus 0.4em\relax The Internet Society, 2021.

\bibitem{DBLP:conf/crypto/KiayiasRDO17}
\BIBentryALTinterwordspacing
A.~Kiayias, A.~Russell, B.~David, and R.~Oliynykov, ``Ouroboros: {A} provably
  secure proof-of-stake blockchain protocol,'' in \emph{Advances in Cryptology
  - {CRYPTO} 2017 - 37th Annual International Cryptology Conference, Santa
  Barbara, CA, USA, August 20-24, 2017, Proceedings, Part {I}}, ser. Lecture
  Notes in Computer Science, J.~Katz and H.~Shacham, Eds., vol. 10401.\hskip
  1em plus 0.5em minus 0.4em\relax Springer, 2017, pp. 357--388. [Online].
  Available: \url{https://doi.org/10.1007/978-3-319-63688-7\_12}
\BIBentrySTDinterwordspacing

\bibitem{larimer2014delegated}
D.~Larimer, ``Delegated proof-of-stake (dpos),'' \emph{Bitshare whitepaper},
  vol.~81, p.~85, 2014.

\bibitem{schuh2017bitshares}
F.~Schuh and D.~Larimer, ``Bitshares 2.0: general overview,'' \emph{accessed
  June-2017.[Online]. Available: http://docs. bitshares.
  org/downloads/bitshares-general. pdf}, 2017.

\bibitem{DBLP:journals/pomacs/HuangWWTLZLHJ20}
\BIBentryALTinterwordspacing
Y.~Huang, H.~Wang, L.~Wu, G.~Tyson, X.~Luo, R.~Zhang, X.~Liu, G.~Huang, and
  X.~Jiang, ``Understanding (mis)behavior on the {EOSIO} blockchain,''
  \emph{Proc. {ACM} Meas. Anal. Comput. Syst.}, vol.~4, no.~2, pp. 37:1--37:28,
  2020. [Online]. Available: \url{https://doi.org/10.1145/3392155}
\BIBentrySTDinterwordspacing

\bibitem{xu2018eos}
B.~Xu, D.~Luthra, Z.~Cole, and N.~Blakely, ``Eos: An architectural,
  performance, and economic analysis,'' \emph{Retrieved June}, vol.~11, p.
  2019, 2018.

\bibitem{DBLP:journals/access/GuidiMR20}
\BIBentryALTinterwordspacing
B.~Guidi, A.~Michienzi, and L.~Ricci, ``Steem blockchain: Mining the inner
  structure of the graph,'' \emph{{IEEE} Access}, vol.~8, pp.
  210\,251--210\,266, 2020. [Online]. Available:
  \url{https://doi.org/10.1109/ACCESS.2020.3038550}
\BIBentrySTDinterwordspacing

\bibitem{Sun2021DTDPoSAD}
Y.~Sun, B.~Yan, Y.~Yao, and J.~Yu, ``Dt-dpos: A delegated proof of stake
  consensus algorithm with dynamic trust,'' \emph{Procedia Computer Science},
  vol. 187, pp. 371--376, 2021.

\bibitem{DBLP:journals/isci/LiuXCMG21}
\BIBentryALTinterwordspacing
J.~Liu, M.~Xie, S.~Chen, C.~Ma, and Q.~Gong, ``An improved dpos consensus
  mechanism in blockchain based on {PLTS} for the smart autonomous multi-robot
  system,'' \emph{Inf. Sci.}, vol. 575, pp. 528--541, 2021. [Online].
  Available: \url{https://doi.org/10.1016/j.ins.2021.06.046}
\BIBentrySTDinterwordspacing

\bibitem{DBLP:journals/access/YangZWLXZ19}
\BIBentryALTinterwordspacing
F.~Yang, W.~Zhou, Q.~Wu, R.~Long, N.~N. Xiong, and M.~Zhou, ``Delegated proof
  of stake with downgrade: {A} secure and efficient blockchain consensus
  algorithm with downgrade mechanism,'' \emph{{IEEE} Access}, vol.~7, pp.
  118\,541--118\,555, 2019. [Online]. Available:
  \url{https://doi.org/10.1109/ACCESS.2019.2935149}
\BIBentrySTDinterwordspacing

\bibitem{DBLP:conf/mobiquitous/FanC18}
\BIBentryALTinterwordspacing
X.~Fan and Q.~Chai, ``Roll-dpos: {A} randomized delegated proof of stake scheme
  for scalable blockchain-based internet of things systems,'' in
  \emph{Proceedings of the 15th {EAI} International Conference on Mobile and
  Ubiquitous Systems: Computing, Networking and Services, MobiQuitous 2018, 5-7
  November 2018, New York City, NY, {USA}}, H.~Schulzrinne and P.~Li,
  Eds.\hskip 1em plus 0.5em minus 0.4em\relax {ACM}, 2018, pp. 482--484.
  [Online]. Available: \url{https://doi.org/10.1145/3286978.3287023}
\BIBentrySTDinterwordspacing

\bibitem{Luo2018ANE}
Y.~Luo, Y.~Chen, Q.~Chen, and Q.~Liang, ``A new election algorithm for dpos
  consensus mechanism in blockchain,'' \emph{2018 7th International Conference
  on Digital Home (ICDH)}, pp. 116--120, 2018.

\bibitem{DBLP:journals/tii/XuLK20}
\BIBentryALTinterwordspacing
G.~Xu, Y.~Liu, and P.~W. Khan, ``Improvement of the dpos consensus mechanism in
  blockchain based on vague sets,'' \emph{{IEEE} Trans. Ind. Informatics},
  vol.~16, no.~6, pp. 4252--4259, 2020. [Online]. Available:
  \url{https://doi.org/10.1109/TII.2019.2955719}
\BIBentrySTDinterwordspacing

\bibitem{castro1999practical}
M.~Castro, B.~Liskov \emph{et~al.}, ``Practical byzantine fault tolerance,'' in
  \emph{OsDI}, vol.~99, no. 1999, 1999, pp. 173--186.

\bibitem{DBLP:journals/ppna/LiQL21}
\BIBentryALTinterwordspacing
Y.~Li, L.~Qiao, and Z.~Lv, ``An optimized byzantine fault tolerance algorithm
  for consortium blockchain,'' \emph{Peer-to-Peer Netw. Appl.}, vol.~14, no.~5,
  pp. 2826--2839, 2021. [Online]. Available:
  \url{https://doi.org/10.1007/s12083-021-01103-8}
\BIBentrySTDinterwordspacing

\bibitem{DBLP:journals/tpds/LiFZXCI21}
\BIBentryALTinterwordspacing
W.~Li, C.~Feng, L.~Zhang, H.~Xu, B.~Cao, and M.~A. Imran, ``A scalable
  multi-layer {PBFT} consensus for blockchain,'' \emph{{IEEE} Trans. Parallel
  Distributed Syst.}, vol.~32, no.~5, pp. 1146--1160, 2021. [Online].
  Available: \url{https://doi.org/10.1109/TPDS.2020.3042392}
\BIBentrySTDinterwordspacing

\bibitem{DBLP:journals/iacr/LuuNBZGS15}
\BIBentryALTinterwordspacing
``{SCP:} {A} computationally-scalable byzantine consensus protocol for
  blockchains,'' \emph{{IACR} Cryptol. ePrint Arch.}, p. 1168, 2015, withdrawn.
  [Online]. Available: \url{http://eprint.iacr.org/2015/1168}
\BIBentrySTDinterwordspacing

\bibitem{DBLP:journals/ijnsec/LiHGJF19}
\BIBentryALTinterwordspacing
Z.~Li, J.~Huang, D.~Gao, Y.~Jiang, and L.~Fan, ``{ISCP:} an improved blockchain
  consensus protocol,'' \emph{Int. J. Netw. Secur.}, vol.~21, no.~3, pp.
  359--367, 2019. [Online]. Available:
  \url{http://ijns.jalaxy.com.tw/contents/ijns-v21-n3/ijns-2019-v21-n3-p359-367.pdf}
\BIBentrySTDinterwordspacing

\bibitem{DBLP:conf/icdcs/AmiriAA19}
\BIBentryALTinterwordspacing
M.~J. Amiri, D.~Agrawal, and A.~E. Abbadi, ``Parblockchain: Leveraging
  transaction parallelism in permissioned blockchain systems,'' in \emph{39th
  {IEEE} International Conference on Distributed Computing Systems, {ICDCS}
  2019, Dallas, TX, USA, July 7-10, 2019}.\hskip 1em plus 0.5em minus
  0.4em\relax {IEEE}, 2019, pp. 1337--1347. [Online]. Available:
  \url{https://doi.org/10.1109/ICDCS.2019.00134}
\BIBentrySTDinterwordspacing

\bibitem{DBLP:conf/dasfaa/GaiWDP18}
\BIBentryALTinterwordspacing
F.~Gai, B.~Wang, W.~Deng, and W.~Peng, ``Proof of reputation: {A}
  reputation-based consensus protocol for peer-to-peer network,'' in
  \emph{Database Systems for Advanced Applications - 23rd International
  Conference, {DASFAA} 2018, Gold Coast, QLD, Australia, May 21-24, 2018,
  Proceedings, Part {II}}, ser. Lecture Notes in Computer Science, J.~Pei,
  Y.~Manolopoulos, S.~W. Sadiq, and J.~Li, Eds., vol. 10828.\hskip 1em plus
  0.5em minus 0.4em\relax Springer, 2018, pp. 666--681. [Online]. Available:
  \url{https://doi.org/10.1007/978-3-319-91458-9\_41}
\BIBentrySTDinterwordspacing

\bibitem{DBLP:journals/iotj/AyazSTG21}
\BIBentryALTinterwordspacing
F.~Ayaz, Z.~Sheng, D.~Tian, and Y.~L. Guan, ``A proof-of-quality-factor
  (poqf)-based blockchain and edge computing for vehicular message
  dissemination,'' \emph{{IEEE} Internet Things J.}, vol.~8, no.~4, pp.
  2468--2482, 2021. [Online]. Available:
  \url{https://doi.org/10.1109/JIOT.2020.3026731}
\BIBentrySTDinterwordspacing

\bibitem{DBLP:journals/access/GuoLNS20}
\BIBentryALTinterwordspacing
H.~Guo, W.~Li, M.~M. Nejad, and C.~Shen, ``Proof-of-event recording system for
  autonomous vehicles: {A} blockchain-based solution,'' \emph{{IEEE} Access},
  vol.~8, pp. 182\,776--182\,786, 2020. [Online]. Available:
  \url{https://doi.org/10.1109/ACCESS.2020.3029512}
\BIBentrySTDinterwordspacing

\bibitem{DBLP:conf/icdcs/BahriG19}
\BIBentryALTinterwordspacing
L.~Bahri and S.~Girdzijauskas, ``Trust mends blockchains: Living up to
  expectations,'' in \emph{39th {IEEE} International Conference on Distributed
  Computing Systems, {ICDCS} 2019, Dallas, TX, USA, July 7-10, 2019}.\hskip 1em
  plus 0.5em minus 0.4em\relax {IEEE}, 2019, pp. 1358--1368. [Online].
  Available: \url{https://doi.org/10.1109/ICDCS.2019.00136}
\BIBentrySTDinterwordspacing

\bibitem{DBLP:conf/sigmod/DangDLCLO19}
\BIBentryALTinterwordspacing
H.~Dang, T.~T.~A. Dinh, D.~Loghin, E.~Chang, Q.~Lin, and B.~C. Ooi, ``Towards
  scaling blockchain systems via sharding,'' in \emph{Proceedings of the 2019
  International Conference on Management of Data, {SIGMOD} Conference 2019,
  Amsterdam, The Netherlands, June 30 - July 5, 2019}, P.~A. Boncz,
  S.~Manegold, A.~Ailamaki, A.~Deshpande, and T.~Kraska, Eds.\hskip 1em plus
  0.5em minus 0.4em\relax {ACM}, 2019, pp. 123--140. [Online]. Available:
  \url{https://doi.org/10.1145/3299869.3319889}
\BIBentrySTDinterwordspacing

\bibitem{DBLP:conf/ccs/LuuNZBGS16}
\BIBentryALTinterwordspacing
L.~Luu, V.~Narayanan, C.~Zheng, K.~Baweja, S.~Gilbert, and P.~Saxena, ``A
  secure sharding protocol for open blockchains,'' in \emph{Proceedings of the
  2016 {ACM} {SIGSAC} Conference on Computer and Communications Security,
  Vienna, Austria, October 24-28, 2016}, E.~R. Weippl, S.~Katzenbeisser,
  C.~Kruegel, A.~C. Myers, and S.~Halevi, Eds.\hskip 1em plus 0.5em minus
  0.4em\relax {ACM}, 2016, pp. 17--30. [Online]. Available:
  \url{https://doi.org/10.1145/2976749.2978389}
\BIBentrySTDinterwordspacing

\bibitem{DBLP:journals/iacr/HanYLCV21}
\BIBentryALTinterwordspacing
R.~Han, J.~Yu, H.~Lin, S.~Chen, and P.~J.~E. Ver{\'{\i}}ssimo, ``On the
  security and performance of blockchain sharding,'' \emph{{IACR} Cryptol.
  ePrint Arch.}, p. 1276, 2021. [Online]. Available:
  \url{https://eprint.iacr.org/2021/1276}
\BIBentrySTDinterwordspacing

\bibitem{DBLP:conf/ccs/ZamaniM018}
\BIBentryALTinterwordspacing
M.~Zamani, M.~Movahedi, and M.~Raykova, ``Rapidchain: Scaling blockchain via
  full sharding,'' in \emph{Proceedings of the 2018 {ACM} {SIGSAC} Conference
  on Computer and Communications Security, {CCS} 2018, Toronto, ON, Canada,
  October 15-19, 2018}, D.~Lie, M.~Mannan, M.~Backes, and X.~Wang, Eds.\hskip
  1em plus 0.5em minus 0.4em\relax {ACM}, 2018, pp. 931--948. [Online].
  Available: \url{https://doi.org/10.1145/3243734.3243853}
\BIBentrySTDinterwordspacing

\bibitem{DBLP:conf/sac/XuH20}
\BIBentryALTinterwordspacing
Y.~Xu and Y.~Huang, ``An n/2 byzantine node tolerate blockchain sharding
  approach,'' in \emph{{SAC} '20: The 35th {ACM/SIGAPP} Symposium on Applied
  Computing, online event, [Brno, Czech Republic], March 30 - April 3, 2020},
  C.~Hung, T.~Cern{\'{y}}, D.~Shin, and A.~Bechini, Eds.\hskip 1em plus 0.5em
  minus 0.4em\relax {ACM}, 2020, pp. 349--352. [Online]. Available:
  \url{https://doi.org/10.1145/3341105.3374069}
\BIBentrySTDinterwordspacing

\bibitem{DBLP:conf/icpp/ZhangHQZ0C20}
\BIBentryALTinterwordspacing
J.~Zhang, Z.~Hong, X.~Qiu, Y.~Zhan, S.~Guo, and W.~Chen, ``Skychain: {A} deep
  reinforcement learning-empowered dynamic blockchain sharding system,'' in
  \emph{{ICPP} 2020: 49th International Conference on Parallel Processing,
  Edmonton, AB, Canada, August 17-20, 2020}, J.~N. Amaral, L.~K. John, and
  X.~Shen, Eds.\hskip 1em plus 0.5em minus 0.4em\relax {ACM}, 2020, pp.
  3:1--3:11. [Online]. Available: \url{https://doi.org/10.1145/3404397.3404460}
\BIBentrySTDinterwordspacing

\bibitem{DBLP:conf/wcnc/BandaraSRML22}
\BIBentryALTinterwordspacing
E.~Bandara, S.~Shetty, A.~Rahman, R.~Mukkamala, and X.~Liang, ``Moose: {A}
  scalable blockchain architecture for 5g enabled iot with sharding and network
  slicing,'' in \emph{{IEEE} Wireless Communications and Networking Conference,
  {WCNC} 2022, Austin, TX, USA, April 10-13, 2022}.\hskip 1em plus 0.5em minus
  0.4em\relax {IEEE}, 2022, pp. 1194--1199. [Online]. Available:
  \url{https://doi.org/10.1109/WCNC51071.2022.9771885}
\BIBentrySTDinterwordspacing

\bibitem{DBLP:conf/bsci/MadillNLR22}
\BIBentryALTinterwordspacing
E.~Madill, B.~Nguyen, C.~K. Leung, and S.~Rouhani, ``Scalesfl: {A} sharding
  solution for blockchain-based federated learning,'' in \emph{{BSCI} 2022:
  Proceedings of the 4th {ACM} International Symposium on Blockchain and Secure
  Critical Infrastructure, Nagasaki, Japan, May 30, 2022}, K.~Gai and K.~R.
  Choo, Eds.\hskip 1em plus 0.5em minus 0.4em\relax {ACM}, 2022, pp. 95--106.
  [Online]. Available: \url{https://doi.org/10.1145/3494106.3528680}
\BIBentrySTDinterwordspacing

\bibitem{DBLP:journals/access/XuH20a}
\BIBentryALTinterwordspacing
Y.~Xu and Y.~Huang, ``Segment blockchain: {A} size reduced storage mechanism
  for blockchain,'' \emph{{IEEE} Access}, vol.~8, pp. 17\,434--17\,441, 2020.
  [Online]. Available: \url{https://doi.org/10.1109/ACCESS.2020.2966464}
\BIBentrySTDinterwordspacing

\bibitem{DBLP:journals/tkde/QiZJZ21}
\BIBentryALTinterwordspacing
X.~Qi, Z.~Zhang, C.~Jin, and A.~Zhou, ``A reliable storage partition for
  permissioned blockchain,'' \emph{{IEEE} Trans. Knowl. Data Eng.}, vol.~33,
  no.~1, pp. 14--27, 2021. [Online]. Available:
  \url{https://doi.org/10.1109/TKDE.2020.3012668}
\BIBentrySTDinterwordspacing

\bibitem{tran2022internet}
H.~Tran-Dang, N.~Krommenacker, P.~Charpentier, and D.-S. Kim, ``The internet of
  things for logistics: Perspectives, application review, and challenges,''
  \emph{IETE Technical Review}, vol.~39, no.~1, pp. 93--121, 2022.

\bibitem{alkhateeb2022hybrid}
A.~Alkhateeb, C.~Catal, G.~Kar, and A.~Mishra, ``Hybrid blockchain platforms
  for the internet of things (iot): A systematic literature review,''
  \emph{Sensors}, vol.~22, no.~4, p. 1304, 2022.

\bibitem{DBLP:journals/iotj/Novo18}
\BIBentryALTinterwordspacing
O.~Novo, ``Blockchain meets iot: An architecture for scalable access management
  in iot,'' \emph{{IEEE} Internet Things J.}, vol.~5, no.~2, pp. 1184--1195,
  2018. [Online]. Available: \url{https://doi.org/10.1109/JIOT.2018.2812239}
\BIBentrySTDinterwordspacing

\bibitem{DBLP:conf/wd/MarchangIW19}
\BIBentryALTinterwordspacing
J.~Marchang, G.~Ibbotson, and P.~Wheway, ``Will blockchain technology become a
  reality in sensor networks?'' in \emph{2019 Wireless Days, {WD} 2019,
  Manchester, United Kingdom, April 24-26, 2019}.\hskip 1em plus 0.5em minus
  0.4em\relax {IEEE}, 2019, pp. 1--4. [Online]. Available:
  \url{https://doi.org/10.1109/WD.2019.8734268}
\BIBentrySTDinterwordspacing

\bibitem{DBLP:journals/jnca/AggarwalCAKCZ19}
\BIBentryALTinterwordspacing
S.~Aggarwal, R.~Chaudhary, G.~S. Aujla, N.~Kumar, K.~R. Choo, and A.~Y. Zomaya,
  ``Blockchain for smart communities: Applications, challenges and
  opportunities,'' \emph{J. Netw. Comput. Appl.}, vol. 144, pp. 13--48, 2019.
  [Online]. Available: \url{https://doi.org/10.1016/j.jnca.2019.06.018}
\BIBentrySTDinterwordspacing

\bibitem{DBLP:conf/icdcs/HuangKC0WL19}
\BIBentryALTinterwordspacing
J.~Huang, L.~Kong, G.~Chen, L.~Cheng, K.~Wu, and X.~Liu, ``B-iot: Blockchain
  driven internet of things with credit-based consensus mechanism,'' in
  \emph{39th {IEEE} International Conference on Distributed Computing Systems,
  {ICDCS} 2019, Dallas, TX, USA, July 7-10, 2019}.\hskip 1em plus 0.5em minus
  0.4em\relax {IEEE}, 2019, pp. 1348--1357. [Online]. Available:
  \url{https://doi.org/10.1109/ICDCS.2019.00135}
\BIBentrySTDinterwordspacing

\bibitem{DBLP:journals/iotj/BiswasSLMMW20}
\BIBentryALTinterwordspacing
S.~Biswas, K.~Sharif, F.~Li, S.~Maharjan, S.~P. Mohanty, and Y.~Wang, ``Pobt:
  {A} lightweight consensus algorithm for scalable iot business blockchain,''
  \emph{{IEEE} Internet Things J.}, vol.~7, no.~3, pp. 2343--2355, 2020.
  [Online]. Available: \url{https://doi.org/10.1109/JIOT.2019.2958077}
\BIBentrySTDinterwordspacing

\bibitem{DBLP:journals/fgcs/WangLCKK20}
\BIBentryALTinterwordspacing
E.~K. Wang, Z.~Liang, C.~Chen, S.~Kumari, and M.~K. Khan, ``Porx: {A}
  reputation incentive scheme for blockchain consensus of iiot,'' \emph{Future
  Gener. Comput. Syst.}, vol. 102, pp. 140--151, 2020. [Online]. Available:
  \url{https://doi.org/10.1016/j.future.2019.08.005}
\BIBentrySTDinterwordspacing

\bibitem{DBLP:journals/iotj/AiC22}
\BIBentryALTinterwordspacing
Z.~Ai and W.~Cui, ``A proof-of-transactions blockchain consensus protocol for
  large-scale iot,'' \emph{{IEEE} Internet Things J.}, vol.~9, no.~11, pp.
  7931--7943, 2022. [Online]. Available:
  \url{https://doi.org/10.1109/JIOT.2021.3108627}
\BIBentrySTDinterwordspacing

\bibitem{DBLP:journals/jpdc/DorriKJG19}
\BIBentryALTinterwordspacing
A.~Dorri, S.~S. Kanhere, R.~Jurdak, and P.~Gauravaram, ``{LSB:} {A} lightweight
  scalable blockchain for iot security and anonymity,'' \emph{J. Parallel
  Distributed Comput.}, vol. 134, pp. 180--197, 2019. [Online]. Available:
  \url{https://doi.org/10.1016/j.jpdc.2019.08.005}
\BIBentrySTDinterwordspacing

\bibitem{DBLP:journals/iotj/BiswasSLNW19}
\BIBentryALTinterwordspacing
S.~Biswas, K.~Sharif, F.~Li, B.~Nour, and Y.~Wang, ``A scalable blockchain
  framework for secure transactions in iot,'' \emph{{IEEE} Internet Things J.},
  vol.~6, no.~3, pp. 4650--4659, 2019. [Online]. Available:
  \url{https://doi.org/10.1109/JIOT.2018.2874095}
\BIBentrySTDinterwordspacing

\bibitem{DBLP:journals/tnse/DingGLW21}
\BIBentryALTinterwordspacing
X.~Ding, J.~Guo, D.~Li, and W.~Wu, ``An incentive mechanism for building a
  secure blockchain-based internet of things,'' \emph{{IEEE} Trans. Netw. Sci.
  Eng.}, vol.~8, no.~1, pp. 477--487, 2021. [Online]. Available:
  \url{https://doi.org/10.1109/TNSE.2020.3040446}
\BIBentrySTDinterwordspacing

\bibitem{DBLP:journals/cem/KhanLH22}
\BIBentryALTinterwordspacing
S.~Khan, W.~Lee, and S.~O. Hwang, ``Aechain: {A} lightweight blockchain for iot
  applications,'' \emph{{IEEE} Consumer Electron. Mag.}, vol.~11, no.~2, pp.
  64--76, 2022. [Online]. Available:
  \url{https://doi.org/10.1109/MCE.2021.3060373}
\BIBentrySTDinterwordspacing

\bibitem{DBLP:journals/cem/ZhangZX22}
\BIBentryALTinterwordspacing
C.~Zhang, L.~Zhu, and C.~Xu, ``{BPAF:} blockchain-enabled reliable and
  privacy-preserving authentication for fog-based iot devices,'' \emph{{IEEE}
  Consumer Electron. Mag.}, vol.~11, no.~2, pp. 88--96, 2022. [Online].
  Available: \url{https://doi.org/10.1109/MCE.2021.3061808}
\BIBentrySTDinterwordspacing

\bibitem{DBLP:journals/iotj/GuoDW21}
\BIBentryALTinterwordspacing
J.~Guo, X.~Ding, and W.~Wu, ``A blockchain-enabled ecosystem for distributed
  electricity trading in smart city,'' \emph{{IEEE} Internet Things J.},
  vol.~8, no.~3, pp. 2040--2050, 2021. [Online]. Available:
  \url{https://doi.org/10.1109/JIOT.2020.3015980}
\BIBentrySTDinterwordspacing

\bibitem{DBLP:journals/tgcn/GuoDW22}
\BIBentryALTinterwordspacing
------, ``An architecture for distributed energies trading in byzantine-based
  blockchains,'' \emph{{IEEE} Trans. Green Commun. Netw.}, vol.~6, no.~2, pp.
  1216--1230, 2022. [Online]. Available:
  \url{https://doi.org/10.1109/TGCN.2022.3142438}
\BIBentrySTDinterwordspacing

\bibitem{DBLP:journals/iotj/LiFJXLW22}
\BIBentryALTinterwordspacing
T.~Li, Y.~Fang, Z.~Jian, X.~Xie, Y.~Lu, and G.~Wang, ``{ATOM:} architectural
  support and optimization mechanism for smart contract fast update and
  execution in blockchain-based iot,'' \emph{{IEEE} Internet Things J.},
  vol.~9, no.~11, pp. 7959--7971, 2022. [Online]. Available:
  \url{https://doi.org/10.1109/JIOT.2021.3106942}
\BIBentrySTDinterwordspacing

\bibitem{DBLP:journals/iotj/ZhouFW22}
\BIBentryALTinterwordspacing
J.~Zhou, G.~Feng, and Y.~Wang, ``Optimal deployment mechanism of blockchain in
  resource-constrained iot systems,'' \emph{{IEEE} Internet Things J.}, vol.~9,
  no.~11, pp. 8168--8177, 2022. [Online]. Available:
  \url{https://doi.org/10.1109/JIOT.2021.3106355}
\BIBentrySTDinterwordspacing

\bibitem{DBLP:journals/iotj/YangLSYSZ21}
\BIBentryALTinterwordspacing
L.~Yang, M.~Li, P.~Si, R.~Yang, E.~Sun, and Y.~Zhang, ``Energy-efficient
  resource allocation for blockchain-enabled industrial internet of things with
  deep reinforcement learning,'' \emph{{IEEE} Internet Things J.}, vol.~8,
  no.~4, pp. 2318--2329, 2021. [Online]. Available:
  \url{https://doi.org/10.1109/JIOT.2020.3030646}
\BIBentrySTDinterwordspacing

\bibitem{DBLP:journals/cn/WuWML21}
\BIBentryALTinterwordspacing
Y.~Wu, Z.~Wang, Y.~Ma, and V.~C.~M. Leung, ``Deep reinforcement learning for
  blockchain in industrial iot: {A} survey,'' \emph{Comput. Networks}, vol.
  191, p. 108004, 2021. [Online]. Available:
  \url{https://doi.org/10.1016/j.comnet.2021.108004}
\BIBentrySTDinterwordspacing

\bibitem{DBLP:journals/iotj/0007YSWZ20}
\BIBentryALTinterwordspacing
M.~Li, F.~R. Yu, P.~Si, W.~Wu, and Y.~Zhang, ``Resource optimization for
  delay-tolerant data in blockchain-enabled iot with edge computing: {A} deep
  reinforcement learning approach,'' \emph{{IEEE} Internet Things J.}, vol.~7,
  no.~10, pp. 9399--9412, 2020. [Online]. Available:
  \url{https://doi.org/10.1109/JIOT.2020.3007869}
\BIBentrySTDinterwordspacing

\bibitem{DBLP:journals/tii/LiuYTLS19}
\BIBentryALTinterwordspacing
M.~Liu, F.~R. Yu, Y.~Teng, V.~C.~M. Leung, and M.~Song, ``Performance
  optimization for blockchain-enabled industrial internet of things (iiot)
  systems: {A} deep reinforcement learning approach,'' \emph{{IEEE} Trans. Ind.
  Informatics}, vol.~15, no.~6, pp. 3559--3570, 2019. [Online]. Available:
  \url{https://doi.org/10.1109/TII.2019.2897805}
\BIBentrySTDinterwordspacing

\bibitem{DBLP:journals/iotj/YunGC21}
\BIBentryALTinterwordspacing
J.~Yun, Y.~Goh, and J.~Chung, ``Dqn-based optimization framework for secure
  sharded blockchain systems,'' \emph{{IEEE} Internet Things J.}, vol.~8,
  no.~2, pp. 708--722, 2021. [Online]. Available:
  \url{https://doi.org/10.1109/JIOT.2020.3006896}
\BIBentrySTDinterwordspacing

\bibitem{ding2022pricing}
X.~Ding, J.~Guo, D.~Li, and W.~Wu, ``Pricing and budget allocation for iot
  blockchain with edge computing,'' \emph{IEEE Transactions on Cloud
  Computing}, 2022.

\bibitem{V2022BlockchainTF}
A.~K. V, A.~K. Tyagi, and S.~Kumar, ``Blockchain technology for securing
  internet of vehicle: Issues and challenges,'' \emph{2022 International
  Conference on Computer Communication and Informatics (ICCCI)}, pp. 1--6,
  2022.

\bibitem{DBLP:conf/icbc/ChoCH19}
\BIBentryALTinterwordspacing
S.~Cho, N.~Chen, and X.~Hua, ``Developing a vehicle networking platform based
  on blockchain technology,'' in \emph{Blockchain - {ICBC} 2019 - Second
  International Conference, Held as Part of the Services Conference Federation,
  {SCF} 2019, San Diego, CA, USA, June 25-30, 2019, Proceedings}, ser. Lecture
  Notes in Computer Science, J.~Joshi, S.~Nepal, Q.~Zhang, and L.~Zhang, Eds.,
  vol. 11521.\hskip 1em plus 0.5em minus 0.4em\relax Springer, 2019, pp.
  186--201. [Online]. Available:
  \url{https://doi.org/10.1007/978-3-030-23404-1\_13}
\BIBentrySTDinterwordspacing

\bibitem{DBLP:journals/iotj/YangYLZL19}
\BIBentryALTinterwordspacing
Z.~Yang, K.~Yang, L.~Lei, K.~Zheng, and V.~C.~M. Leung, ``Blockchain-based
  decentralized trust management in vehicular networks,'' \emph{{IEEE} Internet
  Things J.}, vol.~6, no.~2, pp. 1495--1505, 2019. [Online]. Available:
  \url{https://doi.org/10.1109/JIOT.2018.2836144}
\BIBentrySTDinterwordspacing

\bibitem{DBLP:journals/access/WangJGYCL20}
\BIBentryALTinterwordspacing
Q.~Wang, T.~Ji, Y.~Guo, L.~Yu, X.~Chen, and P.~Li, ``Trafficchain: {A}
  blockchain-based secure and privacy-preserving traffic map,'' \emph{{IEEE}
  Access}, vol.~8, pp. 60\,598--60\,612, 2020. [Online]. Available:
  \url{https://doi.org/10.1109/ACCESS.2020.2980298}
\BIBentrySTDinterwordspacing

\bibitem{DBLP:journals/iotj/YinWHDJ20}
\BIBentryALTinterwordspacing
B.~Yin, Y.~Wu, T.~Hu, J.~Dong, and Z.~Jiang, ``An efficient collaboration and
  incentive mechanism for internet of vehicles (iov) with secured information
  exchange based on blockchains,'' \emph{{IEEE} Internet Things J.}, vol.~7,
  no.~3, pp. 1582--1593, 2020. [Online]. Available:
  \url{https://doi.org/10.1109/JIOT.2019.2949088}
\BIBentrySTDinterwordspacing

\bibitem{DBLP:journals/iotj/HuiHSLCXD22}
\BIBentryALTinterwordspacing
Y.~Hui, Y.~Huang, Z.~Su, T.~H. Luan, N.~Cheng, X.~Xiao, and G.~Ding, ``{BCC:}
  blockchain-based collaborative crowdsensing in autonomous vehicular
  networks,'' \emph{{IEEE} Internet Things J.}, vol.~9, no.~6, pp. 4518--4532,
  2022. [Online]. Available: \url{https://doi.org/10.1109/JIOT.2021.3105547}
\BIBentrySTDinterwordspacing

\bibitem{DBLP:journals/tvt/KangXNYKZ19}
\BIBentryALTinterwordspacing
J.~Kang, Z.~Xiong, D.~Niyato, D.~Ye, D.~I. Kim, and J.~Zhao, ``Toward secure
  blockchain-enabled internet of vehicles: Optimizing consensus management
  using reputation and contract theory,'' \emph{{IEEE} Trans. Veh. Technol.},
  vol.~68, no.~3, pp. 2906--2920, 2019. [Online]. Available:
  \url{https://doi.org/10.1109/TVT.2019.2894944}
\BIBentrySTDinterwordspacing

\bibitem{DBLP:conf/sss/ChenXSGLS17}
\BIBentryALTinterwordspacing
L.~Chen, L.~Xu, N.~Shah, Z.~Gao, Y.~Lu, and W.~Shi, ``On security analysis of
  proof-of-elapsed-time (poet),'' in \emph{Stabilization, Safety, and Security
  of Distributed Systems - 19th International Symposium, {SSS} 2017, Boston,
  MA, USA, November 5-8, 2017, Proceedings}, ser. Lecture Notes in Computer
  Science, P.~G. Spirakis and P.~Tsigas, Eds., vol. 10616.\hskip 1em plus 0.5em
  minus 0.4em\relax Springer, 2017, pp. 282--297. [Online]. Available:
  \url{https://doi.org/10.1007/978-3-319-69084-1\_19}
\BIBentrySTDinterwordspacing

\bibitem{DBLP:journals/cm/CebeEAAU18}
\BIBentryALTinterwordspacing
M.~Cebe, E.~Erdin, K.~Akkaya, H.~Aksu, and A.~S. Uluagac, ``Block4forensic: An
  integrated lightweight blockchain framework for forensics applications of
  connected vehicles,'' \emph{{IEEE} Commun. Mag.}, vol.~56, no.~10, pp.
  50--57, 2018. [Online]. Available:
  \url{https://doi.org/10.1109/MCOM.2018.1800137}
\BIBentrySTDinterwordspacing

\bibitem{DBLP:journals/vcomm/ZhangLLACCT19}
\BIBentryALTinterwordspacing
L.~Zhang, M.~Luo, J.~Li, M.~H. Au, K.~R. Choo, T.~Chen, and S.~Tian,
  ``Blockchain based secure data sharing system for internet of vehicles: {A}
  position paper,'' \emph{Veh. Commun.}, vol.~16, pp. 85--93, 2019. [Online].
  Available: \url{https://doi.org/10.1016/j.vehcom.2019.03.003}
\BIBentrySTDinterwordspacing

\bibitem{DBLP:journals/iotj/LuoFYS22}
\BIBentryALTinterwordspacing
L.~Luo, J.~Feng, H.~Yu, and G.~Sun, ``Blockchain-enabled two-way auction
  mechanism for electricity trading in internet of electric vehicles,''
  \emph{{IEEE} Internet Things J.}, vol.~9, no.~11, pp. 8105--8118, 2022.
  [Online]. Available: \url{https://doi.org/10.1109/JIOT.2021.3082769}
\BIBentrySTDinterwordspacing

\bibitem{DBLP:journals/tvt/AbishuSYASL22}
\BIBentryALTinterwordspacing
H.~N. Abishu, A.~M. Seid, Y.~H. Yacob, T.~Ayall, G.~Sun, and G.~Liu,
  ``Consensus mechanism for blockchain-enabled vehicle-to-vehicle energy
  trading in the internet of electric vehicles,'' \emph{{IEEE} Trans. Veh.
  Technol.}, vol.~71, no.~1, pp. 946--960, 2022. [Online]. Available:
  \url{https://doi.org/10.1109/TVT.2021.3129828}
\BIBentrySTDinterwordspacing

\bibitem{DBLP:journals/cem/WangSWNR21}
\BIBentryALTinterwordspacing
S.~Wang, S.~Sun, X.~Wang, Z.~Ning, and J.~J. P.~C. Rodrigues, ``Secure
  crowdsensing in 5g internet of vehicles: When deep reinforcement learning
  meets blockchain,'' \emph{{IEEE} Consumer Electron. Mag.}, vol.~10, no.~5,
  pp. 72--81, 2021. [Online]. Available:
  \url{https://doi.org/10.1109/MCE.2020.3048238}
\BIBentrySTDinterwordspacing

\bibitem{Kim2022ByzantineFaultTolerantCV}
S.~Kim and A.~S. Ibrahim, ``Byzantine-fault-tolerant consensus via
  reinforcement learning for permissioned blockchain-empowered v2x network,''
  \emph{IEEE Transactions on Intelligent Vehicles}, 2022.

\bibitem{DBLP:conf/icc/LiuTYLS19}
\BIBentryALTinterwordspacing
M.~Liu, Y.~Teng, F.~R. Yu, V.~C.~M. Leung, and M.~Song, ``Deep reinforcement
  learning based performance optimization in blockchain-enabled internet of
  vehicle,'' in \emph{2019 {IEEE} International Conference on Communications,
  {ICC} 2019, Shanghai, China, May 20-24, 2019}.\hskip 1em plus 0.5em minus
  0.4em\relax {IEEE}, 2019, pp. 1--6. [Online]. Available:
  \url{https://doi.org/10.1109/ICC.2019.8761206}
\BIBentrySTDinterwordspacing

\bibitem{DBLP:conf/icdcs/JiangL019}
\BIBentryALTinterwordspacing
S.~Jiang, X.~Li, and J.~Wu, ``Hierarchical edge-cloud computing for mobile
  blockchain mining game,'' in \emph{39th {IEEE} International Conference on
  Distributed Computing Systems, {ICDCS} 2019, Dallas, TX, USA, July 7-10,
  2019}.\hskip 1em plus 0.5em minus 0.4em\relax {IEEE}, 2019, pp. 1327--1336.
  [Online]. Available: \url{https://doi.org/10.1109/ICDCS.2019.00133}
\BIBentrySTDinterwordspacing

\bibitem{DBLP:journals/tii/QuCWLD22}
\BIBentryALTinterwordspacing
G.~Qu, N.~Cui, H.~Wu, R.~Li, and Y.~Ding, ``Chainfl: {A} simulation platform
  for joint federated learning and blockchain in edge/cloud computing
  environments,'' \emph{{IEEE} Trans. Ind. Informatics}, vol.~18, no.~5, pp.
  3572--3581, 2022. [Online]. Available:
  \url{https://doi.org/10.1109/TII.2021.3117481}
\BIBentrySTDinterwordspacing

\bibitem{DBLP:journals/sensors/HuGWHL22}
\BIBentryALTinterwordspacing
Z.~Hu, H.~Gao, T.~Wang, D.~Han, and Y.~Lu, ``Joint optimization for mobile edge
  computing-enabled blockchain systems: {A} deep reinforcement learning
  approach,'' \emph{Sensors}, vol.~22, no.~9, p. 3217, 2022. [Online].
  Available: \url{https://doi.org/10.3390/s22093217}
\BIBentrySTDinterwordspacing

\bibitem{DBLP:journals/iotj/FengYPCDZ20}
\BIBentryALTinterwordspacing
J.~Feng, F.~R. Yu, Q.~Pei, X.~Chu, J.~Du, and L.~Zhu, ``Cooperative computation
  offloading and resource allocation for blockchain-enabled mobile-edge
  computing: {A} deep reinforcement learning approach,'' \emph{{IEEE} Internet
  Things J.}, vol.~7, no.~7, pp. 6214--6228, 2020. [Online]. Available:
  \url{https://doi.org/10.1109/JIOT.2019.2961707}
\BIBentrySTDinterwordspacing

\bibitem{DBLP:journals/iotj/ZuoJZZ21}
\BIBentryALTinterwordspacing
Y.~Zuo, S.~Jin, S.~Zhang, and Y.~Zhang, ``Blockchain storage and computation
  offloading for cooperative mobile-edge computing,'' \emph{{IEEE} Internet
  Things J.}, vol.~8, no.~11, pp. 9084--9098, 2021. [Online]. Available:
  \url{https://doi.org/10.1109/JIOT.2021.3056656}
\BIBentrySTDinterwordspacing

\bibitem{DBLP:journals/tcss/ChengCDGY22}
\BIBentryALTinterwordspacing
G.~Cheng, Y.~Chen, S.~Deng, H.~Gao, and J.~Yin, ``A blockchain-based mutual
  authentication scheme for collaborative edge computing,'' \emph{{IEEE} Trans.
  Comput. Soc. Syst.}, vol.~9, no.~1, pp. 146--158, 2022. [Online]. Available:
  \url{https://doi.org/10.1109/TCSS.2021.3056540}
\BIBentrySTDinterwordspacing

\bibitem{DBLP:journals/tcom/XiaoDJHWLP20}
\BIBentryALTinterwordspacing
L.~Xiao, Y.~Ding, D.~Jiang, J.~Huang, D.~Wang, J.~Li, and H.~V. Poor, ``A
  reinforcement learning and blockchain-based trust mechanism for edge
  networks,'' \emph{{IEEE} Trans. Commun.}, vol.~68, no.~9, pp. 5460--5470,
  2020. [Online]. Available: \url{https://doi.org/10.1109/TCOMM.2020.2995371}
\BIBentrySTDinterwordspacing

\bibitem{DBLP:journals/jsac/LiuGPCOH22}
\BIBentryALTinterwordspacing
Y.~Liu, X.~Guan, Y.~Peng, H.~Chen, T.~Ohtsuki, and Z.~Han, ``Blockchain-based
  task offloading for edge computing on low-quality data via distributed
  learning in the internet of energy,'' \emph{{IEEE} J. Sel. Areas Commun.},
  vol.~40, no.~2, pp. 657--676, 2022. [Online]. Available:
  \url{https://doi.org/10.1109/JSAC.2021.3118417}
\BIBentrySTDinterwordspacing

\bibitem{DBLP:conf/percom/BaranwalK22}
\BIBentryALTinterwordspacing
G.~Baranwal and D.~Kumar, ``Posp: {A} novel proof of service placement in
  blockchain-based edge computing,'' in \emph{2022 {IEEE} International
  Conference on Pervasive Computing and Communications Workshops and other
  Affiliated Events, PerCom 2022 Workshops, Pisa, Italy, March 21-25,
  2022}.\hskip 1em plus 0.5em minus 0.4em\relax {IEEE}, 2022, pp. 18--21.
  [Online]. Available:
  \url{https://doi.org/10.1109/PerComWorkshops53856.2022.9767225}
\BIBentrySTDinterwordspacing

\bibitem{DBLP:journals/cem/MaskeyBSK21}
\BIBentryALTinterwordspacing
S.~R. Maskey, S.~Badsha, S.~Sengupta, and I.~Khalil, ``Reputation-based miner
  node selection in blockchain-based vehicular edge computing,'' \emph{{IEEE}
  Consumer Electron. Mag.}, vol.~10, no.~5, pp. 14--22, 2021. [Online].
  Available: \url{https://doi.org/10.1109/MCE.2020.3048312}
\BIBentrySTDinterwordspacing

\bibitem{DBLP:journals/iotj/BalistriCCGRS22}
\BIBentryALTinterwordspacing
E.~Balistri, F.~Casellato, S.~Collura, C.~Giannelli, G.~Riberto, and
  C.~Stefanelli, ``Design guidelines and a prototype implementation for
  cyber-resiliency in {IT/OT} scenarios based on blockchain and edge
  computing,'' \emph{{IEEE} Internet Things J.}, vol.~9, no.~7, pp. 4816--4832,
  2022. [Online]. Available: \url{https://doi.org/10.1109/JIOT.2021.3104624}
\BIBentrySTDinterwordspacing

\bibitem{DBLP:journals/ipm/LiLZWL22}
\BIBentryALTinterwordspacing
C.~Li, S.~Liang, J.~Zhang, Q.~Wang, and Y.~Luo, ``Blockchain-based data trading
  in edge-cloud computing environment,'' \emph{Inf. Process. Manag.}, vol.~59,
  no.~1, p. 102786, 2022. [Online]. Available:
  \url{https://doi.org/10.1016/j.ipm.2021.102786}
\BIBentrySTDinterwordspacing

\bibitem{DBLP:journals/tpds/YuanHCZQXXY22}
\BIBentryALTinterwordspacing
L.~Yuan, Q.~He, F.~Chen, J.~Zhang, L.~Qi, X.~Xu, Y.~Xiang, and Y.~Yang,
  ``Csedge: Enabling collaborative edge storage for multi-access edge computing
  based on blockchain,'' \emph{{IEEE} Trans. Parallel Distributed Syst.},
  vol.~33, no.~8, pp. 1873--1887, 2022. [Online]. Available:
  \url{https://doi.org/10.1109/TPDS.2021.3131680}
\BIBentrySTDinterwordspacing

\bibitem{DBLP:journals/corr/KonecnyMYRSB16}
\BIBentryALTinterwordspacing
J.~Kone{\v{c}}n{\'y}, H.~B. McMahan, F.~X. Yu, P.~Richt{\'{a}}rik, A.~T.
  Suresh, and D.~Bacon, ``Federated learning: Strategies for improving
  communication efficiency,'' \emph{CoRR}, vol. abs/1610.05492, 2016. [Online].
  Available: \url{http://arxiv.org/abs/1610.05492}
\BIBentrySTDinterwordspacing

\bibitem{9025660}
A.~Goel, A.~Agarwal, M.~Vatsa, R.~Singh, and N.~Ratha, ``Deepring: Protecting
  deep neural network with blockchain,'' in \emph{2019 IEEE/CVF Conference on
  Computer Vision and Pattern Recognition Workshops (CVPRW)}, 2019, pp.
  2821--2828.

\bibitem{DBLP:journals/wc/FuYLLZ20}
\BIBentryALTinterwordspacing
Y.~Fu, F.~R. Yu, C.~Li, T.~H. Luan, and Y.~Zhang, ``Vehicular blockchain-based
  collective learning for connected and autonomous vehicles,'' \emph{{IEEE}
  Wirel. Commun.}, vol.~27, no.~2, pp. 197--203, 2020. [Online]. Available:
  \url{https://doi.org/10.1109/MNET.001.1900310}
\BIBentrySTDinterwordspacing

\bibitem{DBLP:journals/icl/SunWWFC22}
\BIBentryALTinterwordspacing
J.~Sun, Y.~Wu, S.~Wang, Y.~Fu, and X.~Chang, ``Permissioned blockchain frame
  for secure federated learning,'' \emph{{IEEE} Commun. Lett.}, vol.~26, no.~1,
  pp. 13--17, 2022. [Online]. Available:
  \url{https://doi.org/10.1109/LCOMM.2021.3121297}
\BIBentrySTDinterwordspacing

\bibitem{DBLP:conf/blockchain2/RamananN20}
\BIBentryALTinterwordspacing
P.~Ramanan and K.~Nakayama, ``{BAFFLE} : Blockchain based aggregator free
  federated learning,'' in \emph{{IEEE} International Conference on Blockchain,
  Blockchain 2020, Rhodes, Greece, November 2-6, 2020}.\hskip 1em plus 0.5em
  minus 0.4em\relax {IEEE}, 2020, pp. 72--81. [Online]. Available:
  \url{https://doi.org/10.1109/Blockchain50366.2020.00017}
\BIBentrySTDinterwordspacing

\bibitem{DBLP:journals/tii/LuHDMZ20a}
\BIBentryALTinterwordspacing
Y.~Lu, X.~Huang, Y.~Dai, S.~Maharjan, and Y.~Zhang, ``Blockchain and federated
  learning for privacy-preserved data sharing in industrial iot,'' \emph{{IEEE}
  Trans. Ind. Informatics}, vol.~16, no.~6, pp. 4177--4186, 2020. [Online].
  Available: \url{https://doi.org/10.1109/TII.2019.2942190}
\BIBentrySTDinterwordspacing

\bibitem{DBLP:journals/iotj/LuHZMZ21}
\BIBentryALTinterwordspacing
Y.~Lu, X.~Huang, K.~Zhang, S.~Maharjan, and Y.~Zhang, ``Communication-efficient
  federated learning and permissioned blockchain for digital twin edge
  networks,'' \emph{{IEEE} Internet Things J.}, vol.~8, no.~4, pp. 2276--2288,
  2021. [Online]. Available: \url{https://doi.org/10.1109/JIOT.2020.3015772}
\BIBentrySTDinterwordspacing

\bibitem{DBLP:journals/tnse/PengXCGYGT22}
\BIBentryALTinterwordspacing
Z.~Peng, J.~Xu, X.~Chu, S.~Gao, Y.~Yao, R.~Gu, and Y.~Tang, ``Vfchain: Enabling
  verifiable and auditable federated learning via blockchain systems,''
  \emph{{IEEE} Trans. Netw. Sci. Eng.}, vol.~9, no.~1, pp. 173--186, 2022.
  [Online]. Available: \url{https://doi.org/10.1109/TNSE.2021.3050781}
\BIBentrySTDinterwordspacing

\bibitem{DBLP:journals/network/LiCLHZY21}
\BIBentryALTinterwordspacing
Y.~Li, C.~Chen, N.~Liu, H.~Huang, Z.~Zheng, and Q.~Yan, ``A blockchain-based
  decentralized federated learning framework with committee consensus,''
  \emph{{IEEE} Netw.}, vol.~35, no.~1, pp. 234--241, 2021. [Online]. Available:
  \url{https://doi.org/10.1109/MNET.011.2000263}
\BIBentrySTDinterwordspacing

\bibitem{DBLP:journals/tpds/QuWHC21}
\BIBentryALTinterwordspacing
X.~Qu, S.~Wang, Q.~Hu, and X.~Cheng, ``Proof of federated learning: {A} novel
  energy-recycling consensus algorithm,'' \emph{{IEEE} Trans. Parallel
  Distributed Syst.}, vol.~32, no.~8, pp. 2074--2085, 2021. [Online].
  Available: \url{https://doi.org/10.1109/TPDS.2021.3056773}
\BIBentrySTDinterwordspacing

\bibitem{DBLP:journals/cn/RuckelSH22}
\BIBentryALTinterwordspacing
T.~R{\"{u}}ckel, J.~Sedlmeir, and P.~Hofmann, ``Fairness, integrity, and
  privacy in a scalable blockchain-based federated learning system,''
  \emph{Comput. Networks}, vol. 202, p. 108621, 2022. [Online]. Available:
  \url{https://doi.org/10.1016/j.comnet.2021.108621}
\BIBentrySTDinterwordspacing

\bibitem{DBLP:journals/jpdc/GaoLCXX22}
\BIBentryALTinterwordspacing
L.~Gao, L.~Li, Y.~Chen, C.~Xu, and M.~Xu, ``{FGFL:} {A} blockchain-based fair
  incentive governor for federated learning,'' \emph{J. Parallel Distributed
  Comput.}, vol. 163, pp. 283--299, 2022. [Online]. Available:
  \url{https://doi.org/10.1016/j.jpdc.2022.01.019}
\BIBentrySTDinterwordspacing

\bibitem{DBLP:journals/tvt/LuHZMZ20}
\BIBentryALTinterwordspacing
Y.~Lu, X.~Huang, K.~Zhang, S.~Maharjan, and Y.~Zhang, ``Blockchain empowered
  asynchronous federated learning for secure data sharing in internet of
  vehicles,'' \emph{{IEEE} Trans. Veh. Technol.}, vol.~69, no.~4, pp.
  4298--4311, 2020. [Online]. Available:
  \url{https://doi.org/10.1109/TVT.2020.2973651}
\BIBentrySTDinterwordspacing

\bibitem{DBLP:journals/tc/FengZGQLY22}
\BIBentryALTinterwordspacing
L.~Feng, Y.~Zhao, S.~Guo, X.~Qiu, W.~Li, and P.~Yu, ``{BAFL:} {A}
  blockchain-based asynchronous federated learning framework,'' \emph{{IEEE}
  Trans. Computers}, vol.~71, no.~5, pp. 1092--1103, 2022. [Online]. Available:
  \url{https://doi.org/10.1109/TC.2021.3072033}
\BIBentrySTDinterwordspacing

\bibitem{DBLP:journals/sensors/WangT22}
\BIBentryALTinterwordspacing
R.~Wang and W.~Tsai, ``Asynchronous federated learning system based on
  permissioned blockchains,'' \emph{Sensors}, vol.~22, no.~4, p. 1672, 2022.
  [Online]. Available: \url{https://doi.org/10.3390/s22041672}
\BIBentrySTDinterwordspacing

\bibitem{DBLP:journals/tii/JiaZLZHL22}
\BIBentryALTinterwordspacing
B.~Jia, X.~Zhang, J.~Liu, Y.~Zhang, K.~Huang, and Y.~Liang,
  ``Blockchain-enabled federated learning data protection aggregation scheme
  with differential privacy and homomorphic encryption in iiot,'' \emph{{IEEE}
  Trans. Ind. Informatics}, vol.~18, no.~6, pp. 4049--4058, 2022. [Online].
  Available: \url{https://doi.org/10.1109/TII.2021.3085960}
\BIBentrySTDinterwordspacing

\bibitem{9170559}
Y.~Zhao, J.~Zhao, L.~Jiang, R.~Tan, D.~Niyato, Z.~Li, L.~Lyu, and Y.~Liu,
  ``Privacy-preserving blockchain-based federated learning for iot devices,''
  \emph{IEEE Internet of Things Journal}, vol.~8, no.~3, pp. 1817--1829, 2021.

\bibitem{DBLP:journals/iotj/OtoumRM22}
\BIBentryALTinterwordspacing
S.~Otoum, I.~A. Ridhawi, and H.~T. Mouftah, ``Securing critical iot
  infrastructures with blockchain-supported federated learning,'' \emph{{IEEE}
  Internet Things J.}, vol.~9, no.~4, pp. 2592--2601, 2022. [Online].
  Available: \url{https://doi.org/10.1109/JIOT.2021.3088056}
\BIBentrySTDinterwordspacing

\bibitem{DBLP:journals/network/FengYGQLY22}
\BIBentryALTinterwordspacing
L.~Feng, Z.~Yang, S.~Guo, X.~Qiu, W.~Li, and P.~Yu, ``Two-layered blockchain
  architecture for federated learning over the mobile edge network,''
  \emph{{IEEE} Netw.}, vol.~36, no.~1, pp. 45--51, 2022. [Online]. Available:
  \url{https://doi.org/10.1109/MNET.011.2000339}
\BIBentrySTDinterwordspacing

\bibitem{DBLP:journals/tvt/AyazSTG22}
\BIBentryALTinterwordspacing
F.~Ayaz, Z.~Sheng, D.~Tian, and Y.~L. Guan, ``A blockchain based federated
  learning for message dissemination in vehicular networks,'' \emph{{IEEE}
  Trans. Veh. Technol.}, vol.~71, no.~2, pp. 1927--1940, 2022. [Online].
  Available: \url{https://doi.org/10.1109/TVT.2021.3132226}
\BIBentrySTDinterwordspacing

\bibitem{DBLP:conf/healthcom/Mettler16}
\BIBentryALTinterwordspacing
M.~Mettler, ``Blockchain technology in healthcare: The revolution starts
  here,'' in \emph{18th {IEEE} International Conference on e-Health Networking,
  Applications and Services, Healthcom 2016, Munich, Germany, September 14-16,
  2016}.\hskip 1em plus 0.5em minus 0.4em\relax {IEEE}, 2016, pp. 1--3.
  [Online]. Available: \url{https://doi.org/10.1109/HealthCom.2016.7749510}
\BIBentrySTDinterwordspacing

\bibitem{DBLP:journals/jnca/McGhinCLH19}
\BIBentryALTinterwordspacing
T.~McGhin, K.~R. Choo, C.~Z. Liu, and D.~He, ``Blockchain in healthcare
  applications: Research challenges and opportunities,'' \emph{J. Netw. Comput.
  Appl.}, vol. 135, pp. 62--75, 2019. [Online]. Available:
  \url{https://doi.org/10.1016/j.jnca.2019.02.027}
\BIBentrySTDinterwordspacing

\bibitem{Yaqoob2021BlockchainFH}
I.~Yaqoob, K.~Salah, R.~Jayaraman, and Y.~Al-Hammadi, ``Blockchain for
  healthcare data management: opportunities, challenges, and future
  recommendations,'' \emph{Neural Computing and Applications}, pp. 1--16, 2021.

\bibitem{DBLP:journals/jms/YueWJLJ16}
\BIBentryALTinterwordspacing
X.~Yue, H.~Wang, D.~Jin, M.~Li, and W.~Jiang, ``Healthcare data gateways: Found
  healthcare intelligence on blockchain with novel privacy risk control,''
  \emph{J. Medical Syst.}, vol.~40, no.~10, pp. 218:1--218:8, 2016. [Online].
  Available: \url{https://doi.org/10.1007/s10916-016-0574-6}
\BIBentrySTDinterwordspacing

\bibitem{agbo2019blockchain}
C.~C. Agbo, Q.~H. Mahmoud, and J.~M. Eklund, ``Blockchain technology in
  healthcare: a systematic review,'' in \emph{Healthcare}, vol.~7, no.~2.\hskip
  1em plus 0.5em minus 0.4em\relax MDPI, 2019, p.~56.

\bibitem{haleem2021blockchain}
A.~Haleem, M.~Javaid, R.~P. Singh, R.~Suman, and S.~Rab, ``Blockchain
  technology applications in healthcare: An overview,'' \emph{International
  Journal of Intelligent Networks}, vol.~2, pp. 130--139, 2021.

\bibitem{rahmadika2018blockchain}
S.~Rahmadika and K.-H. Rhee, ``Blockchain technology for providing an
  architecture model of decentralized personal health information,''
  \emph{International Journal of Engineering Business Management}, vol.~10, p.
  1847979018790589, 2018.

\bibitem{DBLP:journals/access/JaimanU20}
\BIBentryALTinterwordspacing
V.~Jaiman and V.~Urovi, ``A consent model for blockchain-based health data
  sharing platforms,'' \emph{{IEEE} Access}, vol.~8, pp. 143\,734--143\,745,
  2020. [Online]. Available: \url{https://doi.org/10.1109/ACCESS.2020.3014565}
\BIBentrySTDinterwordspacing

\bibitem{DBLP:conf/dasfaa/LiuLAFM22}
\BIBentryALTinterwordspacing
L.~Liu, X.~Li, M.~H. Au, Z.~Fan, and X.~Meng, ``Metadata privacy preservation
  for blockchain-based healthcare systems,'' in \emph{Database Systems for
  Advanced Applications - 27th International Conference, {DASFAA} 2022, Virtual
  Event, April 11-14, 2022, Proceedings, Part {I}}, ser. Lecture Notes in
  Computer Science, A.~Bhattacharya, J.~Lee, M.~Li, D.~Agrawal, P.~K. Reddy,
  M.~K. Mohania, A.~Mondal, V.~Goyal, and R.~U. Kiran, Eds., vol. 13245.\hskip
  1em plus 0.5em minus 0.4em\relax Springer, 2022, pp. 404--412. [Online].
  Available: \url{https://doi.org/10.1007/978-3-031-00123-9\_33}
\BIBentrySTDinterwordspacing

\bibitem{Soni2021BlockchainbasedS}
M.~Soni and D.~K. Singh, ``Blockchain-based security \& privacy for biomedical
  and healthcare information exchange systems,'' \emph{Materials Today:
  Proceedings}, 2021.

\bibitem{DBLP:journals/cn/ZaabarCJAA21}
\BIBentryALTinterwordspacing
B.~Zaabar, O.~Cheikhrouhou, F.~Jamil, M.~Ammi, and M.~Abid, ``Healthblock: {A}
  secure blockchain-based healthcare data management system,'' \emph{Comput.
  Networks}, vol. 200, p. 108500, 2021. [Online]. Available:
  \url{https://doi.org/10.1016/j.comnet.2021.108500}
\BIBentrySTDinterwordspacing

\bibitem{DBLP:journals/tnse/BhattacharyaTBT21}
\BIBentryALTinterwordspacing
P.~Bhattacharya, S.~Tanwar, U.~Bodkhe, S.~Tyagi, and N.~Kumar, ``Bindaas:
  Blockchain-based deep-learning as-a-service in healthcare 4.0 applications,''
  \emph{{IEEE} Trans. Netw. Sci. Eng.}, vol.~8, no.~2, pp. 1242--1255, 2021.
  [Online]. Available: \url{https://doi.org/10.1109/TNSE.2019.2961932}
\BIBentrySTDinterwordspacing

\bibitem{DBLP:journals/cn/ZhangYL22}
\BIBentryALTinterwordspacing
G.~Zhang, Z.~Yang, and W.~Liu, ``Blockchain-based privacy preserving e-health
  system for healthcare data in cloud,'' \emph{Comput. Networks}, vol. 203, p.
  108586, 2022. [Online]. Available:
  \url{https://doi.org/10.1016/j.comnet.2021.108586}
\BIBentrySTDinterwordspacing

\bibitem{DBLP:journals/jaihc/ChelladuraiP22}
\BIBentryALTinterwordspacing
U.~Chelladurai and S.~Pandian, ``A novel blockchain based electronic health
  record automation system for healthcare,'' \emph{J. Ambient Intell. Humaniz.
  Comput.}, vol.~13, no.~1, pp. 693--703, 2022. [Online]. Available:
  \url{https://doi.org/10.1007/s12652-021-03163-3}
\BIBentrySTDinterwordspacing

\bibitem{DBLP:journals/titb/WuWNZ22}
\BIBentryALTinterwordspacing
G.~Wu, S.~Wang, Z.~Ning, and B.~Zhu, ``Privacy-preserved electronic medical
  record exchanging and sharing: {A} blockchain-based smart healthcare
  system,'' \emph{{IEEE} J. Biomed. Health Informatics}, vol.~26, no.~5, pp.
  1917--1927, 2022. [Online]. Available:
  \url{https://doi.org/10.1109/JBHI.2021.3123643}
\BIBentrySTDinterwordspacing

\bibitem{DBLP:journals/hij/UddinSJPE21}
\BIBentryALTinterwordspacing
M.~Uddin, K.~Salah, R.~Jayaraman, S.~Pesic, and S.~Ellahham, ``Blockchain for
  drug traceability: Architectures and open challenges,'' \emph{Health
  Informatics J.}, vol.~27, no.~2, p. 146045822110112, 2021. [Online].
  Available: \url{https://doi.org/10.1177/14604582211011228}
\BIBentrySTDinterwordspacing

\bibitem{DBLP:journals/access/MusamihSJADAE21}
\BIBentryALTinterwordspacing
A.~Musamih, K.~Salah, R.~Jayaraman, J.~Arshad, M.~Debe, Y.~Al{-}Hammadi, and
  S.~Ellahham, ``A blockchain-based approach for drug traceability in
  healthcare supply chain,'' \emph{{IEEE} Access}, vol.~9, pp. 9728--9743,
  2021. [Online]. Available: \url{https://doi.org/10.1109/ACCESS.2021.3049920}
\BIBentrySTDinterwordspacing

\bibitem{DBLP:conf/sac/KambiloZGSKV22}
\BIBentryALTinterwordspacing
E.~K. Kambilo, H.~B. Zghal, C.~G. Guegan, V.~Stankovski, P.~Kochovski, and
  D.~Vodislav, ``A blockchain-based framework for drug traceability:
  Chaindrugtrac,'' in \emph{{SAC} '22: The 37th {ACM/SIGAPP} Symposium on
  Applied Computing, Virtual Event, April 25 - 29, 2022}, J.~Hong, M.~Bures,
  J.~W. Park, and T.~Cern{\'{y}}, Eds.\hskip 1em plus 0.5em minus 0.4em\relax
  {ACM}, 2022, pp. 1900--1907. [Online]. Available:
  \url{https://doi.org/10.1145/3477314.3507118}
\BIBentrySTDinterwordspacing

\bibitem{10.1145/3477314.3507118}
\BIBentryALTinterwordspacing
------, ``A blockchain-based framework for drug traceability: Chaindrugtrac,''
  in \emph{Proceedings of the 37th ACM/SIGAPP Symposium on Applied Computing},
  ser. SAC '22.\hskip 1em plus 0.5em minus 0.4em\relax New York, NY, USA:
  Association for Computing Machinery, 2022, p. 1900–1907. [Online].
  Available: \url{https://doi.org/10.1145/3477314.3507118}
\BIBentrySTDinterwordspacing

\bibitem{Liu2019ABS}
W.~Liu, Q.~Yu, Z.~Li, Z.~Li, Y.~Su, and J.~Zhou, ``A blockchain-based system
  for anti-fraud of healthcare insurance,'' \emph{2019 IEEE 5th International
  Conference on Computer and Communications (ICCC)}, pp. 1264--1268, 2019.

\bibitem{DBLP:conf/riiforum/Mendoza-TelloMM20}
\BIBentryALTinterwordspacing
J.~C. Mendoza{-}Tello, T.~Mendoza{-}Tello, and H.~Mora, ``Blockchain as a
  healthcare insurance fraud detection tool,'' in \emph{Research and Innovation
  Forum 2020 - Disruptive Technologies in Times of Change, {RIIFORUM} 2020,
  Athens, Greece, 15-17 April 2020}, A.~Visvizi, M.~D. Lytras, and N.~R.
  Aljohani, Eds.\hskip 1em plus 0.5em minus 0.4em\relax Springer, 2020, pp.
  545--552. [Online]. Available:
  \url{https://doi.org/10.1007/978-3-030-62066-0\_41}
\BIBentrySTDinterwordspacing

\bibitem{9581059}
M.~J. Hossain~Faruk, H.~Shahriar, M.~Valero, S.~Sneha, S.~I. Ahamed, and
  M.~Rahman, ``Towards blockchain-based secure data management for remote
  patient monitoring,'' in \emph{2021 IEEE International Conference on Digital
  Health (ICDH)}, 2021, pp. 299--308.

\bibitem{DBLP:conf/primelife/PighiniVMMMGCC21}
\BIBentryALTinterwordspacing
C.~Pighini, A.~Vezzoni, S.~Mainini, A.~G. Migliavacca, A.~Montanari, M.~R.
  Guarneri, E.~G. Caiani, and A.~Cesareo, ``Syncare: An innovative remote
  patient monitoring system secured by cryptography and blockchain,'' in
  \emph{Privacy and Identity Management. Between Data Protection and Security -
  16th {IFIP} {WG} 9.2, 9.6/11.7, 11.6/SIG 9.2.2 International Summer School,
  Privacy and Identity 2021, Virtual Event, August 16-20, 2021, Revised
  Selected Papers}, ser. {IFIP} Advances in Information and Communication
  Technology, M.~Friedewald, S.~Krenn, I.~Schiering, and S.~Schiffner, Eds.,
  vol. 644.\hskip 1em plus 0.5em minus 0.4em\relax Springer, 2021, pp. 73--89.
  [Online]. Available: \url{https://doi.org/10.1007/978-3-030-99100-5\_7}
\BIBentrySTDinterwordspacing

\bibitem{Wong2019PrototypeOR}
D.~R. Wong, S.~Bhattacharya, and A.~J. Butte, ``Prototype of running clinical
  trials in an untrustworthy environment using blockchain,'' \emph{Nature
  Communications}, vol.~10, 2019.

\bibitem{DBLP:journals/jaihc/AlbaneseCSC20}
\BIBentryALTinterwordspacing
G.~Albanese, J.~Calbimonte, M.~Schumacher, and D.~Calvaresi, ``Dynamic consent
  management for clinical trials via private blockchain technology,'' \emph{J.
  Ambient Intell. Humaniz. Comput.}, vol.~11, no.~11, pp. 4909--4926, 2020.
  [Online]. Available: \url{https://doi.org/10.1007/s12652-020-01761-1}
\BIBentrySTDinterwordspacing

\bibitem{DBLP:journals/fgcs/SinghRATY22}
\BIBentryALTinterwordspacing
S.~Singh, S.~Rathore, O.~Alfarraj, A.~Tolba, and B.~Yoon, ``A framework for
  privacy-preservation of iot healthcare data using federated learning and
  blockchain technology,'' \emph{Future Gener. Comput. Syst.}, vol. 129, pp.
  380--388, 2022. [Online]. Available:
  \url{https://doi.org/10.1016/j.future.2021.11.028}
\BIBentrySTDinterwordspacing

\bibitem{DBLP:journals/sensors/AliAHPFKTZ22}
\BIBentryALTinterwordspacing
A.~Ali, M.~A. Almaiah, F.~Hajjej, M.~F. Pasha, O.~H. Fang, R.~Khan, J.~Teo, and
  M.~Zakarya, ``An industrial iot-based blockchain-enabled secure searchable
  encryption approach for healthcare systems using neural network,''
  \emph{Sensors}, vol.~22, no.~2, p. 572, 2022. [Online]. Available:
  \url{https://doi.org/10.3390/s22020572}
\BIBentrySTDinterwordspacing

\bibitem{DBLP:journals/comcom/HosseinEDKC21}
\BIBentryALTinterwordspacing
K.~M. Hossein, M.~E. Esmaeili, T.~Dargahi, A.~Khonsari, and M.~Conti,
  ``Bchealth: {A} novel blockchain-based privacy-preserving architecture for
  iot healthcare applications,'' \emph{Comput. Commun.}, vol. 180, pp. 31--47,
  2021. [Online]. Available: \url{https://doi.org/10.1016/j.comcom.2021.08.011}
\BIBentrySTDinterwordspacing

\bibitem{DBLP:conf/icact/AichSKACJ022}
\BIBentryALTinterwordspacing
S.~Aich, N.~K. Sinai, S.~Kumar, M.~Ali, Y.~R. Choi, M.~Joo, and H.~Kim,
  ``Protecting personal healthcare record using blockchain {\&} federated
  learning technologies,'' in \emph{24th International Conference on Advanced
  Communication Technology, {ICACT} 2022, Pyeongchang, Korea, February 13-16,
  2022}.\hskip 1em plus 0.5em minus 0.4em\relax {IEEE}, 2022, pp. 109--112.
  [Online]. Available: \url{https://doi.org/10.23919/ICACT53585.2022.9728772}
\BIBentrySTDinterwordspacing

\bibitem{DBLP:books/sp/21/TorkyDH21}
\BIBentryALTinterwordspacing
M.~Torky, A.~Darwish, and A.~E. Hassanien, ``Blockchain use cases for
  {COVID-19:} management, surveillance, tracking and security,'' in
  \emph{Digital Transformation and Emerging Technologies for Fighting
  {COVID-19} Pandemic: Innovative Approaches}, A.~E. Hassanien and A.~Darwish,
  Eds.\hskip 1em plus 0.5em minus 0.4em\relax Springer, 2021, vol. 322, pp.
  261--274. [Online]. Available:
  \url{https://doi.org/10.1007/978-3-030-63307-3\_17}
\BIBentrySTDinterwordspacing

\bibitem{DBLP:journals/connection/YaoJLC22}
\BIBentryALTinterwordspacing
S.~Yao, P.~Jing, P.~Li, and J.~Chen, ``A multi-dimension traceable
  privacy-preserving prevention and control scheme of the {COVID-19} epidemic
  based on blockchain,'' \emph{Connect. Sci.}, vol.~34, no.~1, pp. 1654--1677,
  2022. [Online]. Available:
  \url{https://doi.org/10.1080/09540091.2022.2077912}
\BIBentrySTDinterwordspacing

\bibitem{DBLP:journals/tnse/TanYSYWL22}
\BIBentryALTinterwordspacing
L.~Tan, K.~Yu, N.~Shi, C.~Yang, W.~Wei, and H.~Lu, ``Towards secure and
  privacy-preserving data sharing for {COVID-19} medical records: {A}
  blockchain-empowered approach,'' \emph{{IEEE} Trans. Netw. Sci. Eng.},
  vol.~9, no.~1, pp. 271--281, 2022. [Online]. Available:
  \url{https://doi.org/10.1109/TNSE.2021.3101842}
\BIBentrySTDinterwordspacing

\bibitem{DBLP:journals/itc/AslanA21}
\BIBentryALTinterwordspacing
B.~Aslan and K.~Atasen, ``{COVID-19} information sharing with blockchain,''
  \emph{Inf. Technol. Control.}, vol.~50, no.~4, pp. 674--685, 2021. [Online].
  Available: \url{https://doi.org/10.5755/j01.itc.50.4.29064}
\BIBentrySTDinterwordspacing

\bibitem{DBLP:journals/spe/AbidCKJ22}
\BIBentryALTinterwordspacing
A.~Abid, S.~Cheikhrouhou, S.~Kallel, and M.~Jmaiel, ``Novidchain:
  Blockchain-based privacy-preserving platform for {COVID-19} test/vaccine
  certificates,'' \emph{Softw. Pract. Exp.}, vol.~52, no.~4, pp. 841--867,
  2022. [Online]. Available: \url{https://doi.org/10.1002/spe.2983}
\BIBentrySTDinterwordspacing

\bibitem{DBLP:journals/jcn/TahirTSRK21}
\BIBentryALTinterwordspacing
S.~Tahir, H.~Tahir, A.~Sajjad, M.~Rajarajan, and F.~Khan, ``Privacy-preserving
  {COVID-19} contact tracing using blockchain,'' \emph{J. Commun. Networks},
  vol.~23, no.~5, pp. 360--373, 2021. [Online]. Available:
  \url{https://doi.org/10.23919/JCN.2021.000031}
\BIBentrySTDinterwordspacing

\bibitem{DBLP:journals/access/RicciMFF21}
\BIBentryALTinterwordspacing
L.~Ricci, D.~D.~F. Maesa, A.~Favenza, and E.~Ferro, ``Blockchains for
  {COVID-19} contact tracing and vaccine support: {A} systematic review,''
  \emph{{IEEE} Access}, vol.~9, pp. 37\,936--37\,950, 2021. [Online].
  Available: \url{https://doi.org/10.1109/ACCESS.2021.3063152}
\BIBentrySTDinterwordspacing

\bibitem{DBLP:conf/wcre/KassabD21}
\BIBentryALTinterwordspacing
M.~Kassab and G.~Destefanis, ``Blockchain and contact tracing applications for
  {COVID-19:} the opportunity and the challenges,'' in \emph{28th {IEEE}
  International Conference on Software Analysis, Evolution and Reengineering,
  {SANER} 2021, Honolulu, HI, USA, March 9-12, 2021}.\hskip 1em plus 0.5em
  minus 0.4em\relax {IEEE}, 2021, pp. 723--730. [Online]. Available:
  \url{https://doi.org/10.1109/SANER50967.2021.00092}
\BIBentrySTDinterwordspacing

\bibitem{DBLP:journals/iotj/XuZOFBI21}
\BIBentryALTinterwordspacing
H.~Xu, L.~Zhang, O.~Onireti, Y.~Fang, W.~J. Buchanan, and M.~A. Imran,
  ``Beeptrace: Blockchain-enabled privacy-preserving contact tracing for
  {COVID-19} pandemic and beyond,'' \emph{{IEEE} Internet Things J.}, vol.~8,
  no.~5, pp. 3915--3929, 2021. [Online]. Available:
  \url{https://doi.org/10.1109/JIOT.2020.3025953}
\BIBentrySTDinterwordspacing

\bibitem{arifeen2020blockchain}
M.~M. Arifeen, A.~Al~Mamun, M.~S. Kaiser, and M.~Mahmud, ``Blockchain-enable
  contact tracing for preserving user privacy during covid-19 outbreak,'' 2020.

\bibitem{DBLP:journals/isjgp/ChoudhuryGG21}
\BIBentryALTinterwordspacing
H.~Choudhury, B.~Goswami, and S.~K. Gurung, ``Covidchain: An anonymity
  preserving blockchain based framework for protection against covid-19,''
  \emph{Inf. Secur. J. {A} Glob. Perspect.}, vol.~30, no.~5, pp. 257--280,
  2021. [Online]. Available:
  \url{https://doi.org/10.1080/19393555.2021.1921315}
\BIBentrySTDinterwordspacing

\bibitem{DBLP:journals/access/HasanSJYOE21}
\BIBentryALTinterwordspacing
H.~R. Hasan, K.~Salah, R.~Jayaraman, I.~Yaqoob, M.~A. Omar, and S.~Ellahham,
  ``{COVID-19} contact tracing using blockchain,'' \emph{{IEEE} Access},
  vol.~9, pp. 62\,956--62\,971, 2021. [Online]. Available:
  \url{https://doi.org/10.1109/ACCESS.2021.3074753}
\BIBentrySTDinterwordspacing

\bibitem{DBLP:journals/informatics/TorkyGSH21}
\BIBentryALTinterwordspacing
M.~Torky, E.~Goda, V.~Sn{\'{a}}sel, and A.~E. Hassanien, ``{COVID-19} contact
  tracing and detection-based on blockchain technology,'' \emph{Informatics},
  vol.~8, no.~4, p.~72, 2021. [Online]. Available:
  \url{https://doi.org/10.3390/informatics8040072}
\BIBentrySTDinterwordspacing

\bibitem{DBLP:conf/sigmod/PengXWHXC21}
\BIBentryALTinterwordspacing
Z.~Peng, C.~Xu, H.~Wang, J.~Huang, J.~Xu, and X.~Chu,
  ``P\({}^{\mbox{2}}\)b-trace: Privacy-preserving blockchain-based contact
  tracing to combat pandemics,'' in \emph{{SIGMOD} '21: International
  Conference on Management of Data, Virtual Event, China, June 20-25, 2021},
  G.~Li, Z.~Li, S.~Idreos, and D.~Srivastava, Eds.\hskip 1em plus 0.5em minus
  0.4em\relax {ACM}, 2021, pp. 2389--2393. [Online]. Available:
  \url{https://doi.org/10.1145/3448016.3459237}
\BIBentrySTDinterwordspacing

\bibitem{DBLP:journals/iotm/NarenTHCKG21}
\BIBentryALTinterwordspacing
N.~Naren, A.~Tahiliani, V.~Hassija, V.~Chamola, S.~S. Kanhere, and M.~Guizani,
  ``Privacy-preserving and incentivized contact tracing for {COVID-19} using
  blockchain,'' \emph{{IEEE} Internet Things Mag.}, vol.~4, no.~3, pp. 72--79,
  2021. [Online]. Available: \url{https://doi.org/10.1109/IOTM.0211.2100002}
\BIBentrySTDinterwordspacing

\bibitem{DBLP:journals/tnse/LvWJCQZ22}
\BIBentryALTinterwordspacing
W.~Lv, S.~Wu, C.~Jiang, Y.~Cui, X.~Qiu, and Y.~Zhang, ``Towards large-scale and
  privacy-preserving contact tracing in {COVID-19} pandemic: {A} blockchain
  perspective,'' \emph{{IEEE} Trans. Netw. Sci. Eng.}, vol.~9, no.~1, pp.
  282--298, 2022. [Online]. Available:
  \url{https://doi.org/10.1109/TNSE.2020.3030925}
\BIBentrySTDinterwordspacing

\bibitem{DBLP:journals/iotj/AlansariBMAAA22}
\BIBentryALTinterwordspacing
S.~A. Alansari, M.~M. Badr, M.~M. E.~A. Mahmoud, W.~Alasmary, F.~Alsolami, and
  A.~M. Ali, ``Efficient and privacy-preserving infection control system for
  covid-19-like pandemics using blockchain,'' \emph{{IEEE} Internet Things J.},
  vol.~9, no.~4, pp. 2744--2760, 2022. [Online]. Available:
  \url{https://doi.org/10.1109/JIOT.2021.3092601}
\BIBentrySTDinterwordspacing

\bibitem{DBLP:journals/jicts/RotbiMG22}
\BIBentryALTinterwordspacing
M.~F. Rotbi, S.~Motahhir, and A.~E. Ghzizal, ``Blockchain technology for a safe
  and transparent covid-19 vaccination,'' \emph{J. {ICT} Stand.}, vol.~10,
  no.~2, pp. 125--144, 2022. [Online]. Available:
  \url{https://doi.org/10.13052/jicts2245-800X.1022}
\BIBentrySTDinterwordspacing

\bibitem{DBLP:journals/access/MusamihJSHYA21}
\BIBentryALTinterwordspacing
A.~Musamih, R.~Jayaraman, K.~Salah, H.~R. Hasan, I.~Yaqoob, and
  Y.~Al{-}Hammadi, ``Blockchain-based solution for distribution and delivery of
  {COVID-19} vaccines,'' \emph{{IEEE} Access}, vol.~9, pp. 71\,372--71\,387,
  2021. [Online]. Available: \url{https://doi.org/10.1109/ACCESS.2021.3079197}
\BIBentrySTDinterwordspacing

\bibitem{DBLP:journals/isci/LaxRF21}
\BIBentryALTinterwordspacing
G.~Lax, A.~Russo, and L.~S. Fasc{\`{\i}}, ``A blockchain-based approach for
  matching desired and real privacy settings of social network users,''
  \emph{Inf. Sci.}, vol. 557, pp. 220--235, 2021. [Online]. Available:
  \url{https://doi.org/10.1016/j.ins.2021.01.004}
\BIBentrySTDinterwordspacing

\bibitem{DBLP:journals/percom/Guidi20}
\BIBentryALTinterwordspacing
B.~Guidi, ``When blockchain meets online social networks,'' \emph{Pervasive
  Mob. Comput.}, vol.~62, p. 101131, 2020. [Online]. Available:
  \url{https://doi.org/10.1016/j.pmcj.2020.101131}
\BIBentrySTDinterwordspacing

\bibitem{DBLP:journals/tcss/JiangZ19}
\BIBentryALTinterwordspacing
L.~Jiang and X.~Zhang, ``{BCOSN:} {A} blockchain-based decentralized online
  social network,'' \emph{{IEEE} Trans. Comput. Soc. Syst.}, vol.~6, no.~6, pp.
  1454--1466, 2019. [Online]. Available:
  \url{https://doi.org/10.1109/TCSS.2019.2941650}
\BIBentrySTDinterwordspacing

\bibitem{DBLP:journals/connection/ZhangYSWLS21}
\BIBentryALTinterwordspacing
S.~Zhang, T.~Yao, V.~K.~A. Sandor, T.~Weng, W.~Liang, and J.~Su, ``A novel
  blockchain-based privacy-preserving framework for online social networks,''
  \emph{Connect. Sci.}, vol.~33, no.~3, pp. 555--575, 2021. [Online].
  Available: \url{https://doi.org/10.1080/09540091.2020.1854181}
\BIBentrySTDinterwordspacing

\bibitem{kiayias2018puff}
A.~Kiayias, B.~Livshits, A.~M. Mosteiro, and O.~S.~T. Litos, ``A puff of steem:
  Security analysis of decentralized content curation,'' \emph{arXiv preprint
  arXiv:1810.01719}, 2018.

\bibitem{DBLP:journals/access/GuWJ19}
\BIBentryALTinterwordspacing
K.~Gu, L.~Wang, and W.~Jia, ``Autonomous resource request transaction framework
  based on blockchain in social network,'' \emph{{IEEE} Access}, vol.~7, pp.
  43\,666--43\,678, 2019. [Online]. Available:
  \url{https://doi.org/10.1109/ACCESS.2019.2908627}
\BIBentrySTDinterwordspacing

\bibitem{DBLP:journals/jpdc/RahmanGB20}
\BIBentryALTinterwordspacing
M.~U. Rahman, B.~Guidi, and F.~Baiardi, ``Blockchain-based access control
  management for decentralized online social networks,'' \emph{J. Parallel
  Distributed Comput.}, vol. 144, pp. 41--54, 2020. [Online]. Available:
  \url{https://doi.org/10.1016/j.jpdc.2020.05.011}
\BIBentrySTDinterwordspacing

\bibitem{DBLP:conf/infocom/ZhangSLNLXZ22}
\BIBentryALTinterwordspacing
M.~Zhang, Z.~Sun, H.~Li, B.~Niu, F.~Li, Y.~Xie, and C.~Zheng, ``A
  blockchain-based privacy-preserving framework for cross-social network photo
  sharing,'' in \emph{{IEEE} {INFOCOM} 2022 - {IEEE} Conference on Computer
  Communications Workshops, {INFOCOM} 2022 - Workshops, New York, NY, USA, May
  2-5, 2022}.\hskip 1em plus 0.5em minus 0.4em\relax {IEEE}, 2022, pp. 1--6.
  [Online]. Available:
  \url{https://doi.org/10.1109/INFOCOMWKSHPS54753.2022.9798330}
\BIBentrySTDinterwordspacing

\bibitem{DBLP:journals/toit/YanPFY21}
\BIBentryALTinterwordspacing
Z.~Yan, L.~Peng, W.~Feng, and L.~T. Yang, ``Social-chain: Decentralized trust
  evaluation based on blockchain in pervasive social networking,'' \emph{{ACM}
  Trans. Internet Techn.}, vol.~21, no.~1, pp. 17:1--17:28, 2021. [Online].
  Available: \url{https://doi.org/10.1145/3419102}
\BIBentrySTDinterwordspacing

\bibitem{DBLP:conf/bsci/GuoXWZ22}
\BIBentryALTinterwordspacing
W.~Guo, J.~Xue, Y.~Wang, and Z.~Zhou, ``Blockchain-based reputation evaluation
  using game theory in social networking,'' in \emph{{BSCI} 2022: Proceedings
  of the 4th {ACM} International Symposium on Blockchain and Secure Critical
  Infrastructure, Nagasaki, Japan, May 30, 2022}, K.~Gai and K.~R. Choo,
  Eds.\hskip 1em plus 0.5em minus 0.4em\relax {ACM}, 2022, pp. 107--114.
  [Online]. Available: \url{https://doi.org/10.1145/3494106.3528681}
\BIBentrySTDinterwordspacing

\bibitem{DBLP:journals/isci/ChenXLWH19}
\BIBentryALTinterwordspacing
Y.~Chen, H.~Xie, K.~Lv, S.~Wei, and C.~Hu, ``{DEPLEST:} {A} blockchain-based
  privacy-preserving distributed database toward user behaviors in social
  networks,'' \emph{Inf. Sci.}, vol. 501, pp. 100--117, 2019. [Online].
  Available: \url{https://doi.org/10.1016/j.ins.2019.05.092}
\BIBentrySTDinterwordspacing

\bibitem{DBLP:conf/sigir/NguyenBAND22}
\BIBentryALTinterwordspacing
H.~H. Nguyen, D.~Bozhkov, Z.~Ahmadi, N.~Nguyen, and T.~Doan, ``Sochaindb: {A}
  database for storing and retrieving blockchain-powered social network data,''
  in \emph{{SIGIR} '22: The 45th International {ACM} {SIGIR} Conference on
  Research and Development in Information Retrieval, Madrid, Spain, July 11 -
  15, 2022}, E.~Amig{\'{o}}, P.~Castells, J.~Gonzalo, B.~Carterette, J.~S.
  Culpepper, and G.~Kazai, Eds.\hskip 1em plus 0.5em minus 0.4em\relax {ACM},
  2022, pp. 3036--3045. [Online]. Available:
  \url{https://doi.org/10.1145/3477495.3531735}
\BIBentrySTDinterwordspacing

\bibitem{DBLP:conf/quatic/OchoaMSGFL19}
\BIBentryALTinterwordspacing
I.~S. Och{\^{o}}a, G.~de~Mello, L.~A. Silva, A.~J.~P. Gomes, A.~M.~R.
  Fernandes, and V.~R.~Q. Leithardt, ``Fakechain: {A} blockchain architecture
  to ensure trust in social media networks,'' in \emph{Quality of Information
  and Communications Technology - 12th International Conference, {QUATIC} 2019,
  Ciudad Real, Spain, September 11-13, 2019, Proceedings}, ser. Communications
  in Computer and Information Science, M.~Piattini, P.~R. da~Cunha, I.~G.~R.
  de~Guzm{\'{a}}n, and R.~P{\'{e}}rez{-}Castillo, Eds., vol. 1010.\hskip 1em
  plus 0.5em minus 0.4em\relax Springer, 2019, pp. 105--118. [Online].
  Available: \url{https://doi.org/10.1007/978-3-030-29238-6\_8}
\BIBentrySTDinterwordspacing

\bibitem{lambert2017issues}
D.~M. Lambert and M.~G. Enz, ``Issues in supply chain management: Progress and
  potential,'' \emph{Industrial Marketing Management}, vol.~62, pp. 1--16,
  2017.

\bibitem{ivanov2019impact}
D.~Ivanov, A.~Dolgui, and B.~Sokolov, ``The impact of digital technology and
  industry 4.0 on the ripple effect and supply chain risk analytics,''
  \emph{International Journal of Production Research}, vol.~57, no.~3, pp.
  829--846, 2019.

\bibitem{DBLP:journals/itpro/KshetriL19}
\BIBentryALTinterwordspacing
N.~Kshetri and E.~Loukoianova, ``Blockchain adoption in supply chain networks
  in asia,'' \emph{{IT} Prof.}, vol.~21, no.~1, pp. 11--15, 2019. [Online].
  Available: \url{https://doi.org/10.1109/MITP.2018.2881307}
\BIBentrySTDinterwordspacing

\bibitem{Queiroz2021BlockchainAI}
M.~M. Queiroz, S.~F. Wamba, M.~de~Bourmont, and R.~Telles, ``Blockchain
  adoption in operations and supply chain management: empirical evidence from
  an emerging economy,'' \emph{International Journal of Production Research},
  vol.~59, pp. 6087 -- 6103, 2021.

\bibitem{Wu2021AnAO}
X.~Wu, Z.-P. Fan, and B.~Cao, ``An analysis of strategies for adopting
  blockchain technology in the fresh product supply chain,''
  \emph{International Journal of Production Research}, pp. 1--18, 2021.

\bibitem{Casino2021BlockchainbasedFS}
F.~Casino, V.~Kanakaris, T.~K. Dasaklis, S.~J. Moschuris, S.~Stachtiaris,
  M.~Pagoni, and N.~P. Rachaniotis, ``Blockchain-based food supply chain
  traceability: a case study in the dairy sector,'' \emph{International Journal
  of Production Research}, vol.~59, pp. 5758 -- 5770, 2021.

\bibitem{DBLP:journals/computers/SekuloskaE22}
\BIBentryALTinterwordspacing
J.~D. Sekuloska and A.~Erceg, ``Blockchain technology toward creating a smart
  local food supply chain,'' \emph{Comput.}, vol.~11, no.~6, p.~95, 2022.
  [Online]. Available: \url{https://doi.org/10.3390/computers11060095}
\BIBentrySTDinterwordspacing

\bibitem{marinello2017development}
F.~Marinello, M.~Atzori, L.~Lisi, D.~Boscaro, and A.~Pezzuolo, ``Development of
  a traceability system for the animal product supply chain based on blockchain
  technology,'' in \emph{Proceedings of the European Conference on Precision
  Livestock Farming (ECPLF), Nantes, France}, 2017, pp. 258--268.

\bibitem{kamble2020modeling}
S.~S. Kamble, A.~Gunasekaran, and R.~Sharma, ``Modeling the blockchain enabled
  traceability in agriculture supply chain,'' \emph{International Journal of
  Information Management}, vol.~52, p. 101967, 2020.

\bibitem{DBLP:journals/candie/OmarDJSOA22}
\BIBentryALTinterwordspacing
I.~A. Omar, M.~Debe, R.~Jayaraman, K.~Salah, M.~A. Omar, and J.~Arshad,
  ``Blockchain-based supply chain traceability for {COVID-19} personal
  protective equipment,'' \emph{Comput. Ind. Eng.}, vol. 167, p. 107995, 2022.
  [Online]. Available: \url{https://doi.org/10.1016/j.cie.2022.107995}
\BIBentrySTDinterwordspacing

\bibitem{DBLP:journals/peerj-cs/YakubuLYKM22}
\BIBentryALTinterwordspacing
B.~M. Yakubu, R.~Latif, A.~Yakubu, M.~I. Khan, and A.~I. Magashi, ``Ricechain:
  secure and traceable rice supply chain framework using blockchain
  technology,'' \emph{PeerJ Comput. Sci.}, vol.~8, p. e801, 2022. [Online].
  Available: \url{https://doi.org/10.7717/peerj-cs.801}
\BIBentrySTDinterwordspacing

\bibitem{Caro2018BlockchainbasedTI}
M.~P. Caro, M.~S. Ali, M.~Vecchio, and R.~Giaffreda, ``Blockchain-based
  traceability in agri-food supply chain management: A practical
  implementation,'' \emph{2018 IoT Vertical and Topical Summit on Agriculture -
  Tuscany (IOT Tuscany)}, pp. 1--4, 2018.

\bibitem{Dasaklis2019DefiningGL}
T.~K. Dasaklis, F.~Casino, and C.~Patsakis, ``Defining granularity levels for
  supply chain traceability based on iot and blockchain,'' \emph{Proceedings of
  the International Conference on Omni-Layer Intelligent Systems}, 2019.

\bibitem{Kouhizadeh2021BlockchainTA}
M.~Kouhizadeh, S.~Saberi, and J.~Sarkis, ``Blockchain technology and the
  sustainable supply chain: Theoretically exploring adoption barriers,''
  \emph{International Journal of Production Economics}, vol. 231, p. 107831,
  2021.

\bibitem{Saberi2019BlockchainTA}
S.~Saberi, M.~Kouhizadeh, J.~Sarkis, and L.~Shen, ``Blockchain technology and
  its relationships to sustainable supply chain management,''
  \emph{International Journal of Production Research}, vol.~57, pp. 2117 --
  2135, 2019.

\bibitem{Rana2021BlockchainTF}
R.~L. Rana, C.~Tricase, and L.~D. Cesare, ``Blockchain technology for a
  sustainable agri-food supply chain,'' \emph{British Food Journal}, 2021.

\bibitem{DBLP:conf/blockchain2/MalikDKJ19}
\BIBentryALTinterwordspacing
S.~Malik, V.~Dedeoglu, S.~S. Kanhere, and R.~Jurdak, ``Trustchain: Trust
  management in blockchain and iot supported supply chains,'' in \emph{{IEEE}
  International Conference on Blockchain, Blockchain 2019, Atlanta, GA, USA,
  July 14-17, 2019}.\hskip 1em plus 0.5em minus 0.4em\relax {IEEE}, 2019, pp.
  184--193. [Online]. Available:
  \url{https://doi.org/10.1109/Blockchain.2019.00032}
\BIBentrySTDinterwordspacing

\bibitem{AlRakhami2021ABT}
M.~S. Al-Rakhami and M.~A. Al-Mashari, ``A blockchain-based trust model for the
  internet of things supply chain management,'' \emph{Sensors (Basel,
  Switzerland)}, vol.~21, 2021.

\bibitem{Song2021ASS}
Q.~Song, Y.~Chen, Y.~Zhong, K.~Lan, S.~J. Fong, and R.~Tang, ``A supply-chain
  system framework based on internet of things using blockchain technology,''
  \emph{ACM Transactions on Internet Technology (TOIT)}, vol.~21, pp. 1 -- 24,
  2021.

\bibitem{Centobelli2021BlockchainTF}
P.~Centobelli, R.~Cerchione, P.~D. Vecchio, E.~Oropallo, and G.~Secundo,
  ``Blockchain technology for bridging trust, traceability and transparency in
  circular supply chain,'' \emph{Information \& Management}, 2021.

\bibitem{DBLP:journals/candie/RajJRP22}
\BIBentryALTinterwordspacing
P.~V. R.~P. Raj, S.~K. Jauhar, M.~Ramkumar, and S.~Pratap, ``Procurement,
  traceability and advance cash credit payment transactions in supply chain
  using blockchain smart contracts,'' \emph{Comput. Ind. Eng.}, vol. 167, p.
  108038, 2022. [Online]. Available:
  \url{https://doi.org/10.1016/j.cie.2022.108038}
\BIBentrySTDinterwordspacing

\bibitem{DBLP:journals/ecra/ZhouLZS22}
\BIBentryALTinterwordspacing
Z.~Zhou, X.~Liu, F.~Zhong, and J.~Shi, ``Improving the reliability of the
  information disclosure in supply chain based on blockchain technology,''
  \emph{Electron. Commer. Res. Appl.}, vol.~52, p. 101121, 2022. [Online].
  Available: \url{https://doi.org/10.1016/j.elerap.2022.101121}
\BIBentrySTDinterwordspacing

\bibitem{DBLP:journals/apjor/SunXS22}
\BIBentryALTinterwordspacing
Z.~Sun, Q.~Xu, and B.~Shi, ``Price and product quality decisions for a
  two-echelon supply chain in the blockchain era,'' \emph{Asia Pac. J. Oper.
  Res.}, vol.~39, no.~1, pp. 2\,140\,016:1--2\,140\,016:31, 2022. [Online].
  Available: \url{https://doi.org/10.1142/S0217595921400169}
\BIBentrySTDinterwordspacing

\end{thebibliography}

\begin{IEEEbiography}[{\includegraphics[width=1in,height=1.25in,clip,keepaspectratio]{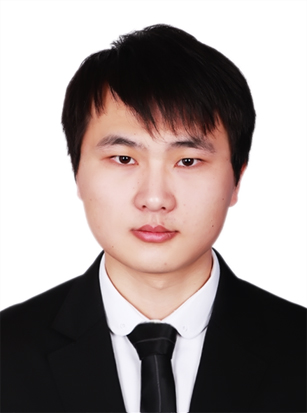}}]{Xiao Li}
received his B.S. and M.S degree in Software Engineering from
Dalian University of Technology, China in 2016 and 2019, respectively. He is
currently pursuing the Ph.D. degree with the Department of Computer Science,
University of Texas at Dallas, Richardson, TX, USA. His current research
interests include data mining and Blockchain.
\end{IEEEbiography}

\begin{IEEEbiography}[{\includegraphics[width=1in,height=1.25in,clip,keepaspectratio]{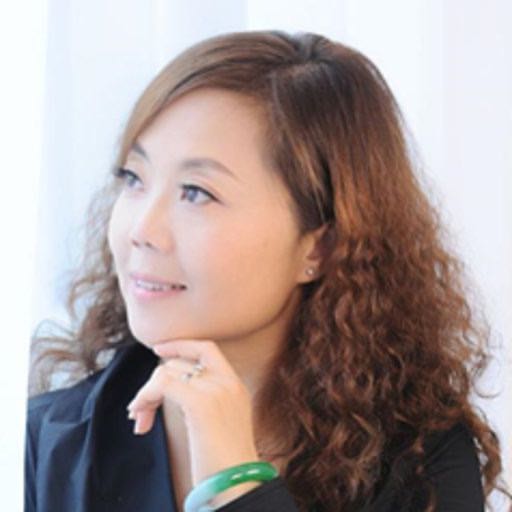}}]{Weili Wu}
(Senior Member, IEEE) received the M.S.
and Ph.D. degrees from the Department of Computer
Science, University of Minnesota, Minneapolis, MN,
USA, in 1998 and 2002, respectively.
She is currently a Full Professor with the
Department of Computer Science, University of
Texas at Dallas, Richardson, TX, USA. Her research
mainly deals in the general research area of data
communication and data management. Her research
focuses on the design and analysis of algorithms
for optimization problems that occur in wireless
networking environments and various database systems.
\end{IEEEbiography}




\end{document}